\documentclass{jfm}
\usepackage{graphicx}
\usepackage{epstopdf}
\usepackage{amssymb}
\usepackage{enumerate}
\usepackage{amsmath}
\usepackage{psfrag}
\usepackage{amsbsy}
\usepackage{amsfonts}
\usepackage{setspace}
\usepackage[usenames, dvipsnames]{color}
\usepackage{float}
\usepackage{tabularx}
\usepackage{epsfig}
\usepackage{bm}
\usepackage{booktabs,array,dcolumn}
\usepackage{arydshln}
\usepackage{dashrule}
\usepackage{wasysym}
\usepackage{multicol}
\usepackage{multirow}
\usepackage{ulem}
\usepackage{nomencl}
\usepackage{multicol}
\usepackage{amsmath}
\usepackage{subfig}
\usepackage{epstopdf}
\usepackage{color, soul}
\usepackage{framed}
\usepackage{nomencl}
\makenomenclature
\usepackage{longtable,tabularx}
\sethlcolor{SkyBlue}

\newcommand{\reffig}[1]{figure \ref{#1}}
\newcommand{\reffigs}[1]{figures \ref{#1}}
\newcommand{\refFig}[1]{Figure \ref{#1}}

\newcommand{\reftab}[1]{table \ref{#1}}

\newcommand{\refeq}[1]{equation \eqref{#1}}
\newcommand{\refeqs}[1]{equations \eqref{#1}}

\newcommand{\refse}[1]{section \ref{#1}}
\graphicspath{{figures/}}
\begin{document}
	
\newtheorem{lemma}{Lemma}
\newtheorem{corollary}{Corollary}

\shorttitle{Propulsive transitions and scaling relations of heaving foil} %for header on odd pages
\shortauthor{G. Li, L. Wang, W. Jiang, H. Liu and R. K. Jaiman} %for header on even pages

\title{Propulsive transitions and scaling relations of a heaving flexible foil in a cylinder wake}
\author
{
	Guojun Li\aff{1,2},
	Lanlan Wang\aff{1,2}
	\corresp{\email{lanlan1900@mail.xjtu.edu.cn}},
	Weitao Jiang\aff{1,2},
	Hongzhong Liu\aff{1,2},
	\and 
	Rajeev Kumar Jaiman\aff{3}
}

\affiliation
{
	\aff{1}
	School of Mechanical Engineering, Xi'an Jiaotong University, Xi'an, Shaanxi, China 710049
	\aff{2}
	State Key Laboratory for Manufacturing Systems Engineering, Xi'an Jiaotong University, Xi'an, Shaanxi, China 710049
	\aff{3}
	Department of Mechanical Engineering, University of British Columbia, Vancouver, BC  Canada V6T 1Z4
}

\maketitle

\begin{abstract}
We numerically investigate the propulsive dynamics of a heaving flexible foil immersed in the wake of a stationary circular cylinder, focusing on the coupled effects of unsteady wake forcing, passive structural flexibility, and prescribed heaving kinematics. The analysis employs a high-fidelity fluid–structure interaction solver based on a partitioned variational formulation with a nonlinear iterative force correction scheme. Systematic simulations are conducted over a broad parameter space of dimensionless heaving amplitude $A^* \in [0.05,0.5]$ and frequency $f^* \in [0.1,0.6]$ at a Reynolds number of 3000. 
Five distinct response modes are identified, namely full-wake, semi-wake, full-wake-flexible, semi-wake-flexible, and vortex-flexible, based on propulsive transitions and associated flow features. 
An empirical boundary plane $A^*$-$f^*$ is discovered, separating regimes where the wake hinders lift performance (wake-dominated) from those where it enhances performance (flapping-dominated). 
Scaling relations for the force and power coefficients are formulated by decomposing the contributions of quasi-steady motion, added-mass effects, structural curvature, wake momentum deficit, and transverse flow gradients.
At sufficiently large $A^* f^*$, a two-way lock-in emerges: the foil not only synchronizes with the cylinder shedding but also modulates it, accelerating the wake and enhancing lift.
 Flexibility is found to be detrimental in fully immersed wakes but beneficial in partial wakes, where it creates extra suction without much extra drag in the semi-wake-flexible mode. 
 These findings elucidate the energy-saving and maneuverability strategies employed by biological propulsors and provide predictive guidelines for the design of bio-inspired energy harvesters and unmanned vehicles in disturbed flows.

\end{abstract}

\section{Introduction} \label{sec1:introduction}
\begin{figure}
	\centering
    \includegraphics[width=0.6 \textwidth]{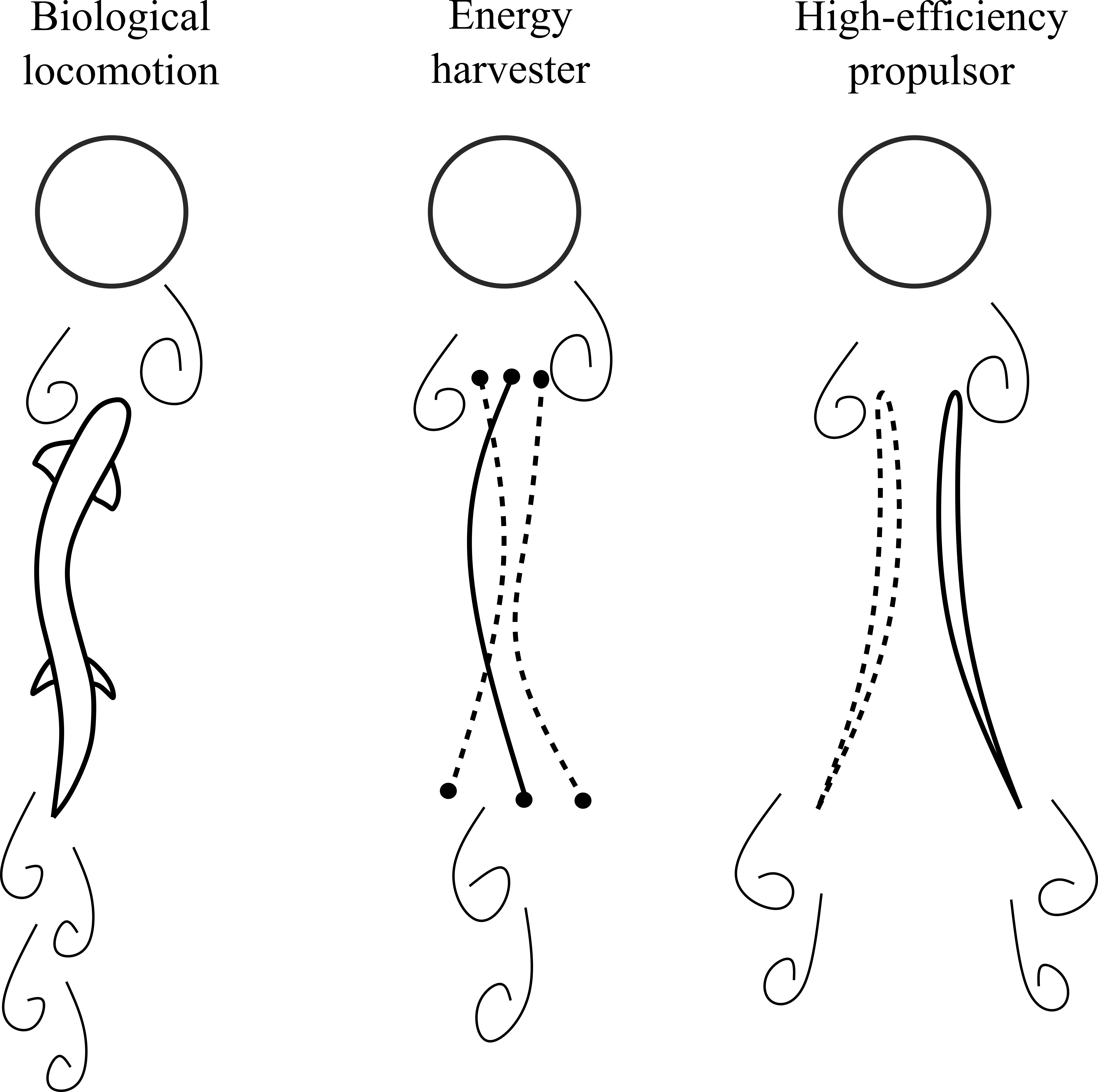}
	\caption{\label{application} Conceptual schematic of a heaving flexible foil immersed in disturbed wake flows. The sketch shows a canonical cylinder–wake forcing case and representative applications including fish exploiting von Kármán streets, flexible energy harvesters, autonomous underwater vehicle propulsors, and wind-assisted sails.}
\end{figure}

% Background
Flapping bodies in nature and engineering generate forces by interacting with unsteady flows. They can extract momentum from ambient disturbances, such as wakes, shear layers, and vortices, to improve propulsion or energy-harvesting efficiency. In these situations, wake flows may impose detrimental velocity deficits and adverse pressures, or they may be exploited to reduce energetic costs and increase maneuverability \citep{fish2006passive}. Similarly, engineered devices such as bio-inspired propulsors, flow-control surfaces, and renewable energy harvesters must operate in disturbed inflows behind struts, within arrays, or near boundaries \citep{zhang2025performance}. However, a clear understanding of how passive structural flexibility is coupled with prescribed heaving kinematics to modulate performance in such flows remains incomplete. \refFig{application} illustrates representative disturbed-wake scenarios, including fish exploiting von Kármán streets, flexible energy harvesters, and autonomous underwater vehicle propulsors. In this study, we investigate the dynamics of a flexible heaving foil immersed in the wake of a stationary circular cylinder. We aim to map performance regimes, reveal two-way lock-in between foil motion and upstream shedding, and derive scaling laws for lift, drag, and power, providing guidelines for energy-efficient bio-inspired and industrial designs.

% biological flying or swimming in disturbed wake flows
In biological locomotion, wake–flexible body interactions are exemplified by fish schooling and bird flocking. Individuals exploit disturbed wakes generated by upstream companions to gain energetic advantages. In fish schools, followers align to harness reverse von Kármán vortices and turbulent structures shed by leaders. This organization reduces tail-beat energy expenditure by up to 56 \% through flow refuging, hydrodynamic drafting, and collective attenuation of turbulence \citep{zhang2024energy,hedrick2024swimming}.
Similarly, birds in V-formations or other orderly flocks draft in the upwash of wingtip vortices from preceding individuals. These particular formations lead to self-organized phase-locking that enhances aerodynamic efficiency and lowers flight costs during migration \citep{newbolt2024flow,ramananarivo2016flow}. Rather than being hindered, animals often exploit complex wakes to conserve energy over long distances or to increase maneuverability for rapid escapes. These behaviors suggest active and optimal synchronization of flexible propulsors with oncoming vortical structures. However, observational and experimental studies have only begun to illuminate these collective behaviors. The precise mechanisms of fluid-structure interaction remain underexplored, especially for flexible foils in controlled wake environments. Systematic numerical investigations are needed to bridge phenomenological insights with predictive models.

% Flapping thin flexible foils, especially for flexible foil and flapping foil
Inspired by biological propulsion, extensive theoretical, experimental and computational studies have examined flexible flapping foils to uncover mechanisms for enhanced thrust and efficiency in unsteady disturbed flows. Compliant foils, unlike rigid or inextensible ones, can self-camber and rotate their edges through extensibility, reorganizing the vortex–pressure field to yield gains in lift, thrust and efficiency \citep{tiomkin2021review,gehrke2025highly,mavroyiakoumou2025sail}. Analytical models coupling thin-airfoil theory with membrane dynamics predict a thrust-to-drag transition near resonance \citep{tzezana2019thrust}. Increased compliance shifts this transition to lower frequencies, with the wake switching from a reverse to a classical von Kármán street lagging the force sign change. Unsteady aerodynamic theories for extensible foils under harmonic motions or gusts extend classical lift functions \citep{tiomkin2022unsteady}, predicting slower initial lift but higher asymptotic lift due to camber induction. Performance gains peak near resonance and are governed by tension, whereas low frequencies yield reduced amplitudes and increased phase lag. These findings motivate the investigation of more realistic articulated morphologies, where segment-level deformation can realize the predicted mechanisms.

% Provide a better connection to this paragraph
Building on these predictions, hovering and forward-flight analogues with bat-inspired articulated foils probe segment-specific deformations that retune leading-edge vortex (LEV) and tip-vortex dynamics, with vortex-induced pressure emerging as the dominant load pathway \citep{kumar2025computational}. This explains why articulation enhances efficiency. Across studies, peak performance arises from moderate camber at optimal aeroelastic numbers and favourable edge alignment, while over-cambering or excessive compliance degrades it, yielding actionable targets for stiffness, pretension and kinematics. Numerical studies of heaving–pitching foils show that optimal stretching coefficients can increase thrust by up to 200 \% and efficiency by 100 \% relative to rigid foils, typically below resonance \citep{zhang2025flapping}. These gains result from concave deformations that weaken detrimental LEVs while avoiding excessive drag. Experiments further demonstrate that at high compliance the LEV is replaced by a bound shear layer, with geometric edge angles mapping the associated topology and performance transitions \citep{gehrke2025highly}. Collectively, these findings indicate that performance depends not only on kinematics but also on resonance, material properties and spatially varying deformation.

% Role of structural flexibility in wake flow and performance/mode transition, frequency lock-in
The interplay between active kinematics and passive flexibility is central to how animals conserve energy or enhance maneuverability in disturbed flows, such as wakes shed by leaders or obstacles. A canonical analogue is a flexible foil immersed in the wake of a bluff body, where periodic vortex shedding imposes unsteady hydrodynamic loads. While the flow past rigid bluff bodies is well characterized, the response of compliant structures to such wake forcing remains less understood. Flexibility fundamentally alters energy-transfer pathways, leading to large-amplitude oscillations, mode transitions and frequency lock-in \citep{furquan2021multiple}. Studies of flexible flags and filaments downstream of cylinders show that compliance can induce constructive resonance and amplify oscillations \citep{ristroph2008anomalous,michelin2008vortex}, but it can also suppress vortex-induced vibration (VIV) depending on mass ratio and bending rigidity \citep{banerjee2015three}. Mode transitions depend on the relationship among the natural frequency, the wake-shedding frequency, and any imposed heaving motion. Active heaving further complicates this interaction, potentially yielding a two-way lock-in: the foil may synchronize with the incoming wake while simultaneously modulating the bluff-body shedding. This feedback has received limited attention. Previous studies indicate that wakes can suppress or enhance force production, producing sharp performance transitions, yet a regime-level framework that unifies wake exploitation, lock-in, and flexibility remains lacking. This motivates a systematic broad-parameter study to unravel the coupled roles of heaving kinematics, compliance, and wake forcing.

% scaling law for flapping structure, coupling mechanism in wake flows
Scaling laws derived from the first principles are central to predicting flapping performance. For rigid foils in uniform flow, thrust and power scale with the Strouhal number and effective angle of attack \citep{floryan2017scaling,gazzola2014scaling}. For flexible foils in isolation, these relations extend to incorporate the coupling between kinematics and structural dynamics, often expressed through dimensionless groups such as effective stiffness and mass ratio \citep{moore2014analytical,mavroyiakoumou2022membrane,mavroyiakoumou2020large}. \cite{zhang2025flapping} proposed a modified scaling for compliant foils in uniform flow by incorporating passive deformation into the pitching amplitude. This framework captures thrust and power growth and the drag-to-thrust wake transition across heave/pitch kinematics, with the Strouhal number and effective angle of attack governing the onset of a reverse von Kármán wake and peak efficiency. In the wake of a bluff body, the inflow is unsteady and modulated by the Kármán street. Wake–body coupling arises when the flapping or natural frequency synchronizes with the shedding frequency, and is further reinforced when the convective phase of the incoming vortices is favorable.

% Let's separate into two paragraphs
When the flapping dynamics synchronizes with the shedding frequency, a downstream flexible foil exhibits distinct small-, moderate-, and large-amplitude modes. The local vortex packing reorganizes and thrust, drag, and energetic cost change sharply. Outside the lock-in window, the foil reverts to self-excited large-amplitude oscillations \citep{shao2011hydrodynamics,parekh2025wake}. Consistent with animal Kármán-gait swimming, experiments and simulations show thrust enhancement or reduced actuation cost when the flapper weaves between, or phase-locks to, incoming vortices. 
Propulsive performance data collapse onto a single curve when the Strouhal number is defined using the local wake velocity and augmented by the ratio of flapping to shedding frequency \citep{wang2024thrust,han2025self}. These results suggest the need for a unified scaling that incorporates wake-momentum deficit, transverse momentum supplied by flapping, passive deformation, and frequency ratio. Such a framework is essential to rationalize efficiency gains and mode transitions of flexible foils in bluff-body wakes, moving beyond phenomenological observation toward predictive design of bio-inspired systems in complex flow environments.

% research gap and questions
In this study, we numerically investigate the fluid–structure interaction of a thin flexible foil undergoing prescribed harmonic heaving in the unsteady wake of a stationary circular cylinder. This configuration serves as a canonical model for wake-modulated propulsion in biological and engineered systems. The problem is solved using a high-fidelity FSI framework based on a partitioned variational formulation with a nonlinear iterative force-correction scheme.
We perform a systematic simulation over heaving amplitude $A^* \in [0.05,0.5]$ and frequency $f^* \in [0.1,0.6]$ at $Re$=3000. This design targets a clear gap: uniform-flow scaling for rigid and flexible flappers is well established \citep{gazzola2014scaling,floryan2017scaling,floryan2018clarifying,quinn2014unsteady,moore2014analytical}, and wake-driven responses behind bluff bodies have mapped synchronization windows for flags and filaments \citep{ristroph2008anomalous,michelin2008vortex,banerjee2015three}. Yet we still lack a regime-level, predictive framework for thin foils in wakes that (i) distinguishes when wake deficit hinders versus enhances performance; (ii) quantifies one- and two-way lock-in between shedding and prescribed heave; and (iii) collapses mean lift/drag/power across isolated and disturbed inflows using minimal physics-based terms that include passive curvature and wake-gradient effects. 

% Key questions 
To close this gap, we pose two guiding questions: How does the cylinder wake reshape foil dynamics and performance, and under what conditions does flexibility enhance rather than degrade it? We answer these questions in four steps. First, we extend Fourier-mode decomposition to a body-moving frame to construct a regime map with five response modes, revealing a sharp boundary between wake-dominated and flapping-dominated performance. Second, we quantify the two-way lock-in at sufficiently large $A^* f^*$, showing that the prescribed heave can reorganize the cylinder shedding and increase the mean lift. Third, we derive unified mechanistic scalings that decompose forces into quasi-steady, added-mass, curvature (passive deformation), and wake terms, enabling data collapse of mean lift/drag/power across uniform and wake-modulated inflows. Fourth, we isolate the role of flexibility by comparing it with a rigid counterpart and by quantifying the curvature contribution within the scaling. Together, these results convert phenomenology into prediction and yield design guidelines for bio-inspired propulsors, energy harvesters, and unmanned vehicles in complex wakes.

% organization
The organization of this manuscript is as follows. The high-fidelity computational framework for the fluid-structure interaction is described in §\ref{sec2:methodology}. The problem setup, including the geometrical configuration, boundary conditions, and dimensionless parameters, is detailed in §\ref{sec3:problem}. The detailed results and discussion are presented in\ref{sec4:results}, beginning with the identification of five distinct flapping modes and a performance phase diagram (§\ref{sec4.1:mode}). This is followed by an analysis of the coupled dynamics and flow features (§\ref{sec4.2:coupled}), the derivation of unified scaling relations for lift, drag, and power coefficients (§\ref{sec4.3:scaling}), and a detailed investigation into the effect of wake flows (§\ref{sec4.4:wake}) and the role of structural flexibility (§\ref{sec4.5:flexible}) on performance. The propulsive aerodynamic performance and its biological implications are discussed in §\ref{sec4.6:performance}. Finally, the key conclusions drawn from this study are summarized in §\ref{sec5:conclusion}.

\section{Computational FSI framework} \label{sec2:methodology}
Flapping flexible foils in unsteady flow are modeled by the incompressible Navier–Stokes equations coupled to flexible multibody structural dynamics. The fluid is discretized using a stabilized Petrov–Galerkin finite element method in an arbitrary Lagrangian–Eulerian (ALE) frame, with temporal integration via the generalized-$\alpha$ scheme. The structure is discretized by a finite-element multibody formulation. A partitioned fluid–structure interaction strategy with nonlinear iterative force correction (NIFC) is employed to mitigate nonphysical added-mass effects and ensure stability. Data transfer across the non-matching fluid–structure interface is performed with radial basis function (RBF) mapping for both kinematics and mesh motion. Each time step proceeds through a predictor–corrector sub-iteration: (i) predict structural displacement and velocity, (ii) transfer interface motion and update the ALE mesh via RBF, (iii) solve the fluid with the current interface, and (iv) correct interface tractions via NIFC. Sub-iterations repeat until convergence. The complete weak variational forms, the numerical discretization details and the partitioned coupling algorithm are given in Appendix A and follow our validated solver framework \citep{li2019novel}. Verification and validation against experiments for 2D flexible foils and 3D multi-segment wings are reported in \cite{li2021flow} and \cite{li2023unsteady}, respectively.

\section{Problem description} \label{sec3:problem}
\begin{figure}
	\centering
	\includegraphics[width=0.9 \textwidth]{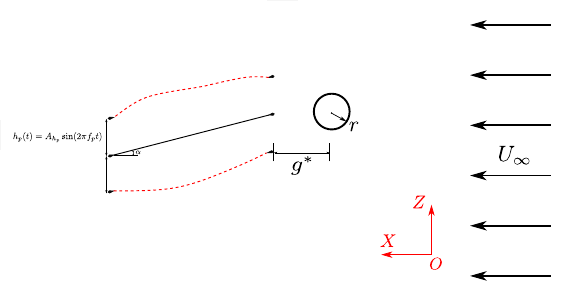}
	\caption{\label{problem} Schematic of a heaving flexible foil immersed in disturbed flows produced by upstream circular cylinder. The heaving structure in the black solid line indicates the initial configuration, while the structure in the red dashed line represents the transient heaving geometry.}
\end{figure}

A thin flexible foil is placed downstream of a stationary circular cylinder with a gap ratio of $g^*$ to model wake-foil interactions. A harmonic heaving motion along the vertical direction is applied at the leading and trailing edges of the flexible foil. This heaving motion has an amplitude of $A_{h_p}$ and a frequency of $f_p$. 
This tandem cylinder–foil configuration represents canonical scenarios such as fish or bats moving behind obstacles and sails interacting in arrays.
A schematic of the wake-heaving flexible foil interaction is illustrated in \reffig{problem}. Uniform incoming fluid flows with a velocity of $U_{\infty}$ are applied for the tandem configuration. The flexible foil is immersed in the wake flows of the upstream cylinder with an angle of attack of $\alpha$. This flexible foil is stretched by the rigid mounts at both ends. The radius of the stationary circular cylinder is set to $r$. The flexible foil has a chord length of $c/r$=4 and a thickness of $h/r$=0.08. The gap ratio between the cylinder and the flexible foil is given as $g^*$=2, where vortices shed from the upstream cylinder can fully interact with the flexible foil. 

The coupled dynamics of the tandem cylinder-heaving flexible foil are simulated by the developed FSI solver. A computational fluid domain of the tandem configuration is built to model the wake-body interaction, which is solved by the Navier-Stokes equations. The computational fluid domain has a size of $40c \times 40c$. We apply uniform fluid flows $U_{\infty}$ along the inlet boundary $\Gamma_{\text{in}}$. A traction-free boundary condition is employed at the outlet boundary $\Gamma_{\text{out}}$. The no-slip wall boundary condition is considered for the surfaces of the cylinder and the flexible foil. The slip wall boundary condition is applied on the top and bottom walls of the computational domain. The computational fluid domain is discretized by unstructured triangular elements. The stretching ratio and the height of the first layer of the finite element mesh are set to 1.15 and $2 \times 10^{-5}$ to ensure the dimensionless wall distance $y^+<$1. 
The structural domain is discretized with structured four-node rectangular elements: the cylinder and rigid mounts are modeled as rigid bodies, while the foil is represented using geometrically exact co-rotational thin-shell elements. The rigid mounts undergo prescribed harmonic heaving enforced via Dirichlet boundary conditions. The nondimensional parameters governing the coupled dynamics of the flexible foil are defined as follows:
\begin{equation}
	Re=\frac{\rho^f U_{\infty} c}{\mu}=3000, \quad \quad m^*=\frac{\rho^s h}{\rho_f c}=25.5, \quad \quad Ae=\frac{E^s h}{0.5 \rho^f U_{\infty}^2 c} =8163 ,
\end{equation}
where $Re$, $m^*$ and $Ae$ denote the Reynolds number, mass ratio and aeroelastic number, respectively. The kinematics of the flexible heaving structure is governed by the dimensionless heaving frequency $f^*=\frac{ f_p c}{U_{\infty}}$ and the amplitude $A^*=\frac{A_{h_p}}{c}$. The lift and drag coefficients are evaluated by integrating the fluid stress tensor $\boldsymbol{\sigma}^f$ into the first layer of the fluid mesh, which are given as
\begin{equation}
	C_L=\frac{1}{0.5 \rho_f U_{\infty}^2 c} \int_{\Gamma^f} (\boldsymbol{\sigma}^f \cdot \boldsymbol{e}) \cdot \boldsymbol{e}_z \text{d} \Gamma, \quad \quad C_D=\frac{1}{0.5 \rho_f U_{\infty}^2 c} \int_{\Gamma^f} (\boldsymbol{\sigma}^f \cdot \boldsymbol{e}) \cdot \boldsymbol{e}_x \text{d} \Gamma ,
\end{equation}

Before examining the wake-flow effects, the FSI solver is validated against two benchmark cases, as detailed in Appendix B. First, the coupled dynamics of a flexible plate with prescribed flapping at the leading edge are simulated and compared with experimental measurements. The other simulation case is a thin flexible plate attached to a circular cylinder, namely the Turek $\&$ Hron case. Our FSI solver can reasonably predict the coupled dynamics of the flapping flexible foil. A convergence study for the tandem cylinder–heaving foil configuration is presented in Appendix C to determine suitable mesh resolutions for subsequent simulations.

\section{Results and discussion} \label{sec4:results}
We consider a flexible foil configuration as a canonical model for the coupled dynamics of animal locomotion, energy harvesters, and stretched sails in disturbed wakes. A series of wake–foil interaction cases is simulated over a parameter space of heaving amplitudes and frequencies using the developed FSI solver. The dimensionless heaving amplitude ranges from 0.05 to 0.5, which covers the full-wake to semi-wake conditions. The dimensionless heaving frequency is selected within $f^* \in [0.1, 0.6]$, which reflects maneuverability situations. The Reynolds number is fixed at a typical value of 3000, representing common efficient flow environments ($Re= \mathcal{O} (10^3) $) for small-scale fish/bat locomotion \citep{triantafyllou1995efficient,liu2024vortices} and energy harvesters \citep{huang2021fluid,farooq2022numerical}. 

To isolate the effects of wake forcing and structural compliance, two reference configurations are considered: a tandem cylinder–heaving rigid foil and an isolated heaving flexible foil. Flapping modes, flow features, and coupled dynamics of flexible foils in wake flows are analyzed in the parameter space of $f^*$--$A^*$. Unified scaling relations of the lift, drag, and power coefficients for the three examined configurations are derived to predict the performance dynamics. Wake flow impacts on the coupled dynamics of the flexible foil are studied by comparing against the isolated heaving structure. The role of flexibility is investigated for the flexible foil and its rigid counterpart. The optimal kinematics of the heaving motion is suggested for the flexible foil to achieve better performance in disturbed flows.

\subsection{Flapping modes of flexible foil in wake flows} \label{sec4.1:mode}
\begin{figure}
	\centering
	\includegraphics[width=1.0 \textwidth]{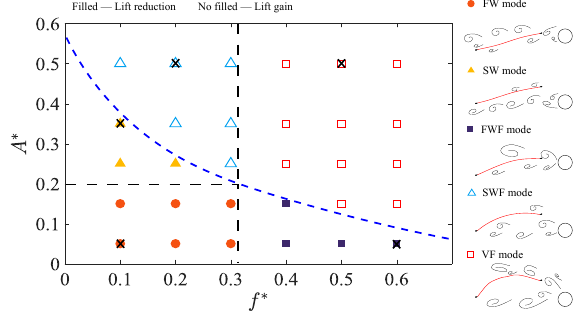}
	\caption{\label{mode} Mode phase diagram of the heaving flexible foil in the $f^*$--$A^*$ parameter space. The blue boundary line corresponding to lift change is expressed as $A^*=\gamma (f^*)^{-\frac{1}{n}}+\delta$. The lift change is defined as the lift coefficient of the heaving flexible foil behind a cylinder relative to its rigid counterpart.}

\end{figure}

The coupled dynamics of the flexible foil in a disturbed wake is based on heaving kinematics to adjust its propulsive performance. Five distinct flapping modes, namely (i) full-wake (FW) mode, (ii) semi-wake (SW) mode, (iii) full-wake-flexible (FWF) mode, (iv) semi-wake-flexible (SWF) mode, and (v) vortex-flexible (VF) mode, are classified by its propulsive performance and dominant flow features as a criterion. \refFig{mode} illustrates the mode phase diagram of the heaving flexible foil in the $f^*$--$A^*$ parameter space. Five distinct flapping modes are represented by labels with different shapes and colors. Compared to the rigid heaving structure behind a cylinder, the mode phase diagram can be divided into two regimes: the lift reduction regime represented by filled labels and the lift gain regime indicated by no-filled shapes. The blue boundary line that separates lift variations can be expressed as a power function of heaving amplitude and frequency, which is given as
\begin{equation}
	A^*=\gamma (f^*)^{-\frac{1}{n}}+\delta ,
\end{equation}
where $\gamma$, $n$ and $\delta$ are fitting coefficients. This boundary line as a threshold criterion indicates a transition from wake-dominated to flapping-dominated modes when the transverse-induced velocity from heave compensates for the axial momentum deficit of the incoming wake over the leading‑edge control volume. Thus, the flexible heaving structure performs better lift performance in wake flows at higher heaving amplitudes and frequencies than its rigid counterpart. The underpinning mechanisms for producing better propulsive performance are explained in the following sections. 

\begin{figure}
	\centering
	\subfloat[]{\includegraphics[width=1.0 \textwidth]{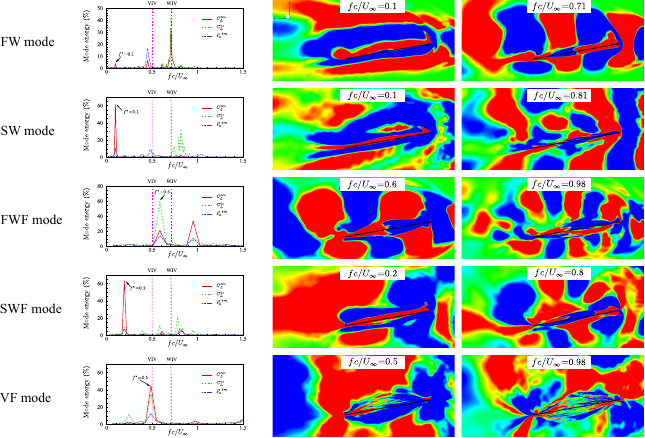}\label{fre_modea}}
	\\
	\subfloat[]{\includegraphics[width=0.8 \textwidth]{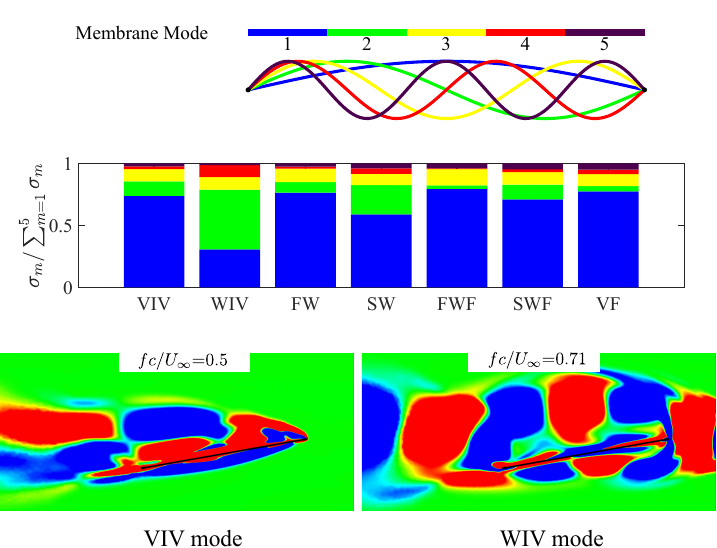}\label{fre_modeb}}
	\caption{\label{fre_mode} Illustrations of (a) frequency characteristics and dominant FMD $y$-vorticity modes of the heaving motion and cylinder wake flows in the flapping-body reference frame corresponding to five distinct modes. (b) mode contributions of the first five-order foil vibration modes for five distinct, VIV and WIV modes.}
\end{figure}

% FMD analysis
Flow characteristics associated with the five response modes are summarized from instantaneous $y$-vorticity fields and schematically illustrated in the right panel of \reffig{mode}. The global Fourier mode decomposition (FMD) method is employed to extract frequency spectra and dominant flow modes of the coupled wake–foil system.  \refFig{fre_mode} \subref{fre_modea} presents the frequency characteristics of the lift performance for the tandem foil structure $C^{tm}_L$ and cylinder $C^{tc}_L$ together with the vibration fluctuations of the tandem foil ${\delta^{\prime}_n}^{tm}$ for five selected flapping modes, which are labeled by black cross lines in \reffig{mode}. There are three factors that can cause vibrations for the tandem flexible foil, namely vortex-induced (VIV), wake-induced (WIV), and heaving-induced vibrations. Two additional cases, the so-called isolated and tandem flexible foils without heaving motion, are simulated to examine the VIV mode excited by shed vortices of the flexible foil itself, and the WIV mode induced by disturbed wake flows of the upstream cylinder. The dominant dimensionless frequencies of the VIV and WIV modes are superimposed on the spectra in \reffig{fre_mode}\subref{fre_modea} as magenta and black dashed lines, respectively.

% FMD details
In the presence of heaving motion, conventional Eulerian-frame decomposition methods are ill-posed for extracting dominant fluid modes near the body. Grid points that lie inside the body at one instant re-enter the fluid at another, so the snapshot sequence is no longer autonomous. This injects geometry artifacts (ghost body outlines, spurious shear layers tied to the changing boundary, and contamination from rigid-body translation/rotation) into the computed decomposition modes and obscures near-body physics where the coupling actually occurs. 
To address this, recent studies have proposed proper orthogonal decomposition and dynamic mode decomposition in body-following reference frames from a Lagrangian perspective \citep{menon2020dynamic,shinde2021lagrangian}.
 Following this coordinate-transformation approach, we extend the global Fourier mode decomposition method from our earlier work \citep{li2022aeroelastic} to extract fluid modes in the heaving-body reference frame. Specifically, we build a diffeomorphic mapping that transports each Eulerian snapshot $\boldsymbol{x}^f$ to a foil-following coordinate system $\boldsymbol{x}_l^f$. Similar to the conventional FMD method, the fluctuation flow field $\boldsymbol{y}^{\prime}_n (\boldsymbol{x}_l^f,t^n)$ is then decomposed into the reference frame of the heaving body. The decomposed spatial fluid mode $\boldsymbol{c}_k$ at a specific discrete frequency $f_k$ is expressed as
\begin{equation}
	\boldsymbol{c}_k = \mathcal{F}(\boldsymbol{y}^{\prime}_n (\boldsymbol{x}_l^f,t^n) )  = \sum_{n=0}^{N-1} \boldsymbol{y}^{\prime}_n (\boldsymbol{x}_l^f,t^n) e^{-i \frac{2 \pi k}{N}n},
	\label{FMD5}
\end{equation}
where $N$ represents the number of snapshots. The proposed FMD method in the reference frame of the heaving body yields objective geometry-consistent fluid modes that isolate coherent fluid structures and wake–foil coupling, without contamination from the changing boundary or from global translation/rotation. Practically, this eliminates remeshing artifacts and domain-occupancy flips, sharpens the spectral content, and provides compact, interpretable modal structures tied to the physics that drive force production. Dominant FMD modes related to heaving motion and disturbed wake flows are extracted from time-varying $y$-vorticity fields in the reference frame of the heaving body, which are presented in \reffig{fre_mode}. 

As shown in \reffig{fre_mode} \subref{fre_modea}, the lift performance of the upstream cylinder and the downstream foil in the FW mode exhibits dominant frequency peaks near the vortex-shedding frequency of the WIV mode. The interaction between the flexible foil and the vortices shed from itself is also evident in the foil vibration frequency spectrum, which is similar to that of the VIV mode. The downstream flexible foil is fully immersed in disturbed wake flows with $fc/U_{\infty}$=0.71 observed from the FMD fluid mode, thus exciting a synchronized wake-induced vibration. The flow pattern associated with wake flows in the FW mode is similar to that of the WIV mode shown in \reffig{fre_mode} \subref{fre_modeb}. The heaving motion with $fc/U_{\infty}$=0.1 causes entire vortices around the entire foil, which induces a sub-frequency peak in the lift spectrum. 

When the flexible foil heaves outside the wake flow region, the heaving motion outweighs the wake effect and primarily governs the coupled fluid-foil dynamics in the SW mode. It can be seen from the frequency spectra that the dominant frequency peaks of the lift and foil vibration transition from the wake-dominated mode to the flapping-dominated mode. The flow pattern associated with the heaving motion in the SW mode is similar to that in the FW mode. The dominant vortex-shedding frequency of the upstream cylinder becomes slightly higher than the WIV mode. As observed from the extracted flow pattern with $fc/U_{\infty}$=0.81, the heaving motion with higher amplitudes is interacted with wake flows, enhancing the vortex-shedding frequency and extending wake regions.

Although the heaving frequency gradually approaches the vortex-shedding frequency, the coupled dynamics in the FWF mode remain primarily governed by the wake, as the foil heaves within the wake region. In the FWF mode, the flexible foil is excited to oscillate severely than in the FW mode because of the flexibility effect. The high frequency heaving motion is interacted with wake flows, resulting in a larger dominant wake frequency at $fc/U_{\infty}$=0.98. In particular, the upstream cylinder lift spectrum exhibits a dominant frequency peak at the heaving frequency of $fc/U_{\infty}$=0.6 rather than the wake frequency, which is different from the FW mode. This indicates that the heaving motion of the downstream structure conversely modulates the vortex-shedding pattern of the upstream cylinder. The vorticity mode excited by the high-frequency heaving motion is similar to that associated with wake flows in the FW mode.

Compared to the SW mode, the flexible foil performs larger foil fluctuations as the heaving amplitude and frequency increase in the SWF mode. As shown in \reffig{mode}, the flexible foil achieves a better lift performance than its rigid counterpart. The reason is attributed to the flexibility effect, which is further analyzed in \refse{sec4.5:flexible}. The heaving-induced mode with $fc/U_{\infty}$=0.2 dominates the surrounding flow characteristics of the flexible heaving foil, while the wake-induced mode is also excited by wake flows at the sub-frequency peak at $fc/U_{\infty}$=0.8. 

The VF mode is excited by the two-way frequency lock-in between the vortex-shedding process and the heaving motion, as well as the large foil deformation caused by the added-mass effect. It can be seen from the frequency spectra in \reffig{fre_mode} \subref{fre_modea} that the heaving motion dominates both the flow characteristics of the flexible foil and the cylinder. The foil vibration fluctuation also exhibits a frequency peak at the heaving frequency $fc/U_{\infty}$=0.5. Some fuzzy and irregular flow features around the flexible foil are observed from the extracted FMD fluid modes at the heaving frequency and the vortex frequency. This is mainly due to the high-frequency, large-amplitude vibrations generated by the flexible foil when forming frequency synchronization.

We further decompose the vibration of the flexible foil $\delta_n^f$ relative to its rigid counterpart $\delta_n^r$ into standard sinusoidal shapes. The purpose is to quantify the contributions of structural vibration modes of $m$-th order to the total flow-induced vibration \citep{li2024enhancing}. The decomposition of vibration fluctuations at each structure node $\xi$ is written as
\begin{equation}
	\frac{\delta_n^f(\xi,t)-\delta_n^r(\xi,t)}{c}  \approx \sum_{m=1}^{M} \beta_m(t) \sin\left( m \pi \frac{\xi}{c} \right) \quad \quad \quad m=1,2,3,\ldots  ,
	\label{mode1}
\end{equation}
where $\beta_m(t)$ is the decomposed coefficient. The contribution of each mode is evaluated by the standard deviation $\sigma_m$ of the decomposed coefficients, which is given as
\begin{equation}
	\sigma_m=\sqrt{\frac{\sum_{k=1}^{K}(\beta_m(t_k)-\overline{\beta}_m)^2}{K-1}}  ,
	\label{mode2}
\end{equation}
where $K$ is the number of decomposed coefficients. In this study, the contributions of the first five structural modes $\sigma_m/{\sum_{m=1}^{5}\sigma_m}$ are calculated for VIV, WIV and five selected modes. The mode contribution contours together with mode shapes are plotted in \reffig{fre_mode} \subref{fre_modeb}. The second-order mode is excited by wake flows and becomes dominant for the non-heaving structure in the WIV mode. The heaving motion and shed vortices excite the first-order mode as the dominant mode. The results indicate that wake flows and the heaving motion redistribute vibration energy and modulate flow features around the flexible foil, thus adjusting propulsive aerodynamic performance. This phenomenon is further confirmed in \refse{sec4.6:performance} by analyzing the frequency characteristics of the propulsive power coefficient.

\subsection{Coupled dynamics and flow features} \label{sec4.2:coupled}
\begin{figure}
	\centering
	\subfloat[]{\includegraphics[width=0.5 \textwidth]{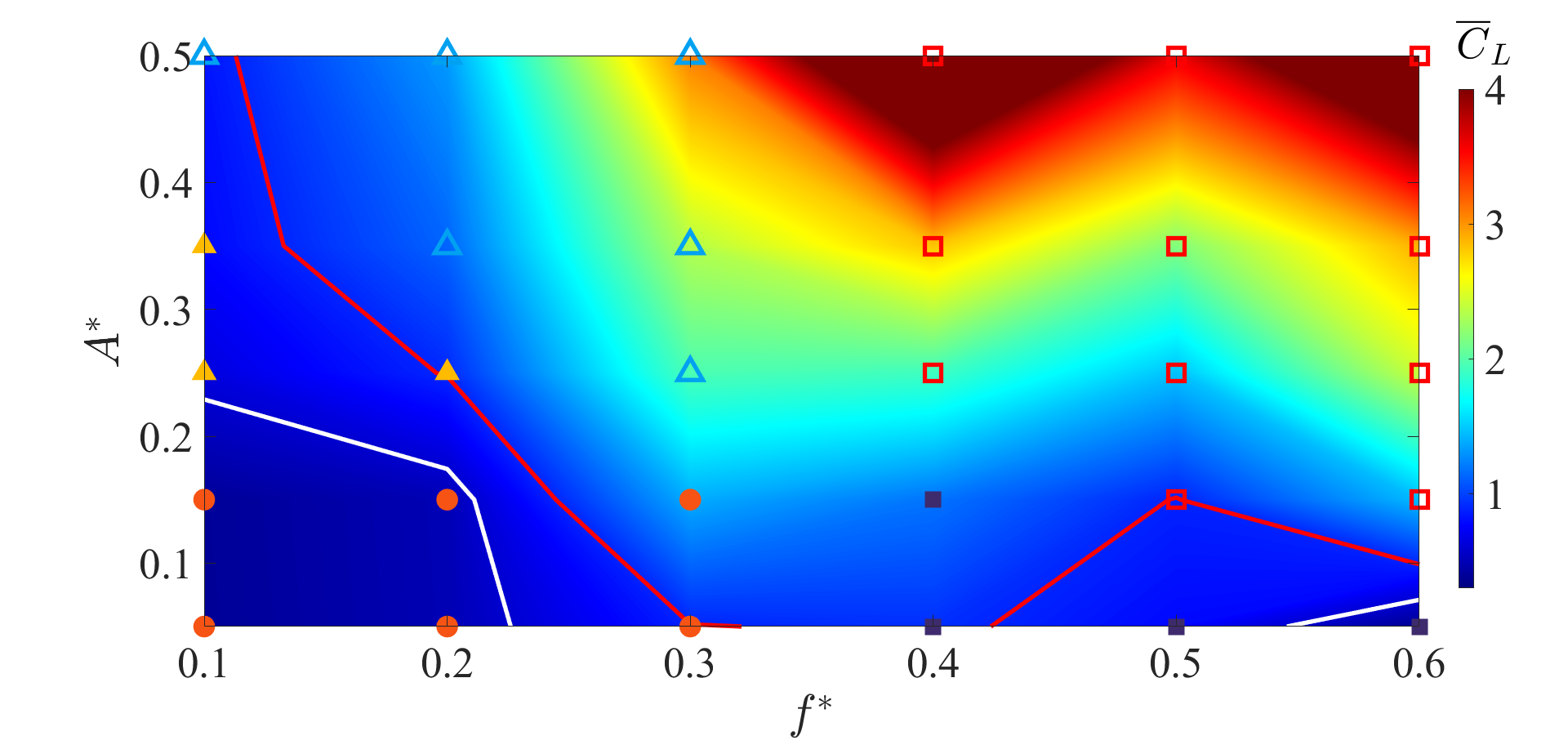}\label{aeroa}}
	\subfloat[]{\includegraphics[width=0.5 \textwidth]{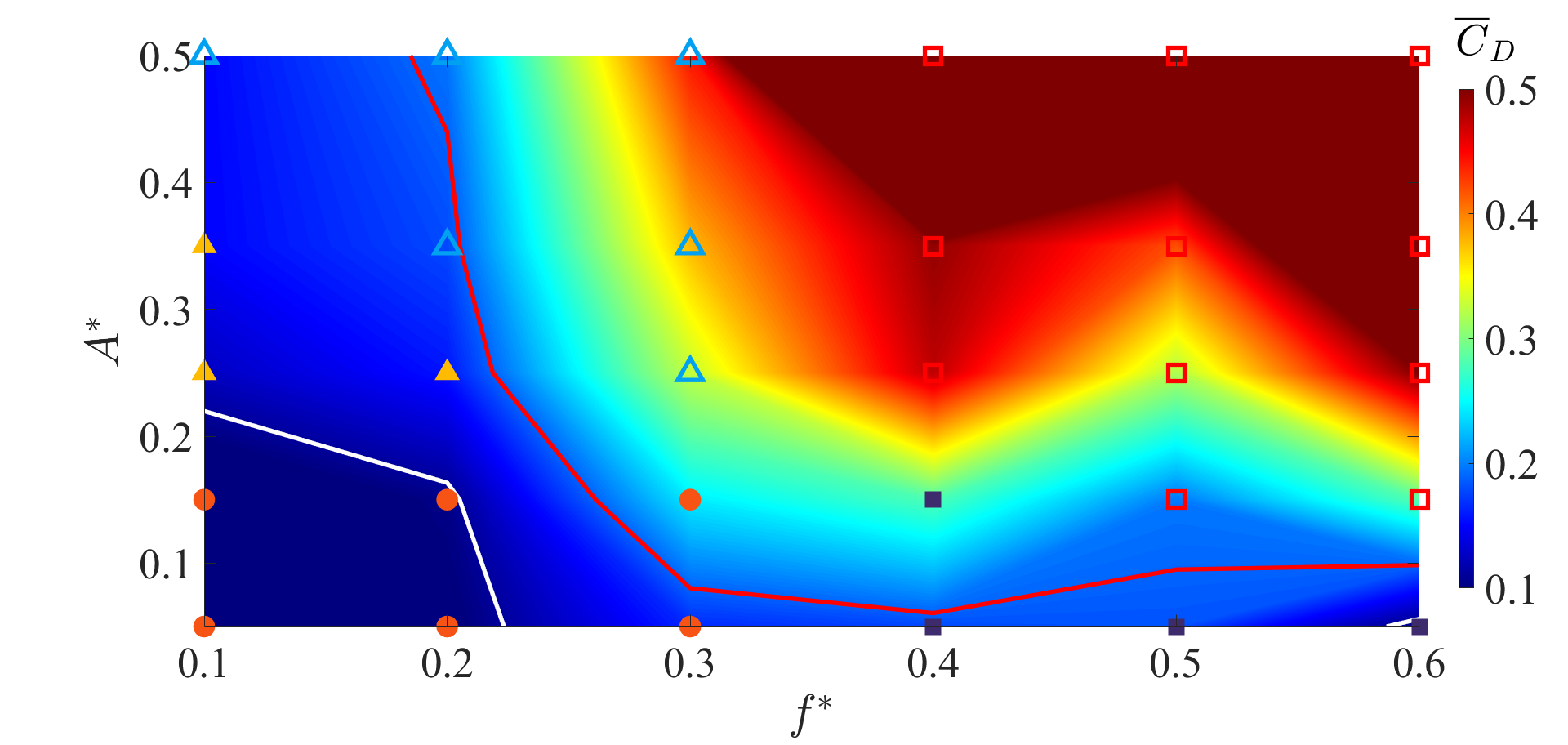}\label{aerob}}
	\\
	\subfloat[]{\includegraphics[width=0.5 \textwidth]{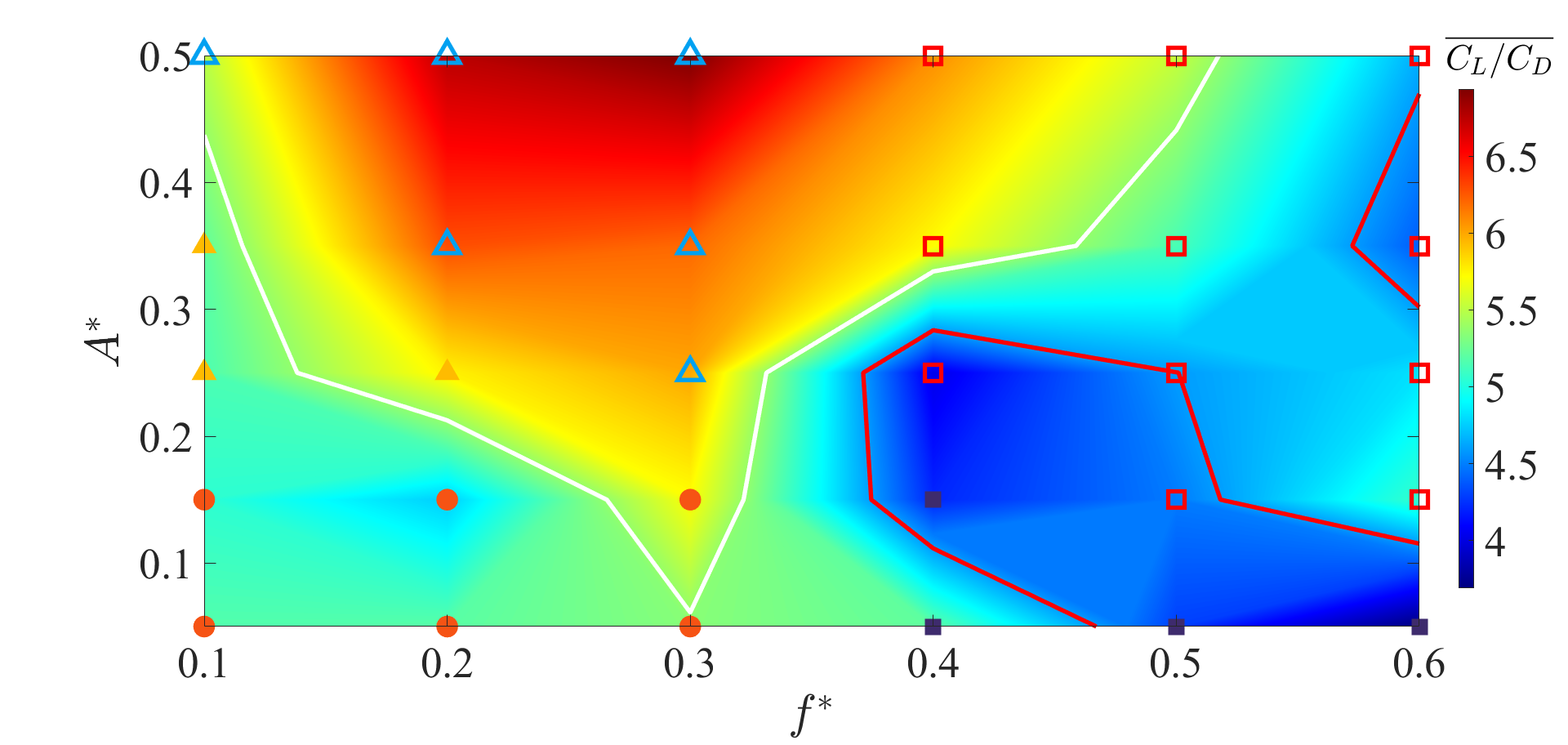}\label{aeroc}}
	\subfloat[]{\includegraphics[width=0.49 \textwidth]{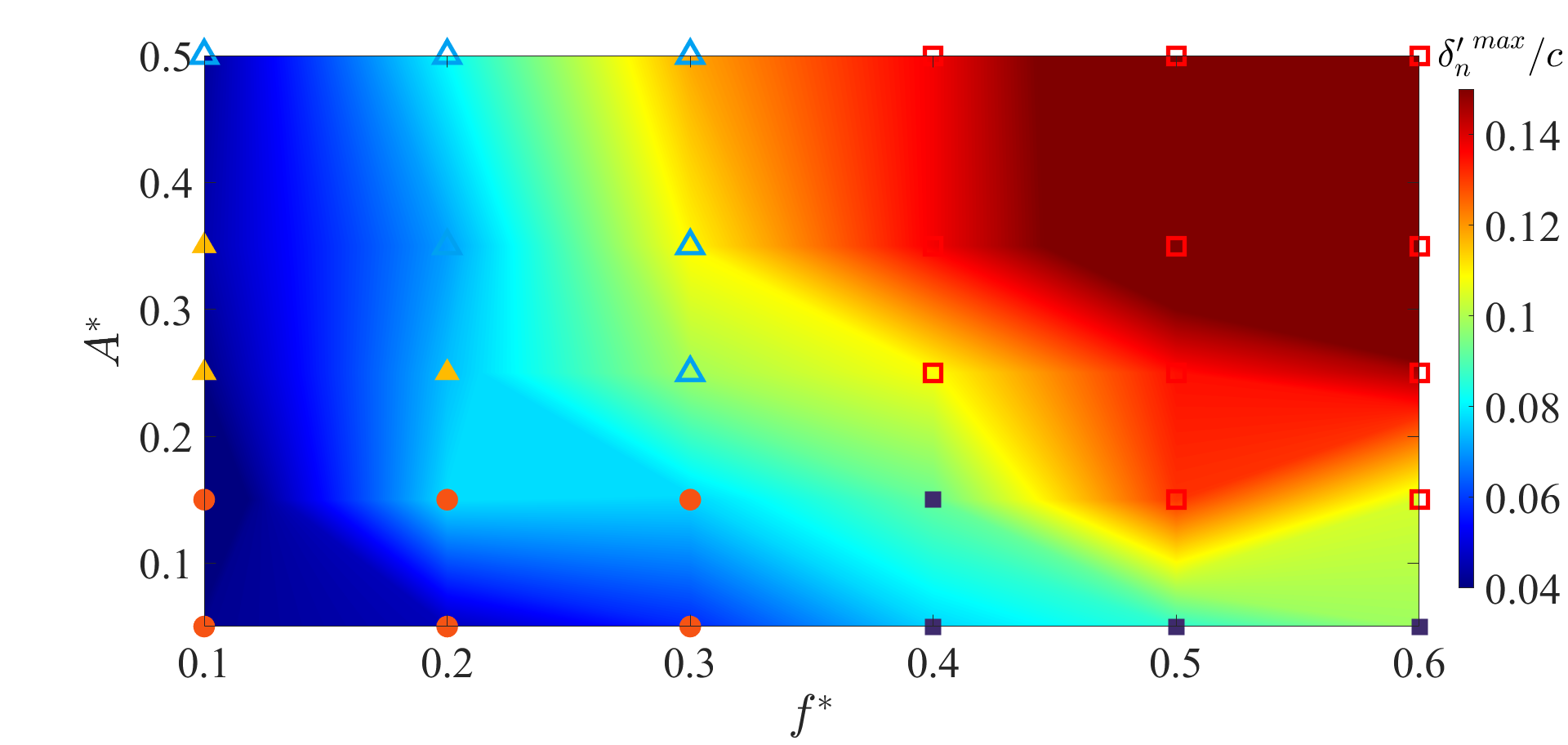}\label{aerod}}
	\caption{\label{aero} Time-averaged (a) lift coefficient, (b) drag coefficient, (c) lift-to-drag ratio and (d) maximum foil deformation relative to its rigid counterpart in the $f^*$--$A^*$ space. The white line indicates the isoline of the non-heaving flexible foil behind a cylinder (WIV mode). The red line represents the isoline of the non-heaving isolated flexible foil (VIV mode).}
\end{figure}

\begin{figure}
	\centering
	\includegraphics[width=1.0 \textwidth]{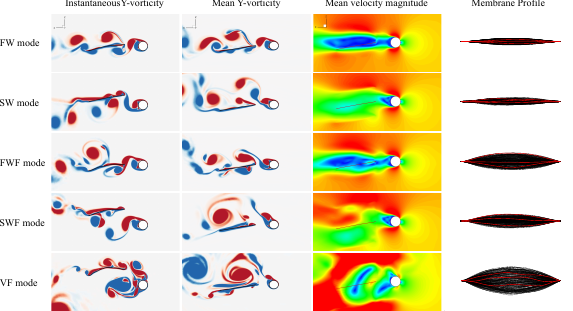}
	\caption{\label{aerodynamic} Illustrations of instantaneous, mean $y$-vorticity and mean velocity magnitude of the heaving flexible foil. Deformation profiles of the heaving structure relative to its rigid counterpart.}
\end{figure}

\begin{figure}
	\centering
	\subfloat[]{\includegraphics[width=0.5 \textwidth]{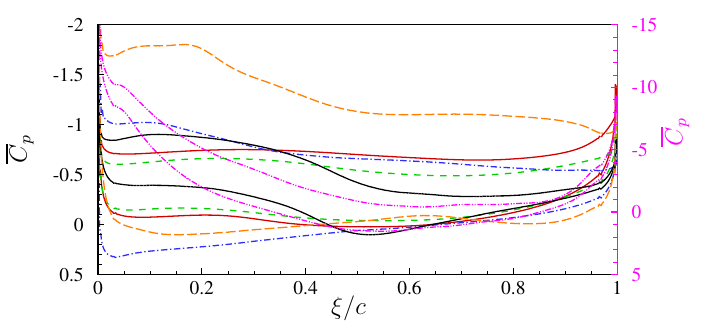}\label{cp_com}}
	\subfloat[]{\includegraphics[width=0.5 \textwidth]{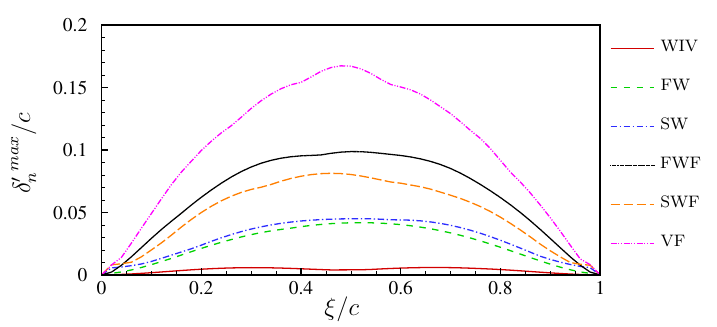}\label{dis_com}}
	\caption{\label{com} Comparison of (a) time-averaged pressure coefficients along the foil chord and (b) maximum foil deformation profile relative to its rigid counterpart.}
\end{figure}

\begin{figure}
	\centering
	\subfloat[]{\includegraphics[width=0.5 \textwidth]{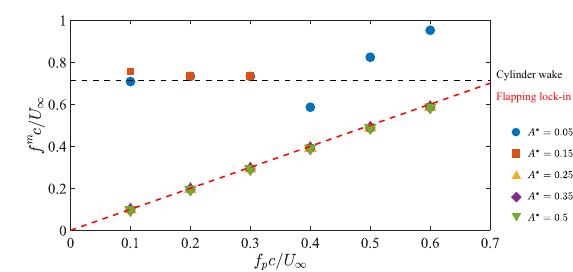}\label{frea}}
	\subfloat[]{\includegraphics[width=0.5 \textwidth]{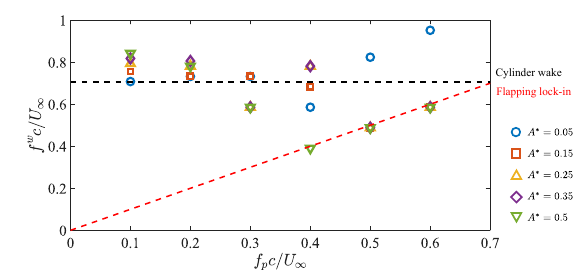}\label{freb}}
	\caption{\label{fre} Dominant flow frequency characteristics of (a) heaving flexible foils and (b) stationary cylinder determined by the FMD method from the vorticity field.}
\end{figure}

The aerodynamic performance and structural deformation of the tandem heaving flexible foil in the $f^*$--$A^*$ space are shown in \reffig{aero}. For comparison, contours also include the performance of non-heaving flexible foils in tandem and isolated configurations, indicated by white and red lines, respectively. This enables assessment of the relative influence of heaving motion and wake forcing on the coupled dynamics. Instantaneous flow fields, structural deformation and time-averaged pressure distributions for the five modes are presented in \reffigs{aerodynamic} and \ref{com} to explain performance variations with amplitude and frequency.
 Dominant flow frequencies are determined from vorticity fields of the cylinder $f^w c / U_{\infty}$ and the flexible foil $f^m c / U_{\infty}$, as depicted in \reffig{fre}. The red dashed line represents the frequency lock-in between the heaving motion $f_p c / U_{\infty}$ and the foil response, while the black dashed line is the vortex-shedding frequency of the isolated cylinder.

The time-averaged lift and drag coefficients change little when the foil heaves within the wake at low frequencies, as shown in \reffigs{aero}\subref{aeroa},\subref{aerob}. In these cases, the propulsive performance approaches that of the non-heaving isolated foil. Such poor-performance regimes are primarily associated with FW, FWF, and SW modes. \reffig{aerodynamic} shows that in FW and FWF the foil is fully immersed in the wake shed by the upstream cylinder. Upon examination of the instantaneous flow characteristics and frequency characteristics plotted in \reffig{fre} \subref{frea}, the vibration of the flexible foil is synchronized with the vortex-shedding frequency of the upstream cylinder. The flexible foil experiences low-velocity flows and the axial momentum deficit of the incoming wake is observed from mean velocity magnitude contours. Therefore, the pressure distribution related to the FW and FWF modes is comparable to that of the WIV mode, as shown in \reffig{com}, leading to poor aerodynamic performance. As the heaving amplitude increases at low heaving frequencies, the structure transitions from the FW mode to the SW mode. Both lift and drag forces slightly increase, caused by additional transverse momentum supplied by heaving motion in wake regions and enhanced pressure distributions on both structure surfaces. The dominant flow frequency of the flexible foil is synchronized with the heaving frequency, while the locking of the frequency with the wake flows of the cylinder is interrupted, as shown in \reffig{fre}.

The SWF mode is excited when the flexible foil heaves in partial wake regions with an asynchronous frequency of the vortex-shedding process. The lift and drag coefficients of the flexible heaving structures are generally better than those of the isolated non-heaving structures, as shown in \reffig{aero}. The optimal lift-to-drag ratio of the heaving structure is produced in the SWF mode. Compared to the SW mode, the flexible foil in the SWF mode heaves in larger deformation, leading to accelerated flow velocities and induced transverse momentum behind the cylinder observed in \reffig{aerodynamic}. The suction effect on the upper surface is greatly enhanced, thereby producing better aerodynamic performance, as presented in \reffig{com} \subref{cp_com}. The dominant flow frequency locks into the heaving frequency in the SWF mode. As the heaving frequency increases, the vortex-shedding frequency gradually approaches the red dashed line presented in \reffig{fre} \subref{freb}. This means that wake flow features are inversely influenced by the downstream heaving motion through two-way frequency lock-in, which changes the propulsive aerodynamic performance.

Once the heaving structure transitions to the VF mode, both the lift and drag forces shown in \reffigs{aero} \subref{aeroa} and \subref{aerob} are greatly enhanced. Compared with the non-heaving flexible foil with tandem configuration, the lift and drag forces of the heaving case with the largest heaving frequency and amplitude in the parameter space are increased by nearly five times. It can be seen from the flow features plotted in \reffig{aerodynamic} that the heaving structure inversely changes vortex structures behind the cylinder and forms a two-way lock-in. The surrounding flows of the flexible foil are greatly accelerated by the heaving motion, resulting in greater structural deformation and higher flow velocities in the wake regions. In \reffig{com}, the pressure distribution of the VF mode is displayed on another $Y$-axis for direct comparison in one plot. Compared with the SWF mode, the suction force near the leading edge is increased five times, while the structural deformation increases nearly twice. By comparing the frequency characteristics in \reffig{fre}, it is found that the flow features of the cylinder lock into the heaving frequency of the flexible foil. These flapping-dominated flow features and two-way frequency lock-in are the key to greatly improving propulsive performance, which is explained further in \refse{sec4.6:performance}.

\subsection{Scaling relations of coupled foil dynamics} \label{sec4.3:scaling}
By analyzing the coupled dynamics of five distinct flapping modes, the propulsive aerodynamic performance is mainly governed by heaving kinematics, passive structural deformation and wake-flow features. The heaving motion of the flexible foil introduces an additional angle of attack relative to the freestream, compared to the non-heaving case. The aerodynamic forces contributed by the heaving motion can be analyzed employing Theodorsen aerodynamic theory. 
In \reffigs{fre_mode} \subref{fre_modea} and \ref{fre}, we find that the heaving motion gradually governs the dominant flow characteristics of the wake-heaving structure interaction system through frequency synchronization as the heaving frequency increases. A decomposed term that evaluates how the flow frequency locks into the heaving frequency is designed to construct a scaling relation for the aerodynamic force and power coefficients. The passive deformation of the flexible foil shows positive contributions to propulsive performance when it heaves in partial wake flows. Heaving flexible foils suffer from reduced aerodynamic forces as a result of the axial momentum deficit of the incoming wake, while the performance is enhanced in a situation in which wake flows are accelerated by heaving motion. 

In this section, we now derive scaling relations for lift, drag, and power by decomposing the total force into five physical contributions: (i) quasi-steady motion, (ii) added-mass effects, (iii) curvature-induced forces from passive deformation, (iv) velocity loss due to the wake deficit, and (v) transverse momentum gradients across the wake.
Consider the predicted lift force as follows
\begin{equation}
	\overline{C}_L^p=\underbrace{{\overline{C}_L}_0}_{\text{staic}} + \underbrace{a_1 f^* A^*}_{\text{quasi-steady}} + \underbrace{a_2 \sin \left( \pi \frac{f^*}{f_0} \right) {A^*}^2}_{\text{heaving}} + \underbrace{a_3 \kappa^* A^*}_{\text{curvature}} +  \underbrace{a_4 W f^* A^*}_{\text{velocity loss}} + \underbrace{a_5 G A^*}_{\text{wake gradient}}  ,
	\label{eq:cl_non}
\end{equation}
where ${\overline{C}_L}_0$ denotes the static lift coefficient of the non-heaving tandem flexible foil. $\kappa^*$ is the curvature of the deformed structure shape. The velocity-loss coefficient and flow gradient of wake flows are defined as $W$ and $G$. The fitting coefficients $a_1$, $a_2$, $a_3$, $a_4$ and $a_5$ are determined by the nonlinear regression fitting algorithm. The terms in the expression of the lift coefficient are derived as follows. 

We assume that the heaving motion is a simple harmonic oscillation governed by $h_p(t) = A_{h_p} \sin (2 \pi f_p t)$. The imposed heaving velocity introduces an effective angle of attack. According to the unsteady thin wing theory proposed by Theodorsen \citep{1935General}, the variation of circulation $\Gamma^{qs}$ is proportional to the instantaneous transverse velocity $\dot{h}_p(t)$, which is given as
\begin{equation}
	\Gamma^{qs}  \propto  \dot{h}_p(t)= 2 \pi f_p A_{h_p} \cos (2 \pi f_p t)  ,
	\label{eq:cl_non2}
\end{equation}
By introducing two dimensionless parameter $f^*=\frac{ f_p c}{U_{\infty}}$ and $A^*=\frac{A_{h_p}}{c}$, the equation can be rewritten as
\begin{equation}
	\Gamma^{qs} \propto 2 \pi  U_{\infty} \frac{ f_p c}{U_{\infty}} \frac{A_{h_p}}{c}  \cos (2 \pi f_p t) = 2 \pi U_{\infty} f^* A^* \cos (2 \pi f_p t)  ,
	\label{eq:cl_non3}
\end{equation}
Thus, the quasi-steady lift coefficient is calculated by
\begin{equation}
	\overline{C}_L^{qs} = \frac{\overline{L}^{qs}}{\frac{1}{2} \rho^f U^2_{\infty} c} = \frac{\rho^f U_{\infty} \overline{\Gamma}^{qs}}{\frac{1}{2} \rho^f U^2_{\infty} c} = \frac{2 \overline{\Gamma}^{qs}}{U_{\infty} c} \propto  \frac{8 U_{\infty} f^* A^*}{U_{\infty} c}= \frac{8}{c} f^* A^*  ,
	\label{eq:cl_non4}
\end{equation}
The quasi-steady lift coefficient component is contributed by the product of heaving frequency and amplitude, which can be finally written as
\begin{equation}
	\overline{C}_L^{qs} = a_1 f^* A^*  ,
	\label{eq:cl_non5}
\end{equation}

% Added mass effect
The additional lifting force contributed by the inertial effect of the fluid during heaving is proportional to the square of the amplitude \citep{sarpkaya2004critical}. To account for lock-in between heave and shedding, a phase modulation factor $\sin \left(  \pi \frac{f^*}{f_0} \right)$ is introduced to consider the frequency synchronization between heaving motion and vortex-shedding process. The lift force associated with added-mass effect during heaving motion is given by
\begin{equation}
	\overline{F}^{ine} \propto \rho^f U^2_{\infty} A^2_{h_p} \sin \left(  \pi \frac{f^*}{f_0} \right)  ,
	\label{eq:cl_non6}
\end{equation}
Thus, the lift coefficient is written as a function of heaving amplitude and frequency
\begin{equation}
	\overline{C}_L^{hea} = \frac{\overline{F}^{ine}}{\frac{1}{2} \rho^f U^2_{\infty} c}  = a_2 \sin \left(  \pi \frac{f^*}{f_0} \right) {A^*}^2  ,
	\label{eq:cl_non7}
\end{equation}

The lift component contributed by the flexible foil deformation during heaving motion can be regarded as a function of structure curvature and heaving amplitude \citep{waldman2017camber}. The circulation related to flexible deformation can be written as
\begin{equation}
	\overline{\Gamma}^{cur} \propto (\kappa)^m (A_{h_p})^n U^p_{\infty} c^q  ,
	\label{eq:cl_non8}
\end{equation}
Using Buckingham's $\pi$ Theorem, the coefficients are determined as $m=n=p=q$=1.  Thus, the lift coefficient contributed by structure deformation is given as
\begin{equation}
	\overline{C}_L^{cur} = \frac{\overline{L}^{cur}}{\frac{1}{2} \rho^f U^2_{\infty} c} = \frac{\rho^f U_{\infty} \overline{\Gamma}^{cur}}{\frac{1}{2} \rho^f U^2_{\infty} c} = \frac{2 \overline{\Gamma}^{cur}}{U_{\infty} c}   \propto \frac{2 \kappa A_{h_p} U_{\infty} c}{ U_{\infty} c} = 2 \kappa A_{h_p}  ,
	\label{eq:cl_non9}
\end{equation}
The lift coefficient component can be further given as a function of nondimensional curvature and heaving amplitude
\begin{equation}
	\overline{C}_L^{cur} = a_3 \kappa^* A^*  ,
	\label{eq:cl_non10}
\end{equation}
From the dimensional analysis, circulation due to bending curvature scales as $\kappa^* A^*$ which captures the suction generated by cambering deformation.

The wake effect on the lift performance of the flexible foil can be divided into the velocity-loss term in the freestream direction and the wake-gradient term in the transverse direction. The velocity-loss term represents the lift reduction caused by the axial momentum deficit of the incoming wake over the leading‑edge control volume. The circulation related to the velocity-loss effect is written as
\begin{equation}
	\Gamma^{vl} = \pi c (U_{\infty}-U_{eff}) \alpha_{eff} \propto \pi c (U_{\infty}-U_{eff})  \frac{\dot{h}_p(t)}{U_{eff}} = \frac{\pi c (U_{\infty}-U_{eff}) 2 \pi f_p A_{h_p} \cos (2 \pi f_p t) }{U_{eff}}  ,
	\label{eq:cl_non11}
\end{equation}
Thus, the lift coefficient is further expressed as
\begin{equation}
	\overline{C}_L^{vl} =  \frac{\overline{L}^{vl}}{\frac{1}{2} \rho^f U^2_{\infty} c} = \frac{\rho^f U_{eff} \overline{\Gamma}^{vl}}{\frac{1}{2} \rho^f U^2_{\infty} c}  \propto 4 \pi^2 \frac{(U_{\infty}-U_{eff})}{U_{\infty}} \frac{f_p c}{U_{\infty}} \frac{A_{h_p}}{c} = 4 \pi^2 W f^* A^*  ,
	\label{eq:cl_non12}
\end{equation}
The lift coefficient component related to the velocity-loss term $W=\frac{(U_{\infty}-U_{eff})}{U_{\infty}}$ can be rewritten as
\begin{equation}
	\overline{C}_L^{vl} = a_4 W f^* A^*  ,
	\label{eq:cl_non13}
\end{equation}

The lift coefficient component caused by the wake flow gradient in transverse direction $G=\frac{c}{U_{\infty}} \left. \frac{\partial u}{\partial z} \right|_{z=0} $ can be defined as
\begin{equation}
	\overline{C}_L^{wg} = A_5 G A^*  ,
	\label{eq:cl_non14}
\end{equation}
where the wake flow gradient $G$ is inherently a function of heaving frequency.

Similarly, the drag force of the flexible foil is contributed by heaving motion, structural flexibility and wake effect, which is expressed as
\begin{equation}
	\overline{C}_D^p=\underbrace{{\overline{C}_D}_0}_{\text{staic}} + \underbrace{b_1 {f^*}^2 {A^*}^{2}}_{\text{quasi-steady}} + \underbrace{b_2 \frac{f^*}{f^*_c} {A^*}^2}_{\text{vortex shedding}} + \underbrace{b_3 \kappa^* A^*}_{\text{curvature}} +  \underbrace{b_4 W f^* A^*}_{\text{velocity loss}} + \underbrace{b_5 G A^*}_{\text{wake gradient}}  ,
	\label{eq:cd_non}
\end{equation}
where $f^*_c$ represents the vortex-shedding frequency. The fitting coefficients $b_1$, $b_2$, $b_3$, $b_4$ and $b_5$ are determined by the numerical simulation data. The derivation of each term for the drag coefficient is presented as follows.
Previous studies on flapping plates found that the drag/thrust force can be scaled as $(f_p A_{h_p})^2$ \citep{quinn2014unsteady,gazzola2014scaling,floryan2017scaling,floryan2018efficient} based on the Theodorsen theory. The drag coefficient contributed by the quasi-steady term can be written as
\begin{equation}
	\overline{C}_D^{qs} = b_1 {f^*}^2 {A^*}^{2}  ,
\end{equation}

An empirical formula was proposed to describe the relationship between mean drag, vortex-shedding frequency and vibration amplitude under vortex-induced vibration \citep{song2016distribution}. Herein, a similar expression is proposed to reflect the contribution of the heaving frequency relative to the vortex-shedding frequency and the heaving amplitude, which is given as
\begin{equation}
	\overline{C}_D^{vs} = b_2 \frac{f^*}{f^*_c} {A^*}^2  ,
\end{equation}
where the frequency relationship between the heaving frequency $f^*$ and the critical vortex-shedding frequency $f^*_c$ reflects the impact of frequency synchronization on drag enhancement. 
The drag force caused by structure deformation and wake flows have expressions similar to those of lift forces, which are defined as
\begin{equation}
	\overline{C}_D^{cur} = b_3 \kappa^* A^*, \quad \overline{C}_D^{vl} = b_4 W f^* A^*, \quad \overline{C}_D^{wg} = b_5 G A^*  ,
\end{equation}

The power coefficient averaged over time, $\overline{C}_{pow}^p$, is defined by the sum of the lift force times the transverse velocity in a heaving period, divided by the pressure of the dynamics and the area of the structure \citep{upfal2025shape} 
\begin{equation}
	\overline{C}_{pow}^p = \frac{\frac{1}{T}\int_{t_0}^{t_0+T} L \cdot \dot{h}_p  \text{d}t}{0.5 \rho^f U_{\infty}^3 b c} \propto  \frac{\pi (\overline{L}-\overline{L}_{0}) f_p A_{h_p}  }{0.5 \rho^f U_{\infty}^3 b c} =\pi (\overline{C}_L - {\overline{C}_L}_0) f^* A^*  ,
	\label{eq:cpow}
\end{equation}
Substituting \refeq{eq:cl_non} into \refeq{eq:cpow}, the time-averaged power coefficient is rewritten as
\begin{equation}
	\overline{C}_{pow}^p=\underbrace{c_1 {f^*}^2 {A^*}^{2}}_{\text{quasi-steady}} + \underbrace{c_2  \sin \left(  \pi \frac{f^*}{f_0} \right) f^* {A^*}^3}_{\text{heaving}} + \underbrace{c_3 \kappa^* f^* {A^*}^2}_{\text{curvature}} +  \underbrace{c_4 W {f^*}^2 {A^*}^{2}}_{\text{velocity loss}} + \underbrace{c_5 G f^* {A^*}^2}_{\text{wake gradient}}  ,
	\label{eq:cpow_non}
\end{equation}
where $c_1$, $c_2$, $c_3$, $c_4$ and $c_5$ are fitting coefficients of the power coefficient. 

\begin{figure}
	\centering
	\subfloat[]{\includegraphics[width=1.0 \textwidth]{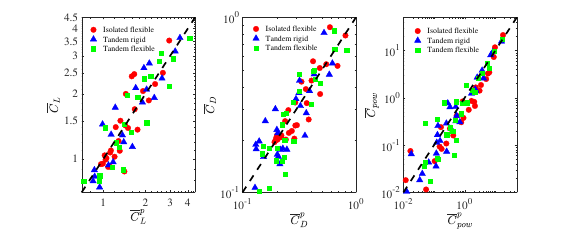}\label{cl_fita}}
	\\
	\subfloat[]{\includegraphics[width=0.9 \textwidth]{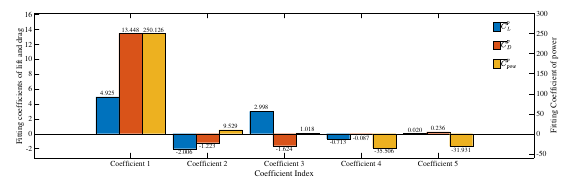}\label{cl_fitb}}
	\caption{\label{cl_fit} (a) Scaling relations of lift, drag and power coefficients for isolated flexible foil, tandem rigid and flexible foils. (b) Comparison of fitting coefficients of lift, drag and power coefficients for tandem flexible foil.}
\end{figure}

Using \refeqs{eq:cl_non}, (\ref{eq:cd_non}), and (\ref{eq:cpow_non}), the predicted lift, drag, and power coefficients are compared with simulation data in \reffig{cl_fit}\subref{cl_fita}. For the isolated flexible configuration, both the velocity-loss and wake-gradient terms are omitted when determining fitting coefficients, while for the tandem rigid case, the curvature terms are set to zero. As shown in \reffig{cl_fit}\subref{cl_fita}, the proposed scaling relations predict the aerodynamic coefficients with good agreement, collapsing the data onto the 1:1 line (black dashed). This confirms that the scaling laws provide a quantitative framework for decomposing and evaluating the relative contributions of heaving, flexibility, and wake effects to overall propulsive performance.

\begin{figure}
	\centering
	\subfloat[]{\includegraphics[width=0.5 \textwidth]{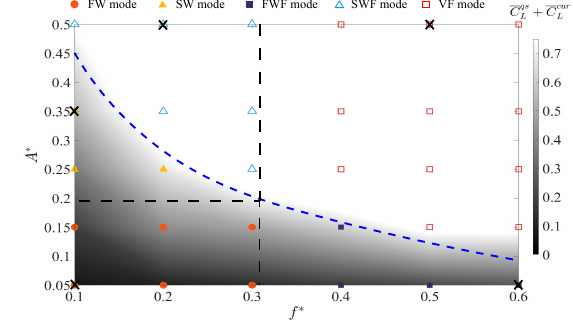}\label{lift_decoma}}
	\subfloat[]{\includegraphics[width=0.5 \textwidth]{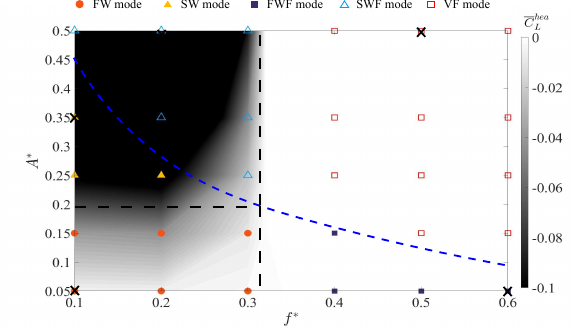}\label{lift_decomb}}
	\caption{\label{lift_decom} Decomposed lift coefficient components contributed by (a) quasi-steady and curvature terms and (b) added-mass effect during heaving motion in the $f^*$--$A^*$ space.}
\end{figure}

The comparisons of the fitted coefficients of the time-averaged lift, drag, and power for the tandem flexible foil is shown in \reffig{cl_fit}\subref{cl_fitb}. Because the power coefficients are significantly larger in magnitude, two $y$-axes are used for clarity. Both quasi-steady and curvature terms contribute positively to lift, as shown in \reffig{lift_decom}\subref{lift_decoma}, where their combined effect is mapped in the $f^*$--$A^*$ space. The critical blue line aligns closely with the threshold boundary separating lift reduction and lift gain in the phase diagram. Compared to the rigid tandem case (with zero curvature), passive deformation in the flexible case generates additional leading-edge suction at higher amplitudes and frequencies (\reffig{com}\subref{cp_com}). In contrast, the velocity-loss term carries a negative coefficient, reflecting reduced axial momentum near the leading edge, whereas the positive wake-gradient coefficient supplies transverse momentum during heaving. The transition from lift reduction to lift gain therefore corresponds to a balance between heave- and flexibility-induced transverse momentum and the axial momentum deficit of the incoming wake. The decomposed added-mass term, $\overline{C}_L^{hea}$, shows negative values on the left part of the vertical black dashed line in \reffig{lift_decom} \subref{lift_decomb}. As the heaving frequency gradually approaches the vortex-shedding frequency of the upstream cylinder, the flexible foil takes advantages of the shed vortices in a synchronized manner and produces positive lift forces.

It can be seen from \reffig{cl_fit} \subref{cl_fitb} that the quasi-steady effect connected with heaving motion exhibits positive influences on drag and power performance. This is mainly caused by the increased local angle of attack near the leading edge introduced by transverse velocity. The frequency lock-in between heaving frequency and vortex-shedding frequency helps in reducing drag force and improving power coefficient. The foil deformation makes it more streamlined in wake flows, resulting in drag reduction along the horizontal direction and power enhancement by resisting fluid damping in the transverse axis. The flexible foil suffers from the axial momentum deficit of the incoming wake, reducing both drag and power coefficients. The transverse momentum supplied by the heave enlarges drag forces but saves power to maintain heaving motion.

It is worth noting from the above analysis that the structural flexibility can turn the wake from a hindrance into a boost from two aspects, namely wake‑deficit compensation and two-way frequency lock-in. The interaction between disturbed wake flows and heaving motion, the structural flexibility are two key factors to modulate flapping mode transition and govern propulsive aerodynamic performance. In the following sections, we reveal the mechanism of how the foil transitions from wake-limited (lift reduction) to self-energized (lift gain) by investigating wake flow effect and role of flexibility for this wake-heaving flexible foil coupling system.

\subsection{Effect of wake flows} \label{sec4.4:wake}
A comparison of aerodynamic performance between the tandem flexible foil and its isolated counterpart is performed to examine wake flow effect. Three dimensionless variables, namely lift-gain $C_{LGF}^w$, drag-penalty $C_{DPF}^w$ and lift-to-drag ratio factors $C_{LDF}^w$, are defined to evaluate relative aerodynamic force changes due to disturbed wake flows
\begin{equation}
	C_{LGF}^w = \frac{\overline{C}_L^{tm} - \overline{C}_L^{im}}{\overline{C}_L^{im}}, \quad  C_{DPF}^w = \frac{\overline{C}_D^{tm} - \overline{C}_D^{im}}{\overline{C}_D^{im}}, \quad  
	C_{LDF}^w = \frac{\overline{{C}_L^{tm}/{C}_D^{tm}} - \overline{{C}_L^{im}/{C}_D^{im}}}{\overline{{C}_L^{im}/{C}_D^{im}}} ,
\end{equation}
where $\overline{C}_L^{tm}$, $\overline{C}_D^{tm}$ and $\overline{{C}_L^{tm}/{C}_D^{tm}}$ are time-averaged lift, drag and lift-to-drag ratio of the tandem foil. $\overline{C}_L^{im}$, $\overline{C}_D^{im}$ and $\overline{{C}_L^{im}/{C}_D^{im}}$ represent the performance for the isolated foil case. 

\begin{figure}
	\centering
	\subfloat[]{\includegraphics[width=0.33 \textwidth]{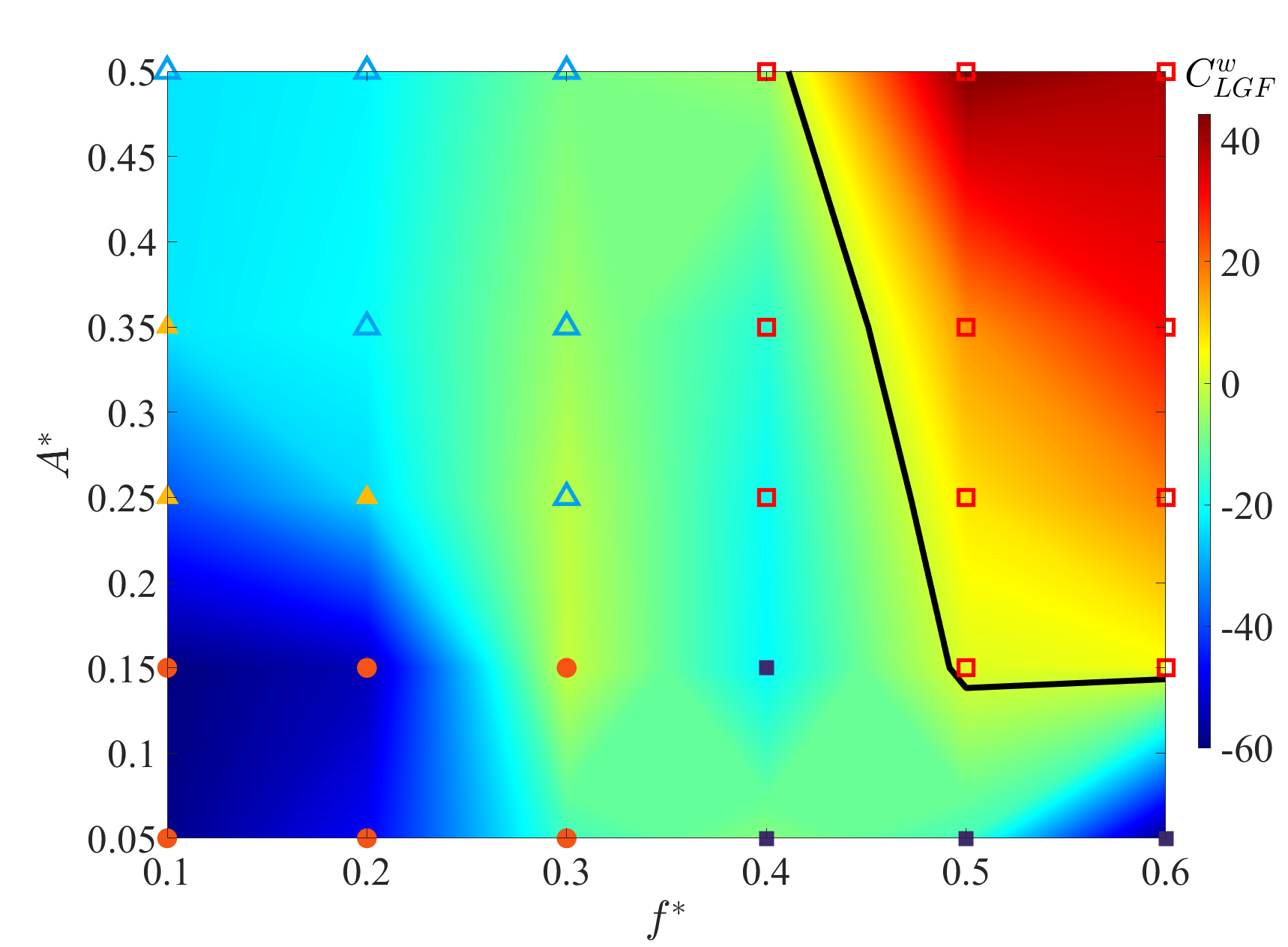}\label{wakea}}
	\subfloat[]{\includegraphics[width=0.33 \textwidth]{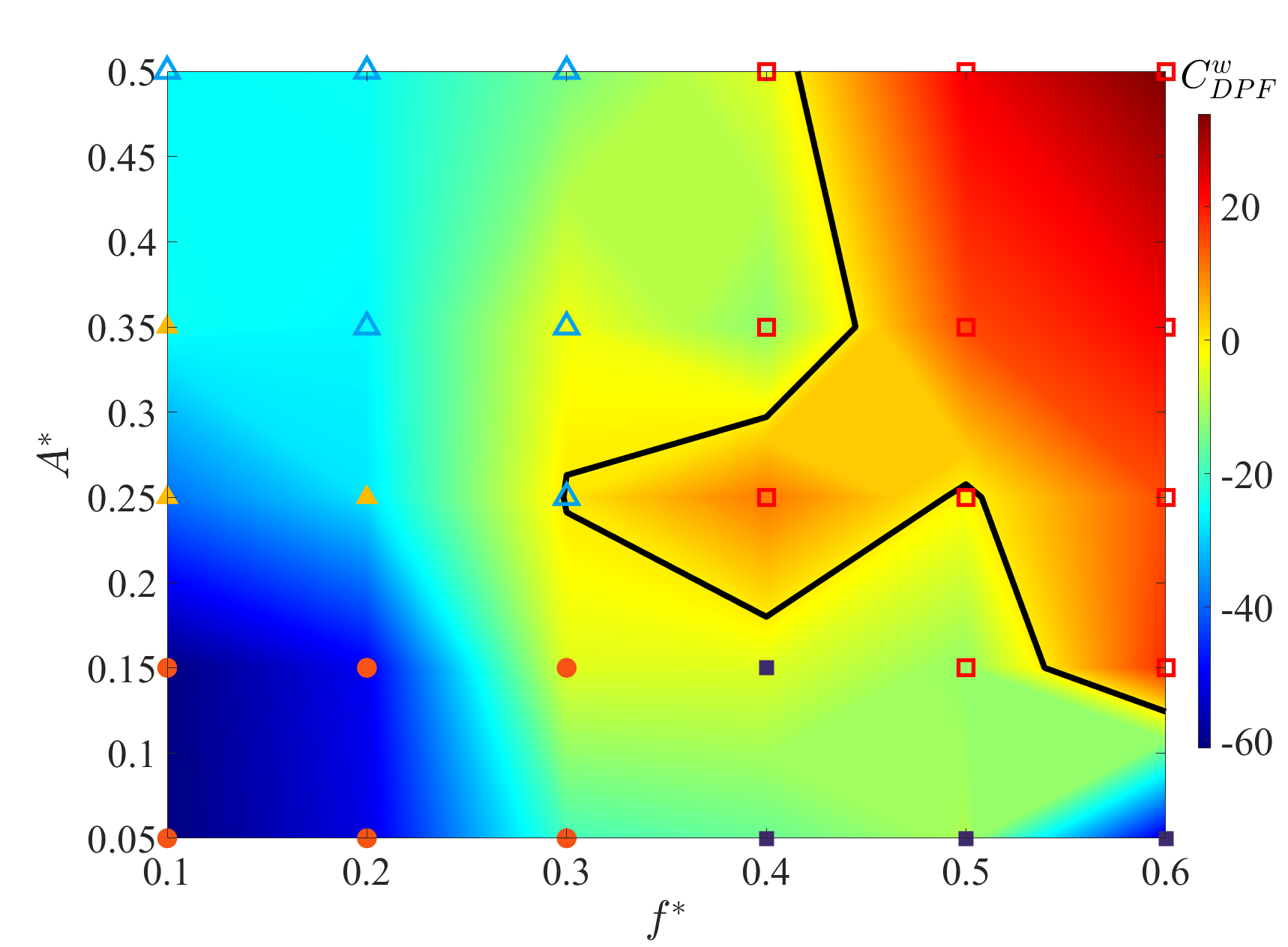}\label{wakeb}}
	\subfloat[]{\includegraphics[width=0.33 \textwidth]{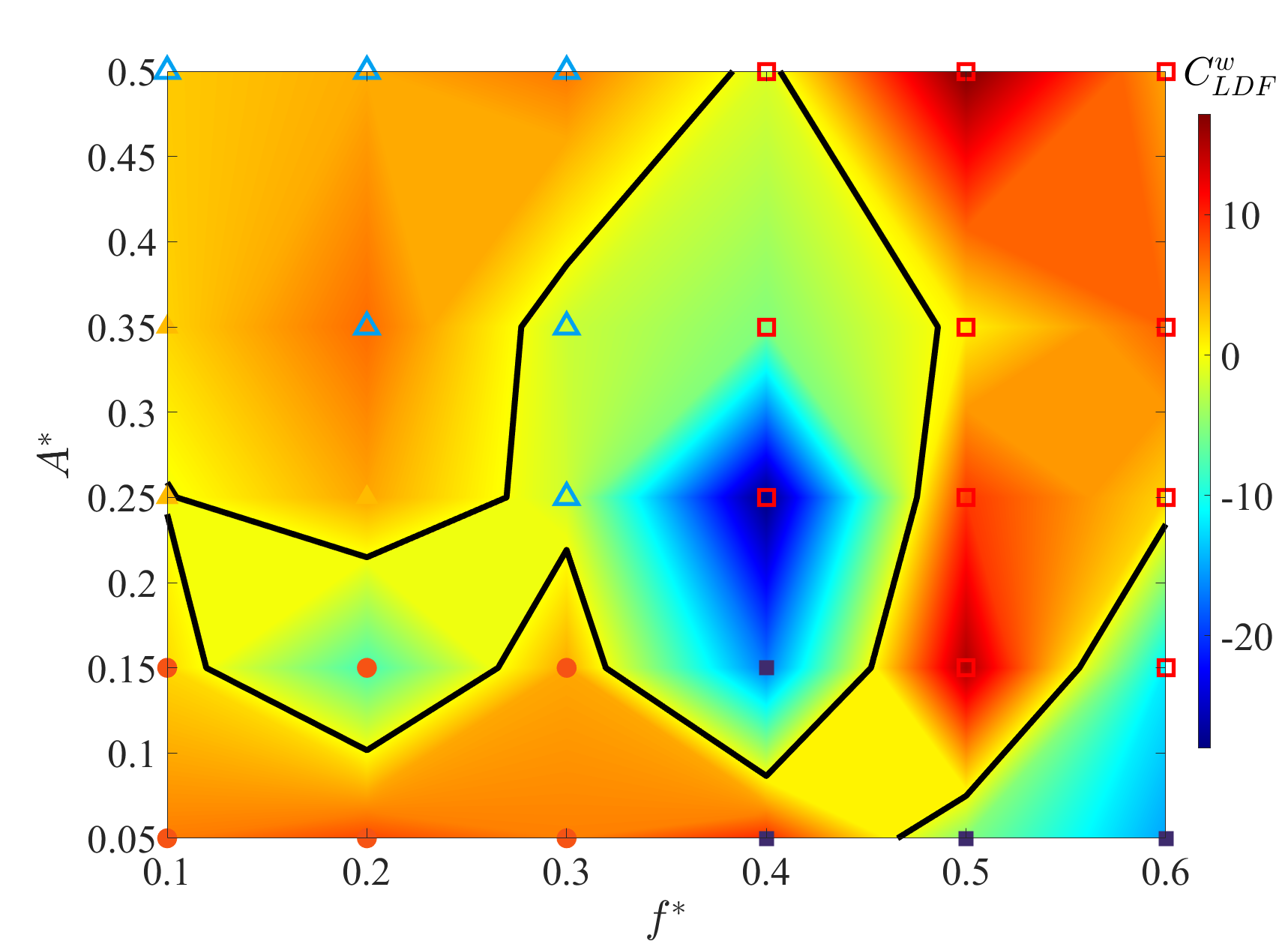}\label{wakec}}
	\\
	\subfloat[]{\includegraphics[width=1.0 \textwidth]{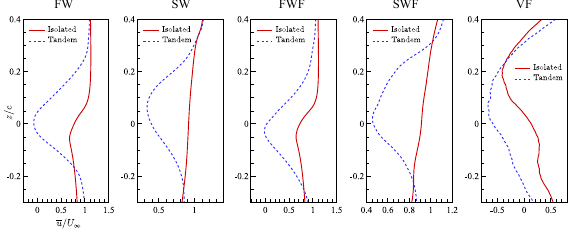}\label{waked}}
	\caption{\label{wake} Phase diagram of (a) lift-gain factor, (b) drag-penalty factor and (c) lift-to-drag ratio factor of the flexible foil due to wake effect in percentage form. The isoline in black color indicates zero value in contours. (d) Comparison of time-averaged horizontal velocity profile along vertical direction at $x/c$=-0.3 between isolated and tandem flexible foils for five flapping modes.}
\end{figure}

\begin{figure}
	\centering
	\subfloat[]{\includegraphics[width=0.33 \textwidth]{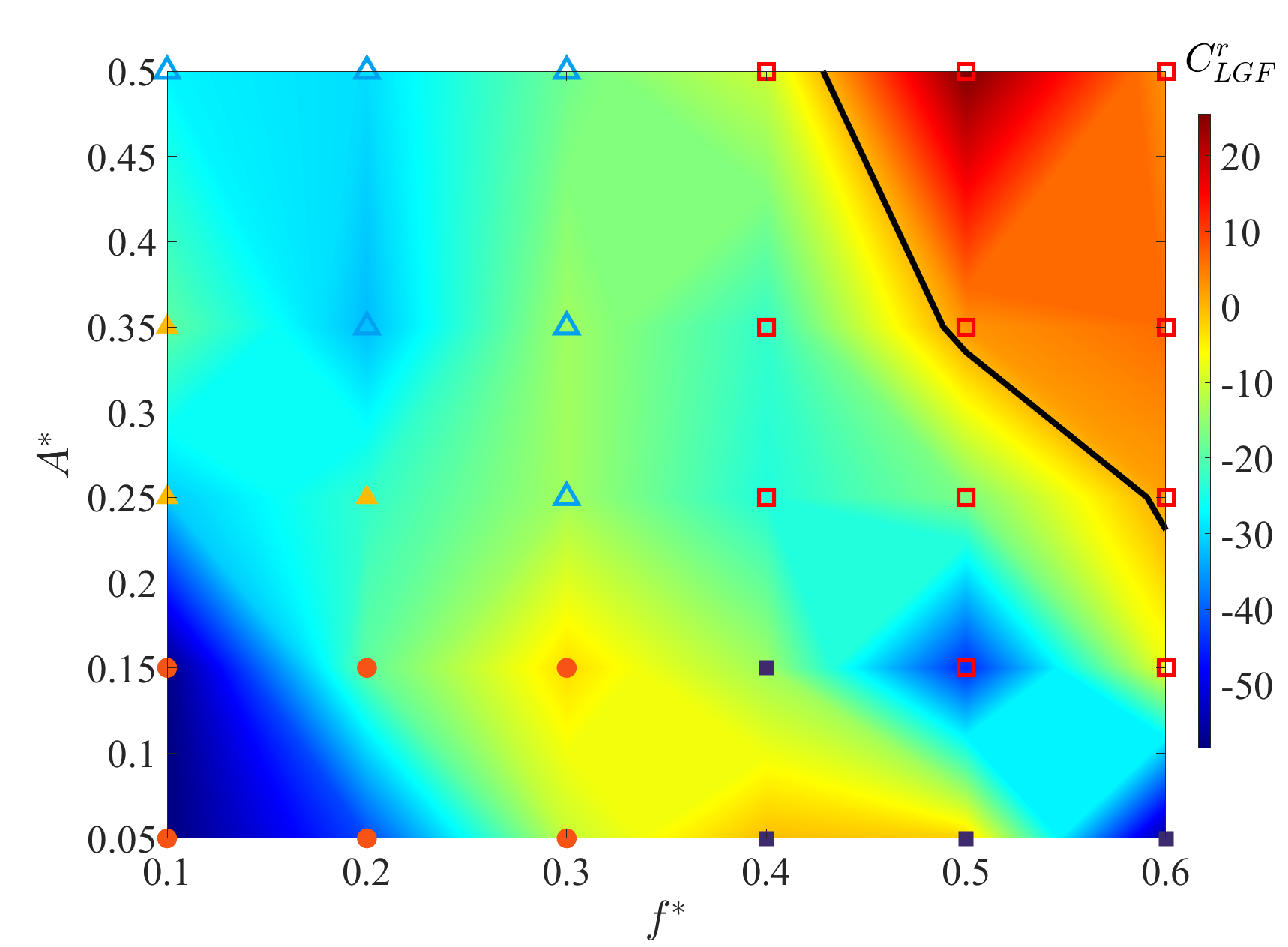}\label{crossa}}
	\subfloat[]{\includegraphics[width=0.33 \textwidth]{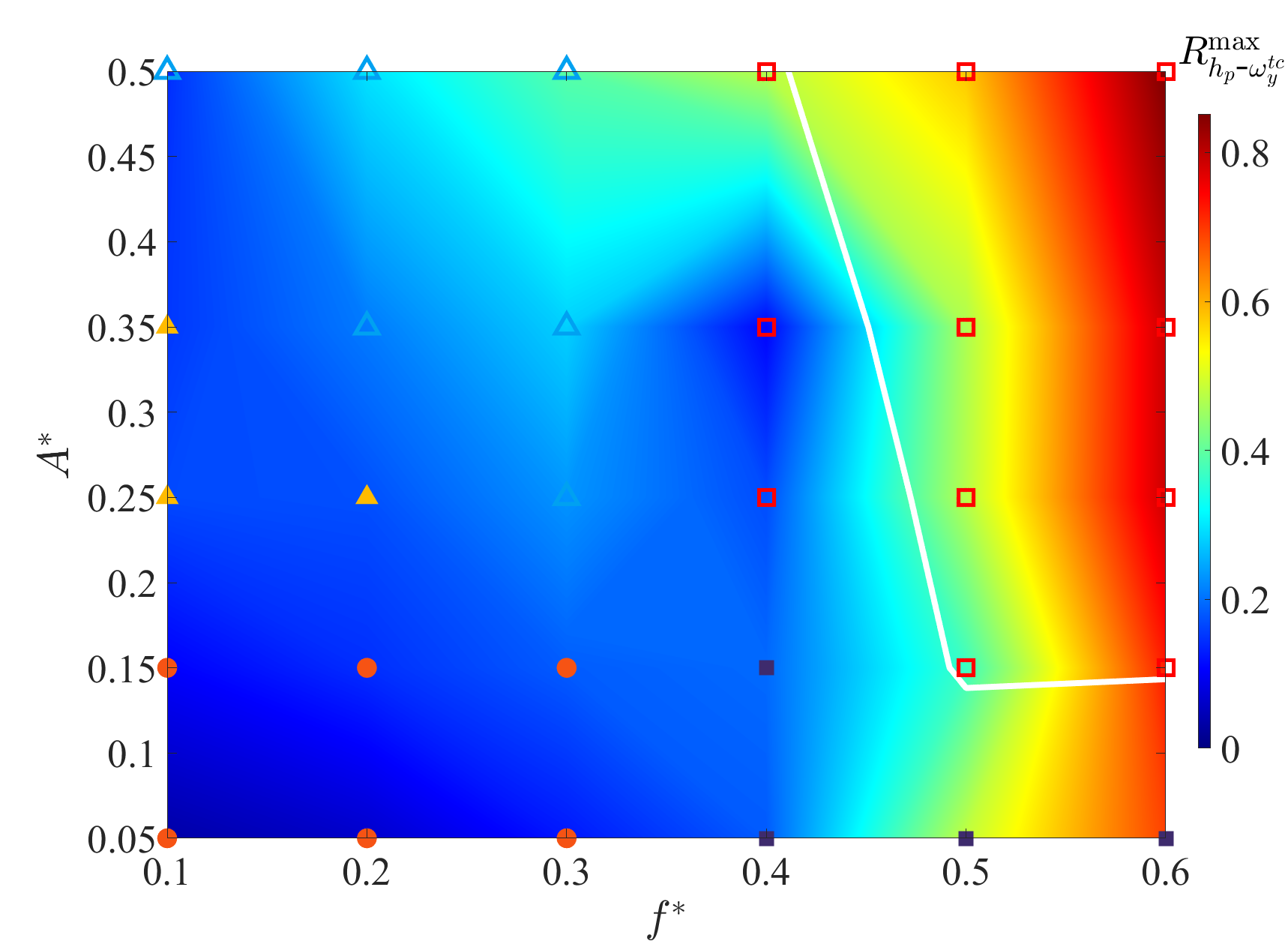}\label{crossb}}
	\subfloat[]{\includegraphics[width=0.33 \textwidth]{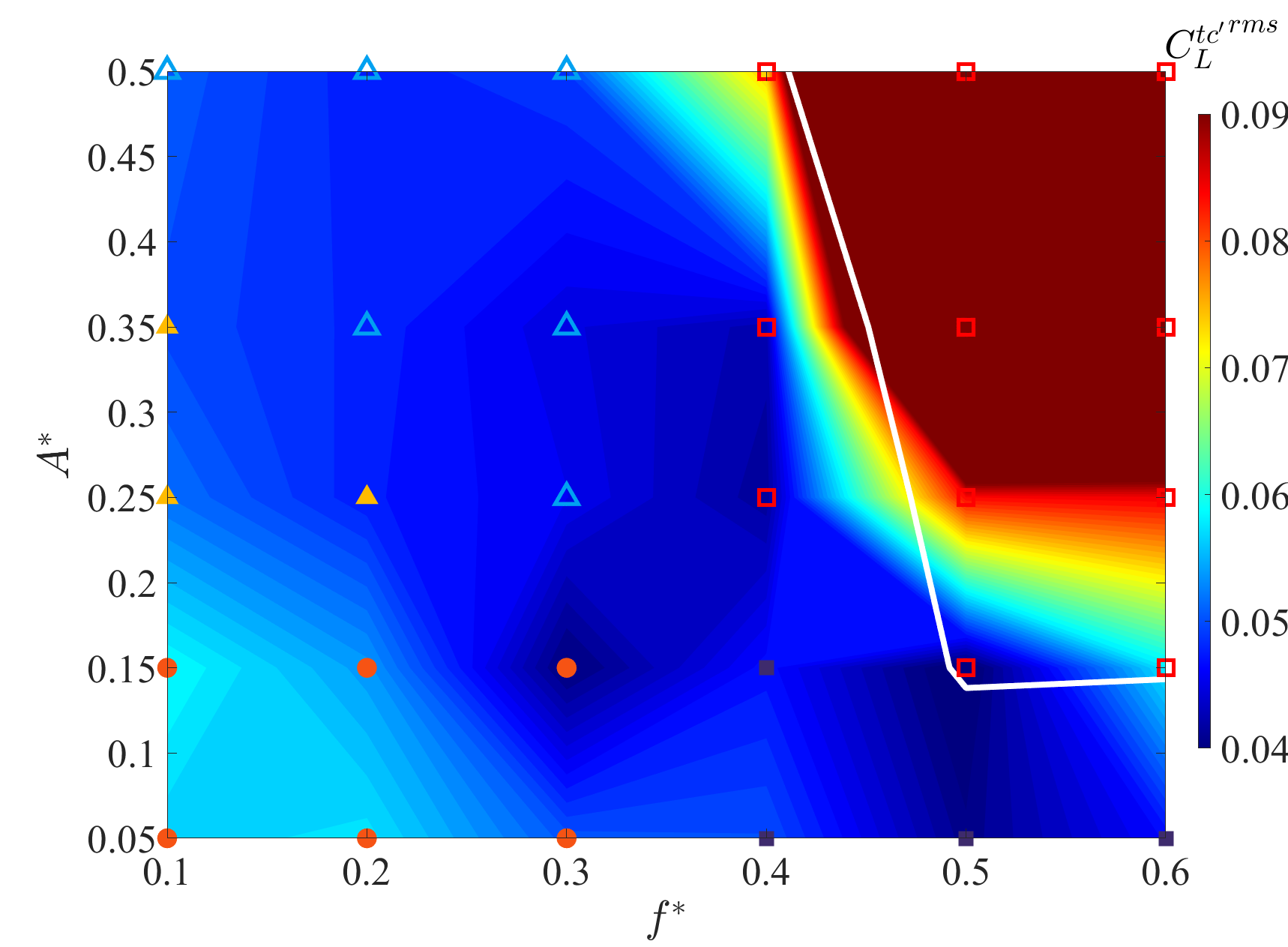}\label{crossc}}
	\caption{\label{cross}  (a) Lift-gain factor between the tandem rigid configuration and the isolated flexible case, (b) maximum cross-correlation coefficient between heaving motion and vortex shedding from upstream cylinder and (c) r.m.s value of the lift fluctuation of the upstream cylinder influenced by downstream heaving flexible foil. The isoline in white color represents the zero lift-gain factor value of the tandem flexible foil. The isoline in black color denotes the zero value of the lift-gain factor of the tandem rigid structure.}
\end{figure}

Contours of these three factors in percentage form in the $f^*$--$A^*$ space are presented in \reffigs{wake} \subref{wakea}-\subref{wakec}, respectively. The black isolines with zero value are added to the contours, which separate the positive and negative contributions for these three factors. It is worth noting that in VF mode, both lift-gain and drag-penalty factors reach positive values. In this situation, the wake plays a boosting role in improving aerodynamics rather than hindering it. However, the flexible foil suffers from the axial momentum deficit of the incoming wake by observing from the horizontal velocity profile comparison in \reffig{wake} \subref{waked}. This velocity-loss effect in the freestream direction degrades the aerodynamic performance, which is also validated by the negative decomposed velocity-loss terms in \refeqs{eq:cl_non}, (\ref{eq:cd_non}) and (\ref{eq:cpow_non}). There must be some underlying mechanisms that can improve aerodynamic performance by modulating entire flow features of the coupled system. To confirm this idea, the flexibility role in improving performance is excluded by calculating the lift-gain factor between the tandem rigid configuration and the isolated flexible case $C_{LGF}^r = ({\overline{C}_L^{tr} - \overline{C}_L^{if}})/{\overline{C}_L^{if}}$. It is observed from \reffig{cross} \subref{crossa} that the rigid structure can still produce better lift performance immersed in wake flows than the isolated flexible case in some VF mode regimes. Thus, the synchronization between heaving motion and wake flows, the so-called two-way lock-in, becomes the key point to enhance performance at large heaving amplitude and frequency.

\begin{figure}
	\centering
	\subfloat[]{\includegraphics[width=0.25 \textwidth]{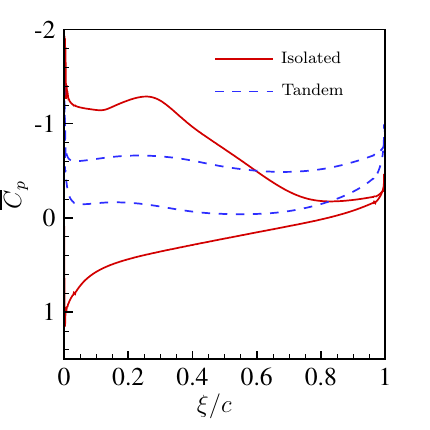}\label{wake_cpa}}
	\subfloat[]{\includegraphics[width=0.45 \textwidth]{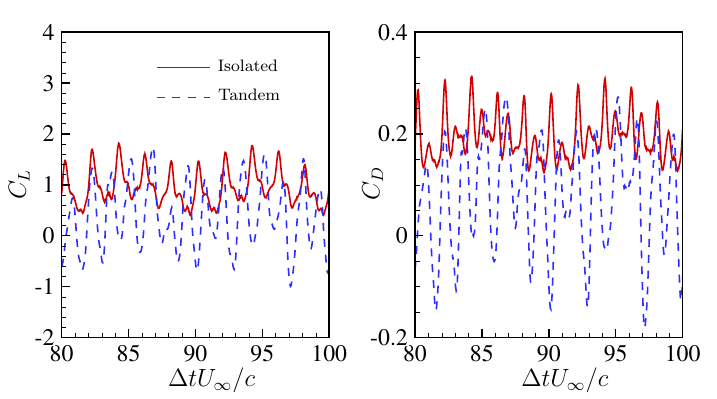}\label{wake_cpb}}
	\
	\subfloat[]{\includegraphics[width=0.24 \textwidth]{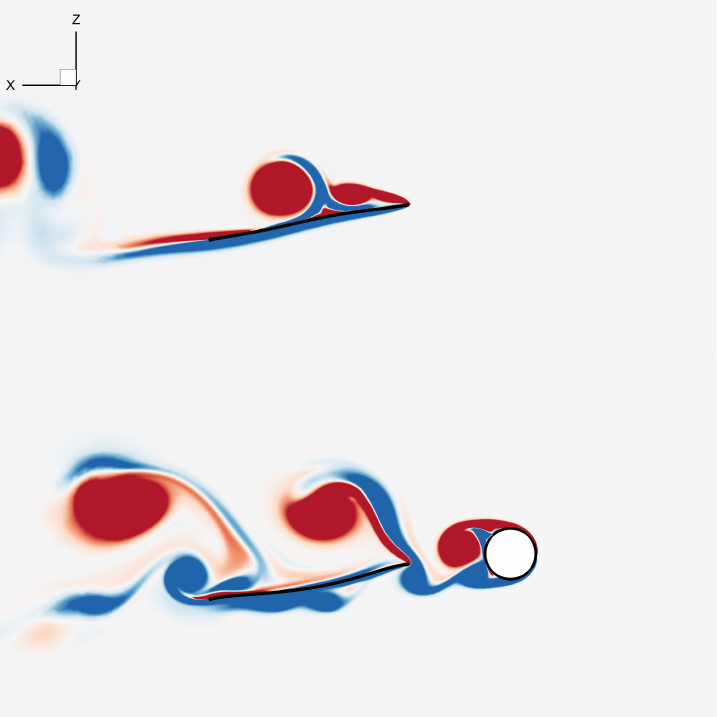}\label{wake_cpc}}
	\\
	\subfloat[]{\includegraphics[width=0.25 \textwidth]{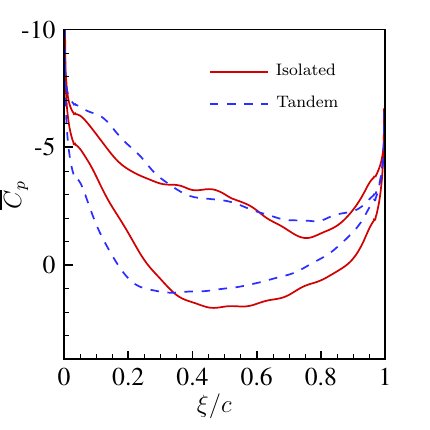}\label{wake_cpd}}
	\subfloat[]{\includegraphics[width=0.45 \textwidth]{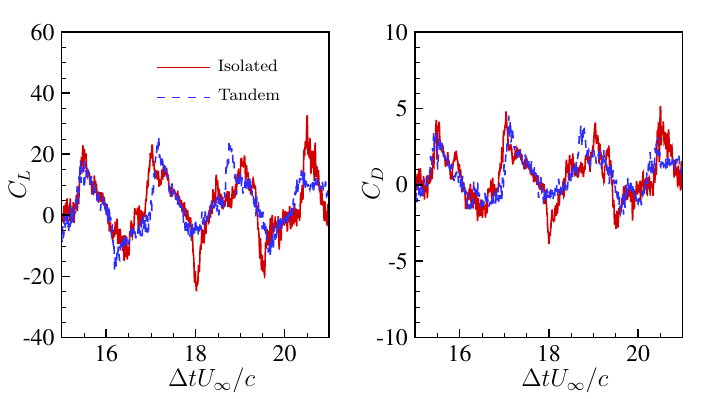}\label{wake_cpe}}
	\
	\subfloat[]{\includegraphics[width=0.24 \textwidth]{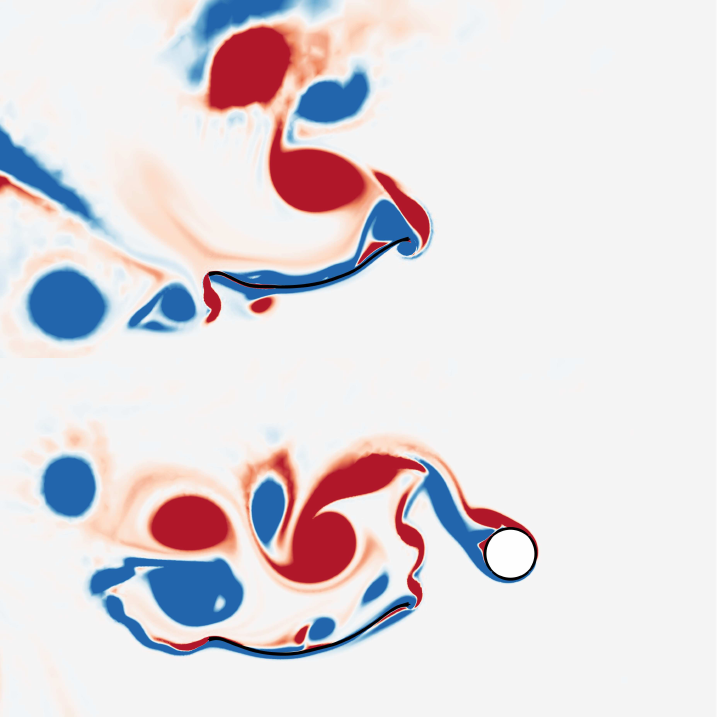}\label{wake_cpf}}
	\\
	\subfloat[]{\includegraphics[width=0.5 \textwidth]{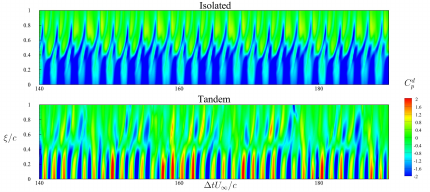}\label{wake_cpg}}
	\subfloat[]{\includegraphics[width=0.5 \textwidth]{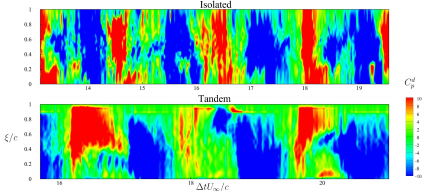}\label{wake_cph}}
	\caption{\label{wake_cp} Comparison of time-averaged (a,d) pressure coefficient distribution along the foil chord, (b,e) lift and drag coefficients, (c,f) instantaneous $y$-vorticity during downstroke  and (g,h) time-varying pressure coefficient difference of upper and lower surfaces between the isolated and tandem configurations at (a,b,c,g) [$f^*$,$A^*$]=[0.1, 0.05] and (d,e,f,h) [$f^*$,$A^*$]=[0.6, 0.25].}
\end{figure}

To further analyze flow characteristics and aerodynamic forces influenced by wake flows, two representative cases of the flexible tandem foil are selected in both the negative and positive regions of the lift-gain factor contour shown in \reffig{wake} \subref{wakea}. The time-averaged pressure coefficient distribution, instantaneous aerodynamic forces and $y$-vorticity contours are compared for isolated and tandem flexible foils in \reffig{wake_cp}. It can be seen from \reffig{wake_cp} \subref{wake_cpc} that low-momentum wake flows generated from the upstream cylinder are merged with the vortices around the foil. This axial momentum deficit of the incoming wake results in a reduced suction effect near the leading edge of the flexible foil, as shown in \reffigs{wake_cp} \subref{wake_cpa} and \subref{wake_cpg}. As a result, the tandem flexible foil generates degraded aerodynamic forces compared to the isolated case at [$f^*$,$A^*$]=[0.1, 0.05], as plotted in \reffig{wake} \subref{wakeb}. 

The tandem flexible foil exhibits enhanced suction near the leading edge at the heaving top position and weakened positive pressure difference during downstroke  at [$f^*$,$A^*$]=[0.6, 0.25], compared to its isolated counterpart, as depicted in \reffigs{wake_cp} \subref{wake_cpa} and \subref{wake_cph}. The overall lift and drag forces of the tandem case plotted in \reffig{wake_cp} \subref{wake_cpe} are improved. The decomposed FMD modes in \reffig{fre_mode} \subref{fre_modea} and the instantaneous vorticity contours in \reffig{wake_cp} \subref{wake_cpf} reveal that the heaving motion of the flexible foil inversely interacts back to the upstream cylinder, altering its vortex-shedding pattern. To understand this two-way frequency lock-in phenomenon, we calculate the cross-correlation between the heaving motion and the vorticity field, and the lift fluctuation of the cylinder to analyze how the heaving motion inversely affect wake flows. 

The cross-correlation $R_{h_p \textrm{-} \omega_y^{tc}}$ of the discrete time signals of heaving motion $h_p(t_n)$ and $y$-vorticity $\omega_y^{tc}(t_{n})$ is defined as
\begin{equation}
	R_{h_p \textrm{-} \omega_y^{tc}}(m) = \frac{1}{N} \sum_{n=0}^{N-m-1} h_p(t_n) \omega_y^{tc}(t_{n+m}) \quad \quad 0 \le m \le N-1  ,
\end{equation}
where $N$ represents the data samples. We selected 4096 samples of the heaving motion and the $y$-vorticity values behind the cylinder to obtain the maximum cross-correlation coefficient, which is displayed in \reffig{cross} \subref{crossb}. The rms value of the lift fluctuation for the upstream cylinder shown in \reffig{cross} \subref{crossc} is defined as
\begin{equation}
	 {C_{L}^{{tc}^{\prime}}}^{rms} = \sqrt{\frac{1}{N} \sum_{n=1}^{N}(C_L^{tc} - \overline{C}_L^{tc})^2},
\end{equation}

In \reffigs{cross} \subref{crossb} and \subref{crossc}, a white line representing the value of the zero lift-gain factor of the tandem flexible foil is added to the plot to examine the positive contribution of the two-way lock-in. It is found that when the flexible foil moves beyond the white line, the vortex structure shed from the upstream cylinder is closely correlated with the heaving motion downstream. Large lift fluctuations of the cylinder are noticed on the right side of the critical lift-gain boundary. This phenomenon indicates that vortices shed from the cylinder are synchronized with the heaving motion, thereby creating extra suction and enhancing aerodynamic forces. The wake then becomes a boost rather than a hindrance on propulsive aerodynamic performance, similar to how fish use the wake of a fish school to swim efficiently \citep{chen2016swimming,fish2006passive}.

\subsection{Role of flexibility} \label{sec4.5:flexible}
\begin{figure}
	\centering
	\subfloat[]{\includegraphics[width=0.33 \textwidth]{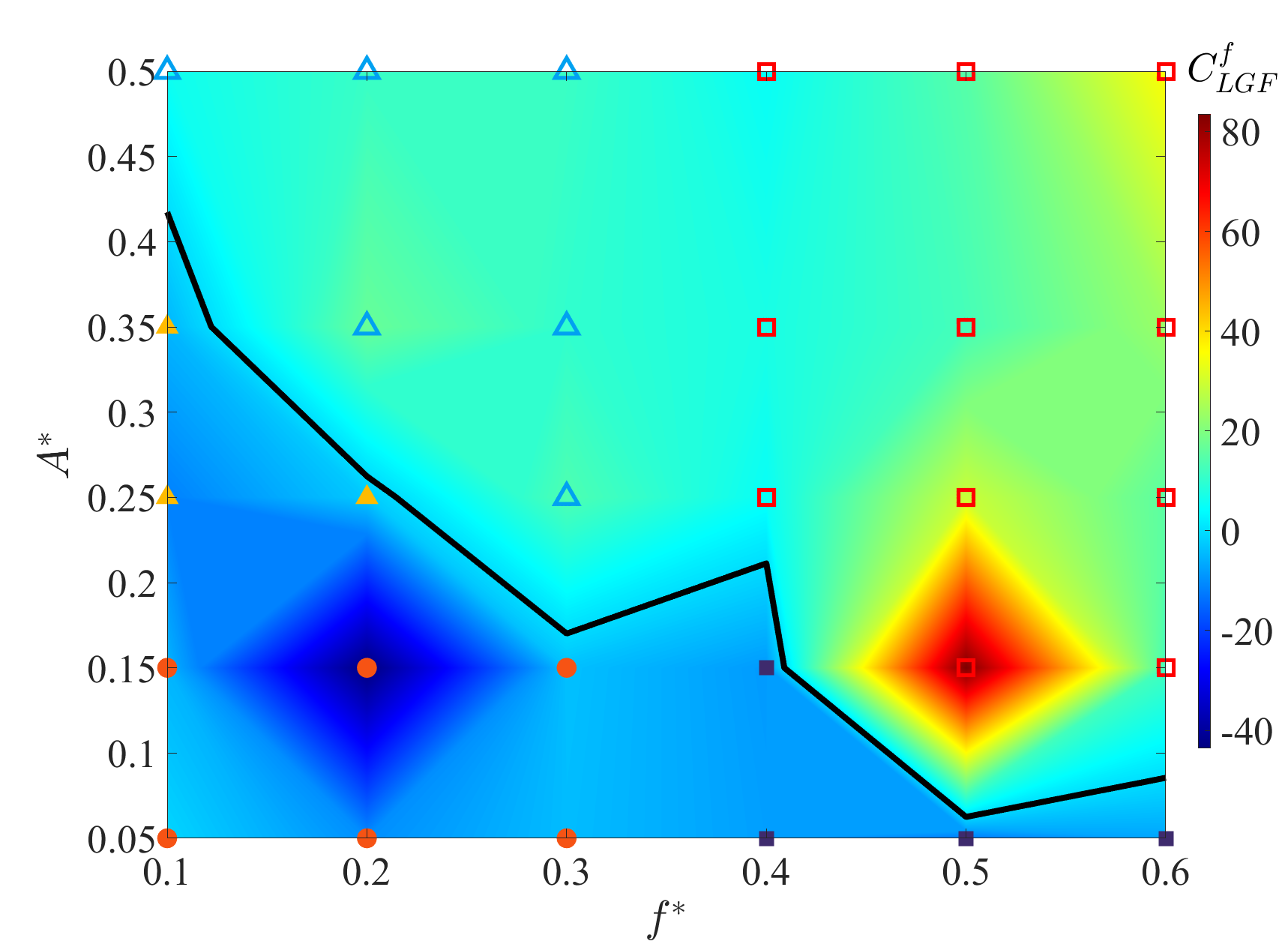}\label{flea}}
	\subfloat[]{\includegraphics[width=0.33 \textwidth]{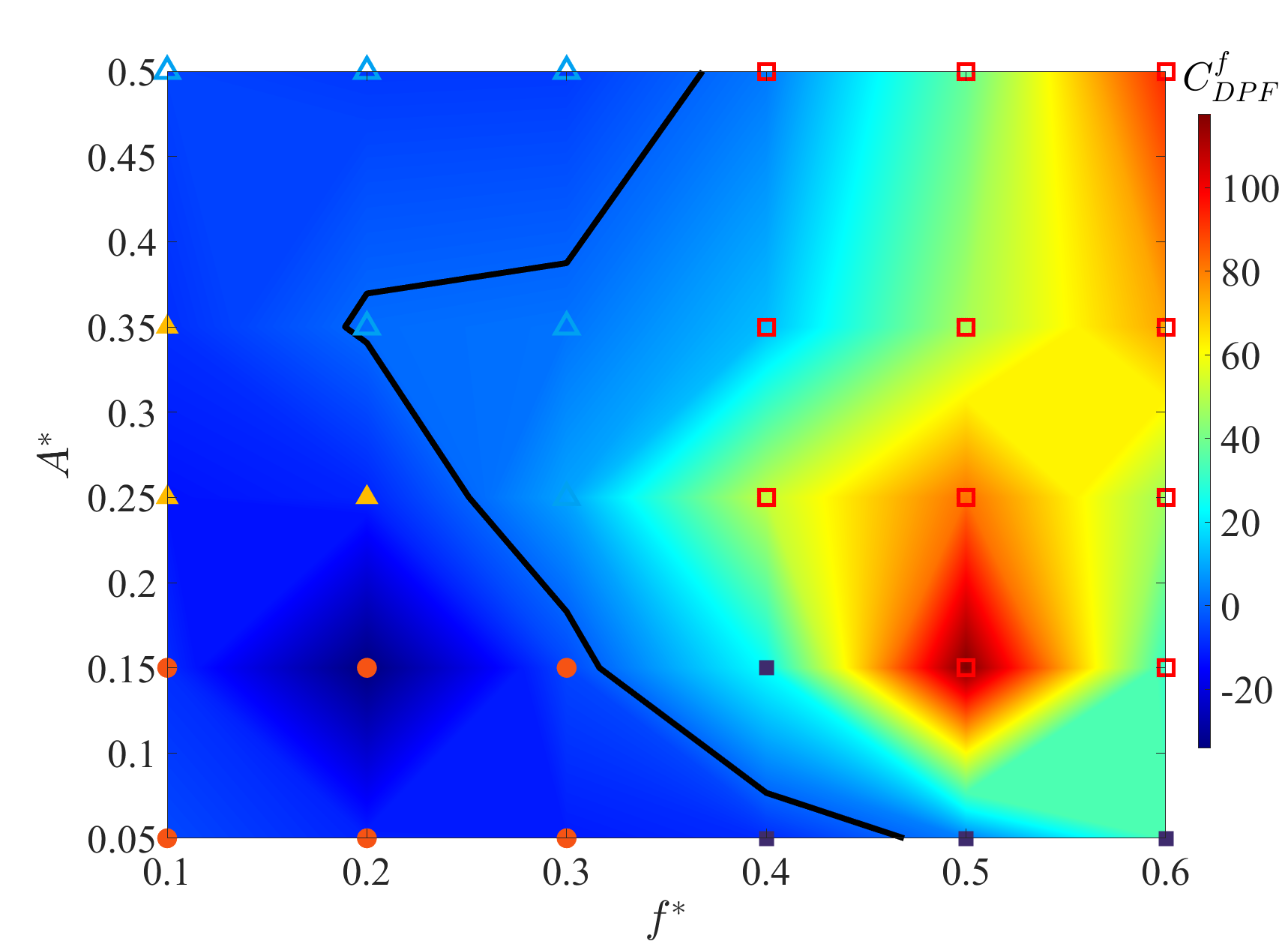}\label{fleb}}
	\subfloat[]{\includegraphics[width=0.33 \textwidth]{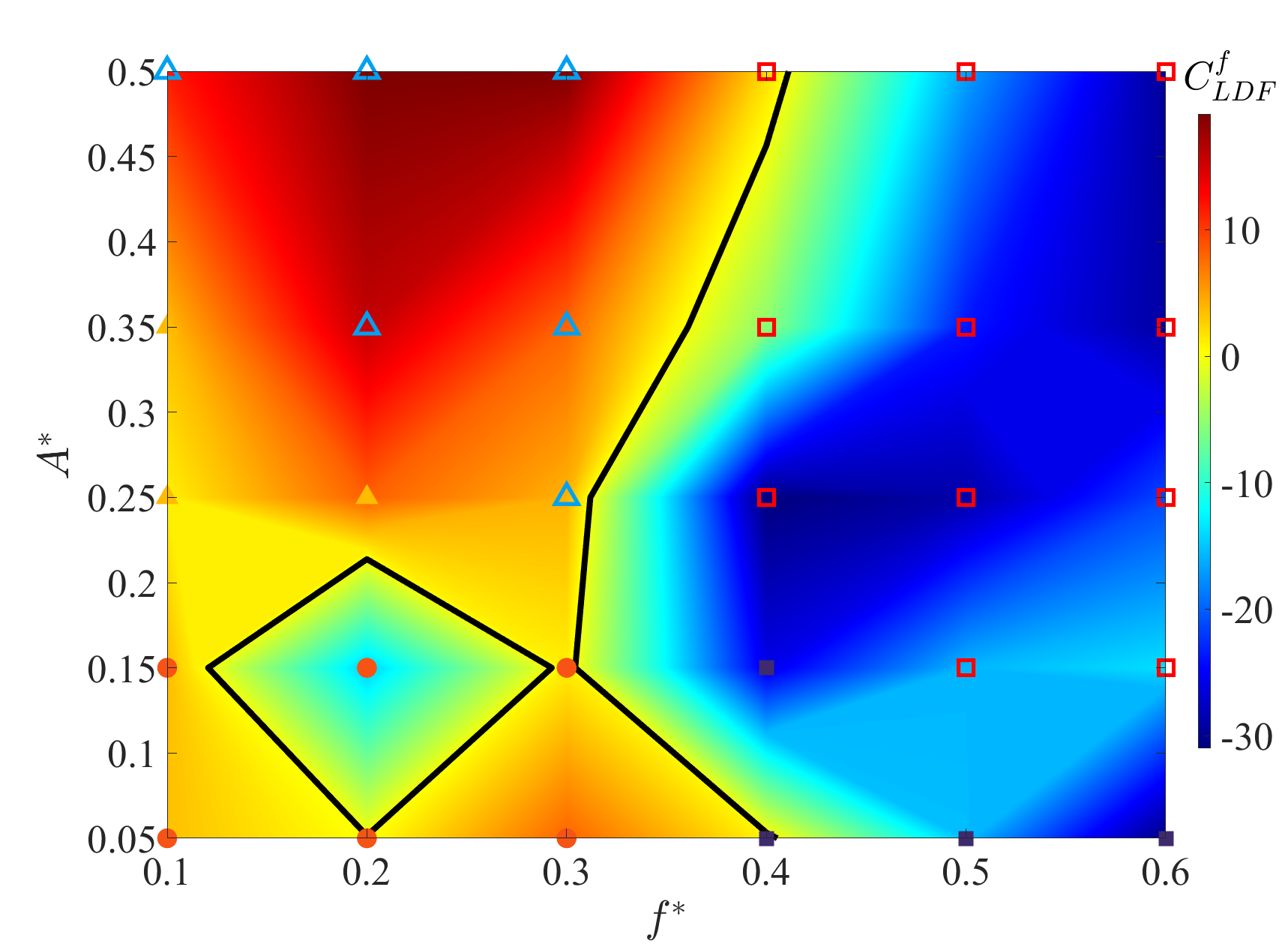}\label{flec}}
	\caption{\label{fle} Phase diagram of (a) lift-gain factor, (b) drag-penalty factor and (c) lift-to-drag factor of the flexible foil due to flexibility effect in percentage form. The isoline in black color indicates the zero value in the phase diagram.}
\end{figure}

The proposed scaling relation in \refeq{eq:cl_non} reveals that the passive deformation of the heaving flexible foil exhibits positive contribution to the lift force. The decomposed lift term associated with curvature has a positive fitting coefficient of 2.998 shown in \reffig{cl_fit} \subref{cl_fitb}. The negative fitting coefficient for the decomposed drag component indicates that the drag force is reduced by the foil's bending. In this section, the propulsive aerodynamic performance of the tandem flexible foil is compared to its rigid counterpart. The purpose is to focus only on the impact of flexibility effect on the propulsive aerodynamic performance in the fluid-structure interaction system. In order to quantitatively evaluate the role of flexibility on the aerodynamic changes relative to the rigid structure, three dimensionless factors are calculated
\begin{equation}
	C_{LGF}^f = \frac{\overline{C}_L^{tf} - \overline{C}_L^{tr}}{\overline{C}_L^{tr}}, \quad C_{DPF}^f = \frac{\overline{C}_D^{tf} - \overline{C}_D^{tr}}{\overline{C}_D^{tr}}, \quad 
	C_{LDF}^f = \frac{\overline{{C}_L^{tf}/{C}_D^{tf}} - \overline{{C}_L^{tr}/{C}_D^{tr}}}{\overline{{C}_L^{tr}/{C}_D^{tr}}} ,
\end{equation}
where $\overline{C}_L^{tf}$, $\overline{C}_D^{tf}$ and $\overline{{C}_L^{tf}/{C}_D^{tf}}$ are the aerodynamic performance of the tandem flexible foil. Variables with a superscript of $tr$ refer to rigid structures. Phase diagrams of these three dimensionless factors are plotted in percentage form in \reffig{fle}. A black isoline with zero value is added in these contour plots to divide the positive and negative regions. 

\begin{figure}
	\centering
	\subfloat[]{\includegraphics[width=0.33 \textwidth]{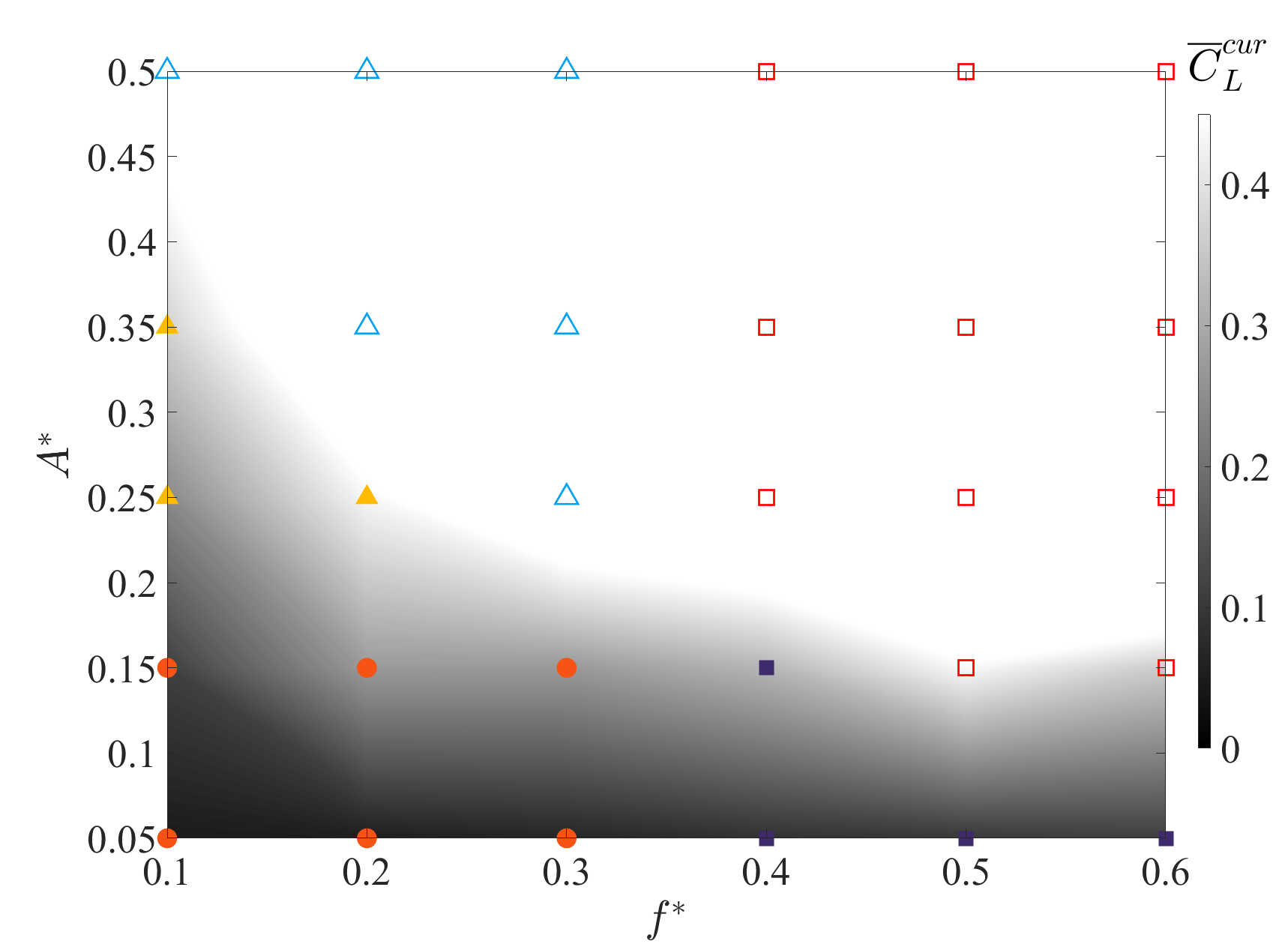}\label{mecha}}
	\subfloat[]{\includegraphics[width=0.33 \textwidth]{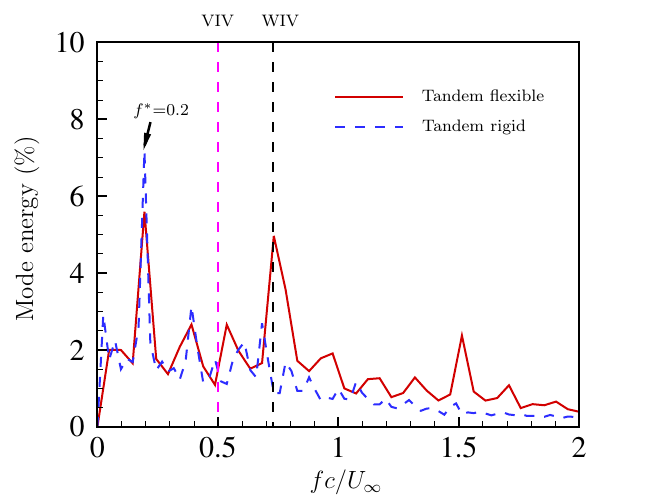}\label{mechb}}
	\subfloat[]{\includegraphics[width=0.32 \textwidth]{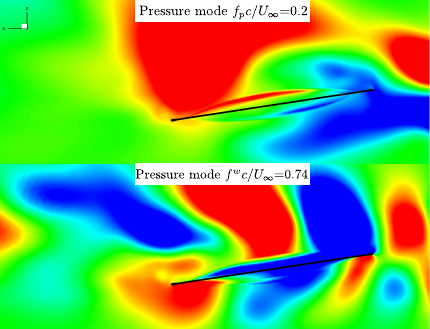}\label{mechc}}
	\caption{\label{mech}  (a) Phase diagram of lift coefficient component contributed by the structural passive deformation. (b) Comparison of frequency characteristics of lift performance between the tandem flexible foil and its rigid case and (c) FMD modes extracted from pressure field for tandem flexible foil at [$f^*$,$A^*$]=[0.2, 0.15].}
\end{figure}

In \reffig{fle} \subref{flea}, the flexible foil performs better lift performance than its rigid counterpart when the coupled system transitions to SWF and VF modes. The lift component contributed by the structural deformation in \reffig{mech} \subref{mecha} is greatly enhanced in these two modes. It is mainly attributed to the enlarged passive deformation shown in \reffig{aero} \subref{aerod} caused by the strong added-mass effect under large heaving amplitude and frequency. The drag penalty it receives is determined more by the frequency of the heaving motion than by the amplitude, as observed in \reffig{fle} \subref{fleb}. It can be seen from \reffig{fle} \subref{flec} that the SWF mode of the flexible foil achieves better lift-to-drag ratio than the rigid case. By comparing instantaneous flow features, pressure distributions plotted in \reffigs{aerodynamic} and \ref{com}, the foil’s bending changes the local flow to create extra suction in the SWF mode, improving lift without much extra drag. 

\begin{figure}
	\centering
	\subfloat[]{\includegraphics[width=0.25 \textwidth]{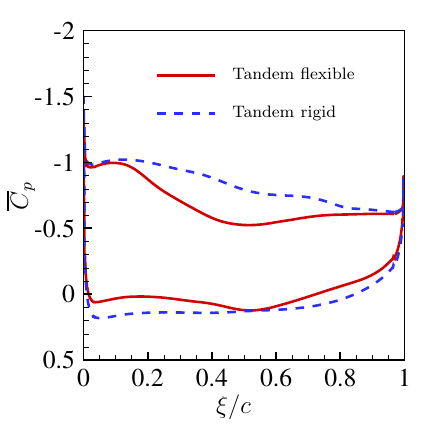}\label{fle_cpa}}
	\subfloat[]{\includegraphics[width=0.45 \textwidth]{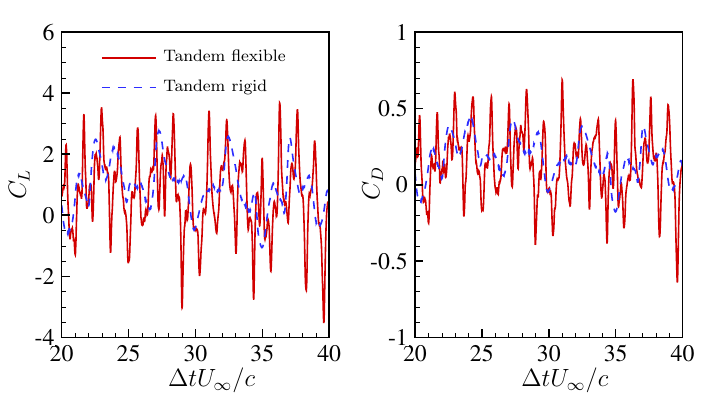}\label{fle_cpb}}
	\
	\subfloat[]{\includegraphics[width=0.24 \textwidth]{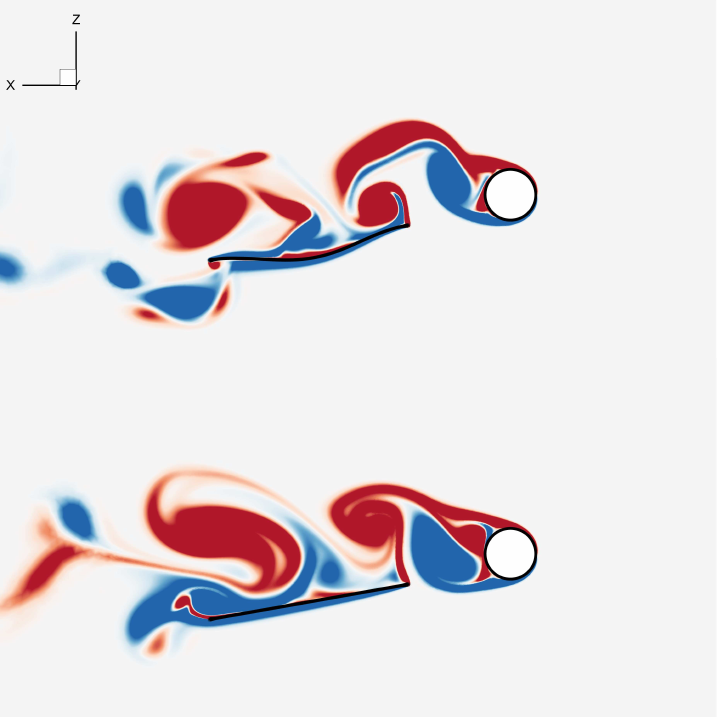}\label{fle_cpc}}
	\\
	\subfloat[]{\includegraphics[width=0.25 \textwidth]{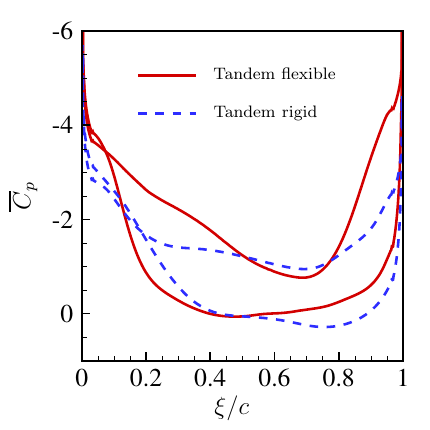}\label{fle_cpd}}
	\subfloat[]{\includegraphics[width=0.45 \textwidth]{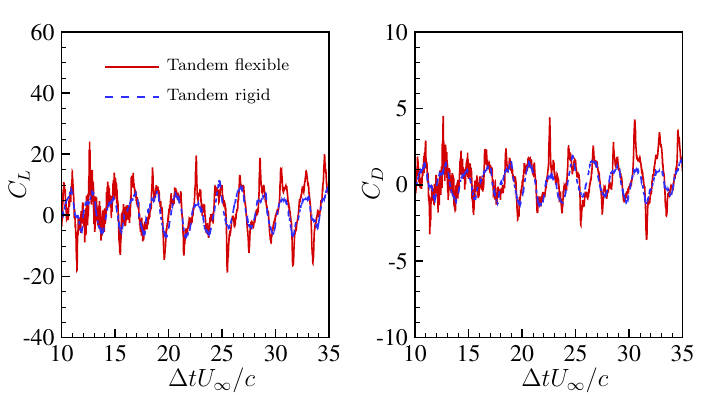}\label{fle_cpe}}
	\
	\subfloat[]{\includegraphics[width=0.24 \textwidth]{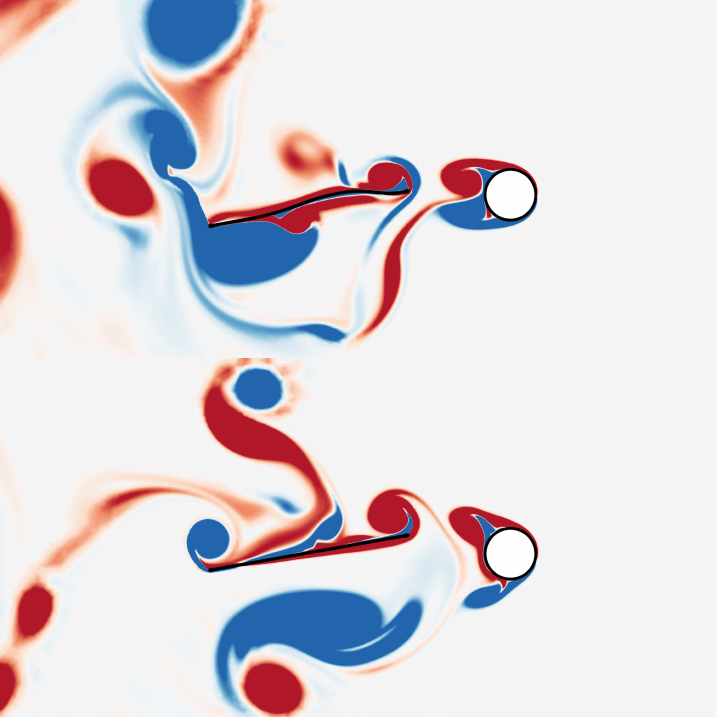}\label{fle_cpf}}
	\\
	\subfloat[]{\includegraphics[width=0.5 \textwidth]{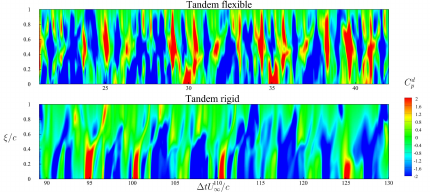}\label{fle_cpg}}
	\subfloat[]{\includegraphics[width=0.5 \textwidth]{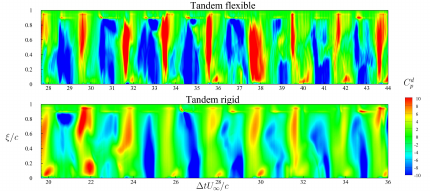}\label{fle_cph}}
	\caption{\label{fle_cp} Comparison of time-averaged (a,d) pressure coefficient distribution along the foil chord, (b,e) lift and drag coefficients and (c,f) instantaneous $y$-vorticity during downstroke and (g,h) time-varying pressure coefficient difference of upper and lower surfaces between the tandem rigid and flexible configurations at (a,b,c,g) [$f^*$,$A^*$]=[0.2, 0.15] and (d,e,f,h) [$f^*$,$A^*$]=[0.5, 0.25].}
\end{figure}

The structural flexibility plays a negative role in lift enhancement in fully immersed wake flows, while it exhibits positive contribution in partial wake. These findings suggest that the interaction between wake, heave motion, and structural flexibility modulates surrounding flow characteristics according to some underlying mechanisms, thereby adjusting lift performance. Two typical cases on both sides of the isoline are selected from \reffig{fle} \subref{flea} to examine how the structural flexibility turns the wake from a hindrance into a boost. A comparison of time-averaged pressure distribution, instantaneous forces, vorticity and pressure difference is displayed in \reffig{fle_cp} for the tandem flexible foil and its rigid case at [$f^*$,$A^*$]=[0.2, 0.15] and [0.5, 0.25]. When the flexible foil moves within wake flows, it deforms downward by merging with detached vortices from the cylinder at the bottom position, as shown in \reffig{fle_cp} \subref{fle_cpc}. This reverse bending shape introduces additional negative angles of attack, producing weaker suction and lower aerodynamic forces, as depicted in \reffigs{fle_cp} \subref{fle_cpa} and \subref{fle_cpb} respectively. High-frequency components are observed in time-varying aerodynamic forces and pressure difference contours in \reffigs{fle_cp} \subref{fle_cpb} and \subref{fle_cpg} for the tandem flexible foil. By examining frequency characteristics in \reffig{mech} \subref{mechb}, the flexible case responds to wake flows with a sub-frequency peak at WIV mode, while the rigid configuration shows less correlations with disturbed wake. 

To gain further insight into the excited WIV mode, FMD modes at the heaving frequency $f_p c/ U_{\infty}=0.2$ and vortex-shedding frequency $f^w c/ U_{\infty}=0.74$ are extracted from time-varying pressure field in the flapping-body reference frame, as plotted in \reffig{mech} \subref{mechc}. These pressure distributions at vortex-shedding frequency is caused by the interaction between the reverse bending deformation and merged wake flows observed in \reffig{fle_cp} \subref{fle_cpc}. From the view point of energy transfer, vibration energies  are redistributed to the vortex-shedding frequency, leading to reduced aerodynamic performance in full wake. This energy redistribution phenomenon is further discussed in \refse{sec4.6:performance}. 

Once the flexible foil heaves exceeding the critical boundary line shown in \reffig{mode}, the passive deformation supplies additional transverse momentum to compensates the axial momentum deficit of the incoming wakes, producing better lift performance than the rigid case. The flexible foil deforms upward at the top position in \reffig{fle_cp} \subref{fle_cpf}, generating an additional positive angle of attack. Moreover, the heaving motion reversely controls vortex-shedding pattern in a two-way frequency-synchronized manner. The bending shape and synchronized leading-edge vortices create extra suction near the leading edge, thereby producing overall better lift performance. Under this situation, the vibration energy is mainly concentrated in the heaving frequency, rather than being attracted by the vortex-shedding frequency.

\subsection{Propulsive performance dynamics} \label{sec4.6:performance}
Aquatic and aerial animals move efficiently in disturbed wakes by appropriately adjusting their flapping kinematics and exploiting vortex patterns \citep{deng2007hydrodynamic,maertens2017optimal}. Depending on the demands of movement, these animals conserve energy for cruising, or thrust hard for explosive starts and high maneuverability. These flexible biological propulsion methods provide scientists with rich experience in designing efficient unmanned aerial/underwater vehicles with optimal flapping kinematics. The scaling relation for the power coefficient proposed in \refeq{eq:cpow_non} indicates that large amplitude, high-frequency heaving motion, and passive deformation consume more power, while wake flows help save energy. Thus, it is important to investigate the connection between flapping kinematics and propulsive aerodynamic performance, revealing the underlying flow control mechanism and summarizing the design guidelines.

\begin{figure}
	\centering
	\subfloat[]{\includegraphics[width=0.33 \textwidth]{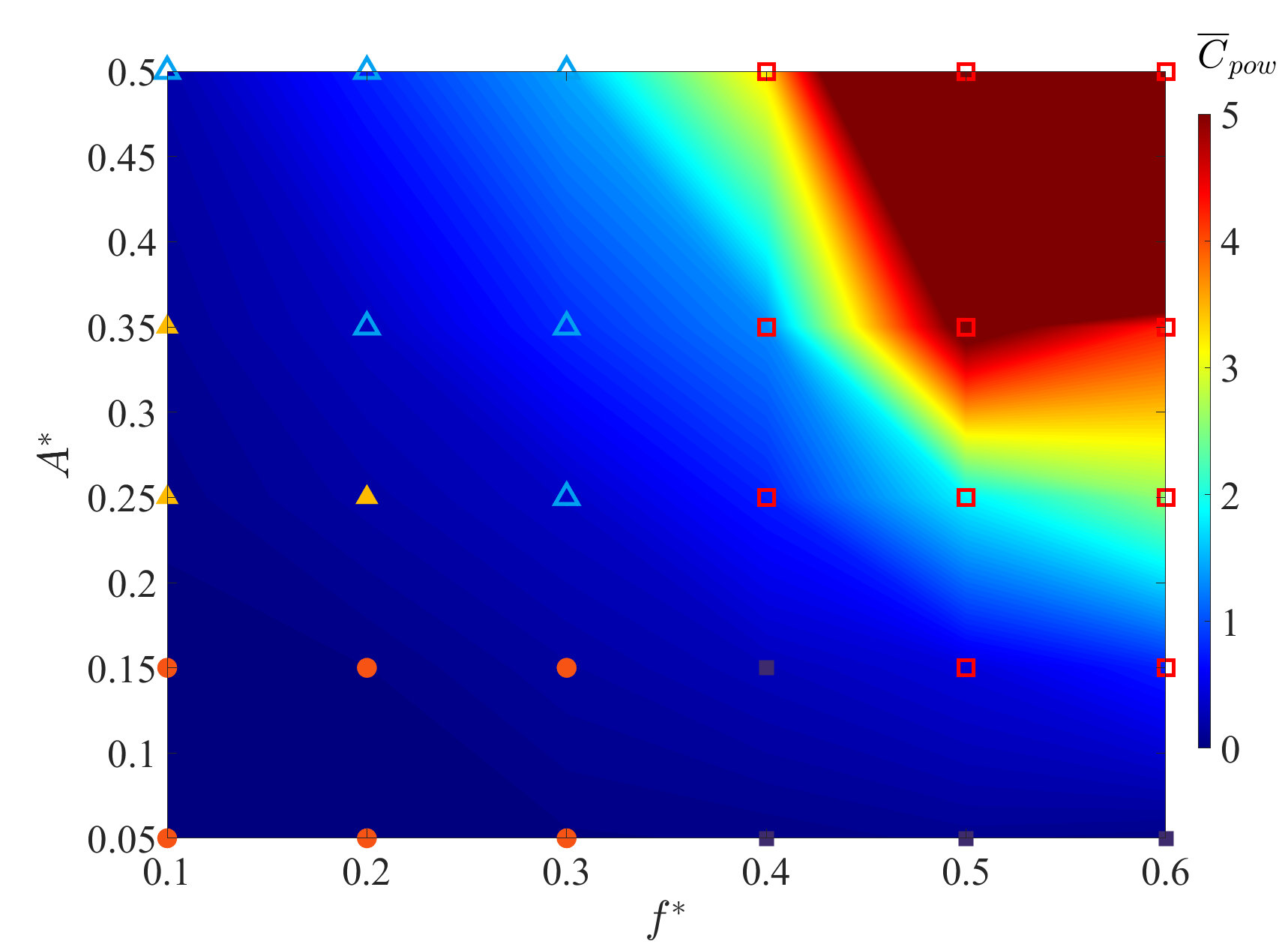}\label{energya}}
	\subfloat[]{\includegraphics[width=0.33 \textwidth]{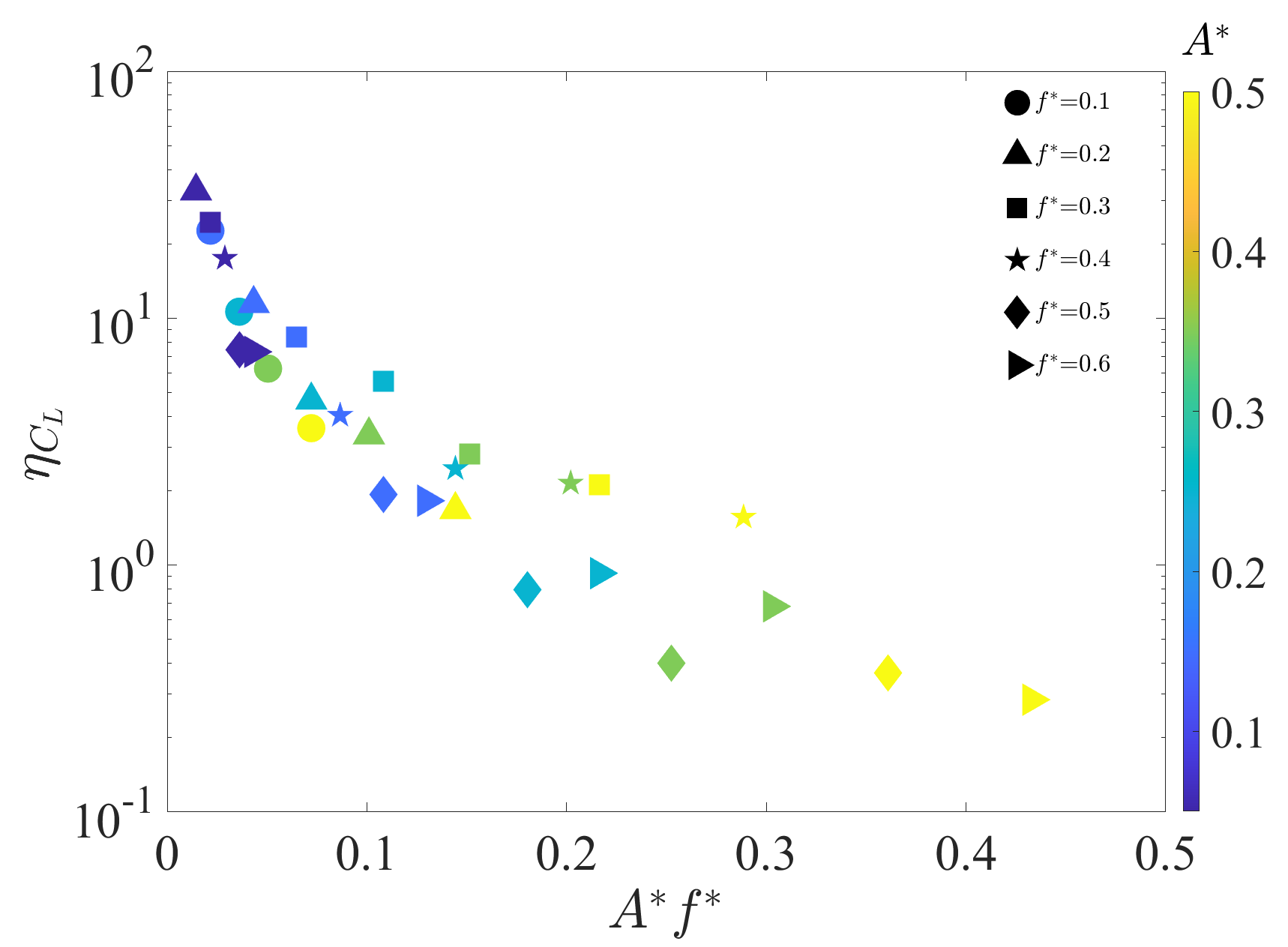}\label{energyb}}
	\subfloat[]{\includegraphics[width=0.33 \textwidth]{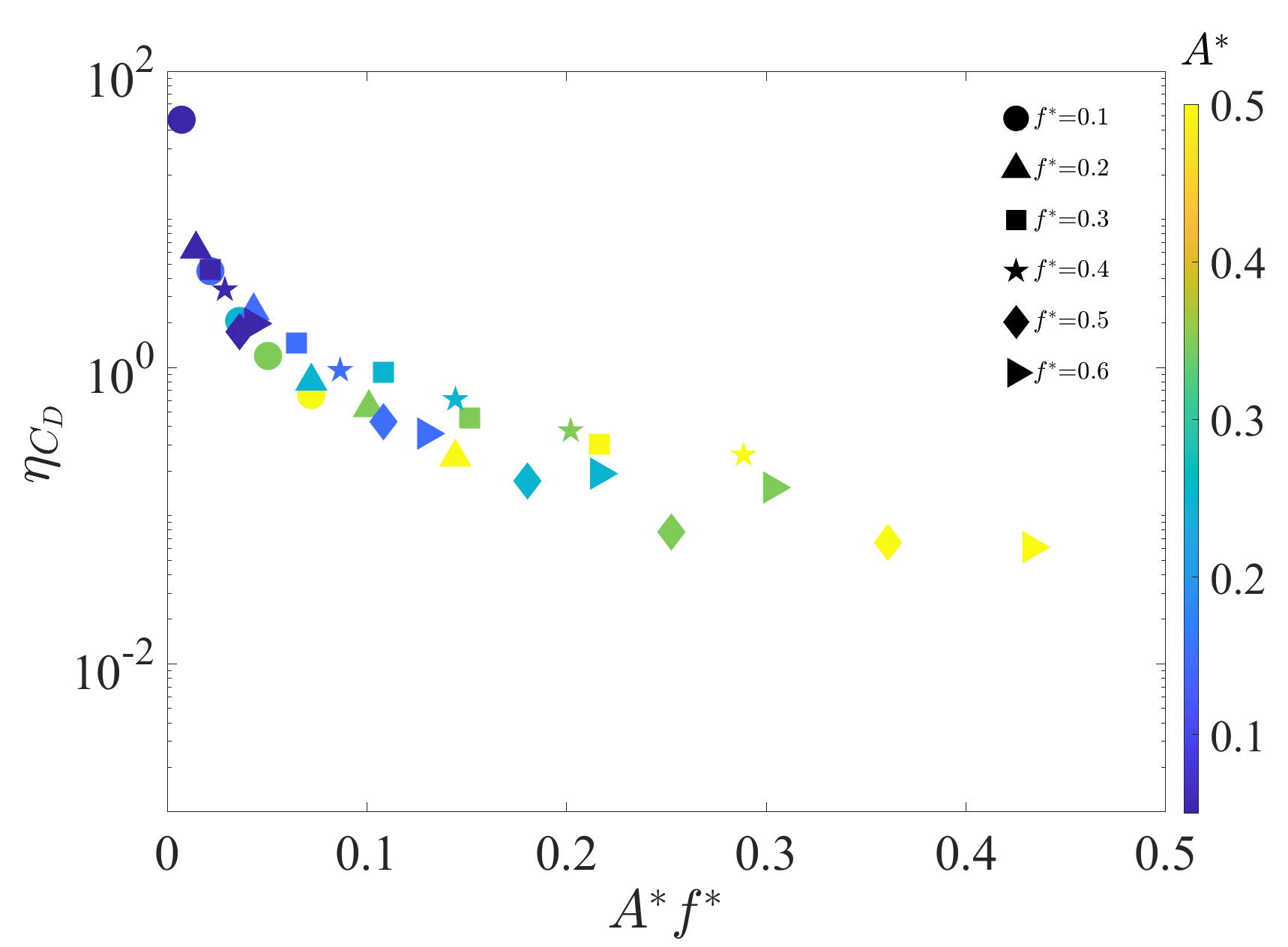}\label{energyc}}
	\caption{\label{energy} Phase diagram of (a) time-averaged power coefficient in the $f^*$--$A^*$ space, (b) propulsive efficiency of lift and (c) propulsive efficiency of drag as a function of dimensionless heaving amplitude and frequency for tandem flexible foil.}
\end{figure}

\refFig{energy} \subref{energya} presents a phase diagram of the time-averaged power coefficient for the tandem flexible foil in the $f^*$--$A^*$ space. The flexible foil consumes low power to maintain propulsion when it moves fully in wake flows or asynchronously with the vortex-shedding frequency. Meanwhile, it produces unfavorable aerodynamic forces even lower than the non-heaving isolated flexible case, as shown in \reffig{aero}. This type of propulsion strategy is not conducive to maneuverability, but is beneficial for energy saving, which is very suitable for efficient long-distance migration. Tightly schooling fish and organized flying birds use similar strategies to allow those behind them to conserve energy in their wake \citep{tian2021research,ligman2023comprehensive,bajec2009organized,beaumont2024aerodynamic}. Propulsive efficiency defined as the ratio between aerodynamic force and power coefficients is calculated to further analyze optimal propulsion manners for different locomotion scenarios. Propulsive efficiencies of lift and drag are expressed as
\begin{equation}
	\eta_{C_L} = \frac{\overline{C}_L}{\overline{C}_{pow}}, \quad \quad \eta_{C_D} = \frac{\overline{C}_D}{\overline{C}_{pow}},
\end{equation}

Previous studies have shown that the propulsion efficiency of flexible foils is proportional to the amplitude and frequency of flapping \citep{floryan2017scaling,lauber2023rapid}. In \reffigs{energy} \subref{energyb} and \subref{energyc}, the propulsive efficiencies of the lift and drag are collapsed onto inverse proportional functions with an overall downward trend. The variation of heaving frequency is represented by different shapes, while the change of heaving amplitude is colored from blue to yellow. It can be seen that the flexible foil maintains optimal propulsive efficiencies in FW, FWF and SW modes. This finding confirms that aquatic and aerial animals can move efficiently in specific organized groups with low flapping frequency or amplitude. As the heaving frequency and amplitude increase, the aerodynamic forces are greatly enhanced in the VF mode, as shown in \reffig{aero}, while consuming more energy (\reffig{energy} \subref{energya}). The propulsive efficiency drops rapidly when heaving in high amplitude and frequency, as depicted in \reffigs{energy} \subref{energyb} and \subref{energyc}. If animals want to escape the influence of low-momentum wake or avoid predators, they need to consume energy at least proportional to $(f^* A^*)^2$  to generate sufficient aerodynamic force. It is difficult for animals to maintain optimal propulsive efficiency while generating aerodynamic forces that provide enough maneuverability.

\begin{figure}
	\centering
	\subfloat[]{\includegraphics[width=0.5 \textwidth]{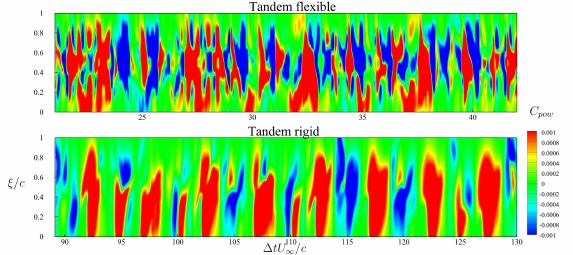}\label{efficiencya}}
	\subfloat[]{\includegraphics[width=0.5 \textwidth]{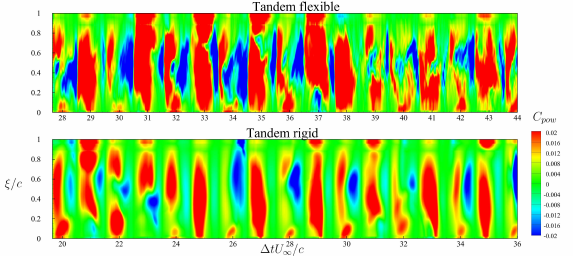}\label{efficiencyb}}
	\\
	\subfloat[]{\includegraphics[width=0.48 \textwidth]{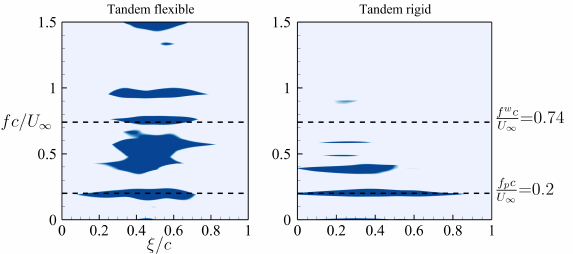}\label{efficiencyc}}
	\quad
	\subfloat[]{\includegraphics[width=0.48 \textwidth]{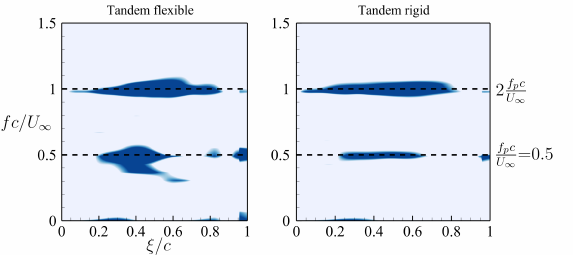}\label{efficiencyd}}
	\caption{\label{efficiency} Comparison of (a,b) space-time contours of propulsive power coefficient and (c,d) frequency spectrum of propulsive power coefficient along the length of heaving structure between tandem flexible configuration and its rigid counterpart at (a,c) [$f^*$,$A^*$]=[0.2, 0.15] and (b,d) [$f^*$,$A^*$]=[0.5, 0.25].}
\end{figure}

As discussed in \refse{sec4.5:flexible}, the tandem flexible foil generates less lift force than its rigid counterpart at low heaving amplitude or frequency conditions. By examining instantaneous flow features, pressure distributions, and decomposed FMD modes, the reason is attributed to the interaction between flexible foil and wake flows and the asynchronous coupling for heaving motion and shed vortices. The WIV mode excited by wake flows redistributes energies absorbed from fluid flows, reconfiguring the structure shapes, and changing pressure distributions. In addition, vortices that drop from the upstream cylinder reach the upper surface of the flexible foil as it bends downward, thus reducing the suction effect. The structural flexibility contributes to better lift performance as the heaving motion is gradually synchronized with the vortex-shedding process. Herein, we further analyze the coupling mechanism between wake flows and structural flexibility on both sides of the critical boundary line shown in \reffig{mode} from the point of view of propulsive energy. \refFig{efficiency} presents a comparison of time-varying propulsive power coefficient along the structure chord and its frequency characteristics between the tandem flexible foil and its rigid case for two typical heaving situations. The flexible foil requires more negative power when heaving fully in wake flows, compared with the tandem rigid case (\reffig{efficiency} \subref{efficiencya}). By examining the frequency characteristics in \reffig{efficiency} \subref{efficiencyc}, except for the dominant heaving frequency of 0.2, the flexible foil also responds to the vortex-shedding frequency with redistributed energies due to structural flexibility. This low-power consuming flexible foil generates poorer aerodynamic forces to maintain balance in disturbed wake. The large passive deformation caused by strong added-mass effect results in more power to resist the fluid damping, as noticed from \reffig{efficiency} \subref{efficiencyb}. The propulsion energy of flexible and rigid structures is concentrated in the heaving frequency and its harmonics (\reffig{efficiency} \subref{efficiencyd}). The flexible foil creates a two-way lock-in between the heaving motion and the vortex-shedding frequency by controlling the wake pattern in reverse.

\begin{figure}
	\centering
	\includegraphics[width=1.0 \textwidth]{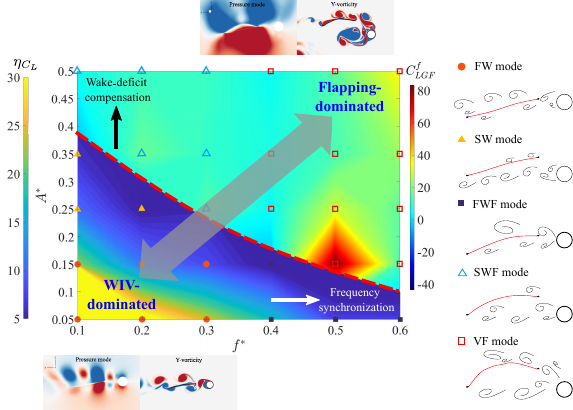}
	\caption{\label{mechanism} Schematic of wake-heaving flexible foil coupling mechanism and propulsive aerodynamic performance in the $f^*$--$A^*$ space. Left: phase diagram of efficiency and lift-gain factor with critical boundary. Right: schematic sketches of the five response modes. The critical boundary line in red color divides contours of propulsive efficiency of lift and lift-gain factor related to flexibility effect. }
\end{figure}

By studying the effects of wake and flexibility on coupled dynamics, we find that disturbed wake can save flapping energy, while structural deformation that synchronously interacts with the oncoming vortex benefits maneuverability. Based on the findings, we summarize a schematic in \reffig{mechanism} to reveal the wake-heaving flexible foil coupling mechanism, thereby suggesting design guidelines and optimal flapping kinematics for human-made flying/swimming vehicles. In \reffig{mechanism}, the propulsive efficiency contour and the lift-gain factor map related to structural flexibility are plotted in the same phase diagram. The critical boundary line in red color governed by $A^*=\gamma (f^*)^{-\frac{1}{n}}+\delta$ is also added in this figure. The positive lift-gain factor contours are displayed on the right top side of the critical boundary line. The optimal propulsive efficiency contours are depicted on the left bottom side. The schematic of five distinct flapping modes is presented in the right panel to build a connection to the coupling mechanism. Along the big gray arrow, the optimal propulsive efficiency is achieved at low heaving amplitude and frequency. In this situation, the coupled dynamics of the flexible foil is mainly governed by vortices shed from the upstream cylinder, stimulating the WIV-dominated mode. By examining the instantaneous vorticity field and the dominant pressure mode in the left bottom corner, the flexible foil encounters the axial momentum deficit of the incoming wake. The proposed scaling relations in \refeqs{eq:cl_non} and (\ref{eq:cpow_non}) indicate that the magnitude of the reduction in the power coefficient by wake is greater than the reduction in lift coefficient, leading to optimal propulsive efficiency in this scenario dominated by WIV.

Follow the large gray arrow diagonally upwards, the foil transitions from wake‑limited (lift reduction) to self‑energized (lift gain) when the transverse induced velocity from heave compensates the axial momentum deficit of the incoming wake over the leading‑edge control volume. The structural flexibility turns the wake from a hindrance into a boost. It can be seen from the flow characteristics displayed on the top that the wake-flexible foil coupling system is dominated by the heaving motion. In SWF mode, the foil’s bending changes the local flow to create extra suction, improving lift without much extra drag. In VF mode, the foil’s motion even talks back to the cylinder, altering its vortex-shedding pattern in a frequency-synchronization manner.

% Implications and discussion of the research results for vehicle design and selection of optimal dynamic characteristics
Based on the findings of this study, the intricate interplay between heaving kinematics, structural flexibility, and wake exploitation mirrors the sophisticated propulsion strategies observed in flying and swimming animals. For instance, the energy-saving WIV-dominated mode, achieved at low heaving amplitude and frequency, is analogous to the practice of birds flying in tight V-formations or fish schooling in dense groups. In these configurations, trailing individuals precisely position themselves in the energy-efficient wake regions shed by leaders, significantly reducing the cost of transport during long-distance migration, a phenomenon directly explained by the axial momentum deficit and optimal propulsive efficiency identified in our FW and SW modes. In contrast, the transition to high-performance flapping-dominated modes (SWF and VF) mirrors explosive escape maneuvers or rapid directional changes seen in bats navigating dense forests or mackerel avoiding predators. These scenarios require animals to break from the wake, actively synchronize their movements with incoming vortical structures, and use their flexible wings or bodies to push back against the flow, thereby generating the high lift forces necessary for agility, a mechanism quantified herein by frequency lock-in, two-way coupling, and the critical transition boundary $A^*=\gamma (f^*)^{-\frac{1}{n}}+\delta$.

These bio-tested strategies provide profound, nature-inspired design guidelines for the next generation of unmanned aerial and underwater vehicles. For endurance-critical missions such as long-range oceanic monitoring, vehicles should operate in a wake-riding mode, leveraging the power-saving benefits of upstream wakes, potentially from other vehicles in a swarm, by employing low-amplitude, low-frequency kinematics. In contrast, for missions that require high maneuverability, such as obstacle avoidance in cluttered environments or rapid acceleration, the vehicle control system should actively transition into a vortex-exploiting mode. This involves increasing heaving amplitude and frequency to cross the critical performance boundary, thereby synchronizing its motion with oncoming vortices, and using flexible appendages to harness energy from the flow, much like a bat modulates its wingbeat. Ultimately, this research advocates for the development of adaptive morphing vehicles capable of intelligently switching between these distinct dynamic regimes, seamlessly blending the energy-saving wisdom of a bird in formation with the agile, vortex-powered escape artistry of a fish.

\section{Conclusions} \label{sec5:conclusion}
This numerical study has systematically elucidated the complex coupled dynamics of a heaving flexible foil interacting with the wake of an upstream circular cylinder. By exploring a wide range of heaving amplitudes and frequencies, we have decoupled the effects of wake flows, active kinematics, and passive flexibility. The system exhibited five distinct dynamical modes—FW, SW, FWF, SWF, and VF—governed by the competition between wake-induced and heaving-induced vibrations. A phase diagram in the $f^*$--$A^*$ parameter space revealed a critical boundary, empirically described by $A^*=\gamma (f^*)^{-\frac{1}{n}}+\delta$, which demarcated a transition from a wake-dominated, lift-reduction regime to a flapping-dominated, lift-gain regime. Unified scaling relations for the time-averaged lift, drag, and power coefficients have been established. These laws quantitatively decomposed the aerodynamic forces into contributions from quasi-steady motion, added-mass effects, structural curvature, wake velocity deficit, and transverse flow gradients, showing reasonable agreement with simulation data. The analysis confirmed that the transition across the critical boundary was fundamentally a balance between the transverse momentum supplied by heaving/flexibility and the axial momentum deficit of the incoming wake.

A novel two-way lock-in phenomenon was identified. While the wake initially dictated the dynamics of the foil at low heaving intensity, the foil reciprocally modulated the vortex shedding of the upstream cylinder at high $f^*$ and $A^*$. This feedback, quantified by high cross-correlation between heaving motion and cylinder vorticity, transformed the wake from a performance hindrance into a boost, leading to significant enhancement in lift and drag forces in the VF mode. The influence of structural flexibility was context-dependent. In fully immersed wakes (FW, FWF modes), flexibility can degrade performance by promoting a reverse bending shape that reduced the effective angle of attack and redistributed energy to unfavorable wake-induced vibration frequencies. In contrast, in partial wakes (SWF mode) and during strong flapping (VF mode), flexibility was highly beneficial. It enhanced the lift-to-drag ratio by streamlining the structure and generating extra leading-edge suction through cambering, without a proportional increase in drag.

The findings mirror sophisticated biological strategies. The energy-saving, WIV-dominated modes at low $A^*$ and $f^*$ are analogous to birds flying in formation or fish schooling, where exploiting upstream wakes reduces the cost of transport. High performance flapping-dominated modes at high $A^*$ and $f^*$ mimic explosive escape maneuvers, where synchronization of motion with oncoming vortices and use of body flexibility generates high forces for agility. This suggests that next-generation morphing aerial and underwater vehicles should employ adaptive kinematics: low-energy, wake-riding modes for endurance, and high-frequency, vortex-exploiting modes for maneuverability. This work provides a foundational framework for understanding and predicting the nonlinear dynamics of flexible foils in complex wake environments. Future studies could extend this approach to three-dimensional configurations, different structural boundary conditions, and turbulent flow regimes to further bridge the gap between fundamental FSI mechanics and real-world engineering applications.

\section*{Acknowledgements}
Guojun Li acknowledges the support from the National Natural Science Foundation of China (NSFC) (Grant Number 12202362), the China Postdoctoral Science Foundation (Grant Number 2023M732798) and the High Performance Computing (HPC) Platform in Xi'an Jiaotong University. This work is funded by the open fund in the State Key Laboratory for Manufacturing Systems Engineering of Xi'an Jiaotong University (Grant Number sklms2023014). Rajeev Kumar Jaiman acknowledges the support of the University of British Columbia and the Natural Sciences and Engineering Research Council of Canada (NSERC).

\section*{Declaration of interests}
The authors report no conflict of interest.

\bibliographystyle{jfm}
\bibliography{reference}

\begin{thebibliography}{55}
\expandafter\ifx\csname natexlab\endcsname\relax\def\natexlab#1{#1}\fi
\def\au#1{#1} \def\ed#1{#1} \def\yr#1{#1}\def\at#1{#1}\def\jt#1{\textit{#1}}
  \def\bt#1{#1}\def\bvol#1{\textbf{#1}} \def\vol#1{#1} \def\pg#1{#1}
  \def\publ#1{#1}\def\arxiv#1{#1}\def\org#1{#1}\def\st#1{\textit{#1}}

\bibitem[Bajec \& Heppner(2009)]{bajec2009organized}
{\sc \au{Bajec, I.~L. } \& \au{Heppner, F.~H. }} \yr{2009}  \at{Organized
  flight in birds}.  \jt{Animal Behaviour}  \bvol{78}~(4),  \pg{777--789}.

\bibitem[Banerjee {\em et~al.\/}(2015)Banerjee, Connell \&
  Yue]{banerjee2015three}
{\sc \au{Banerjee, S. }, \au{Connell, B.~S. } \& \au{Yue, D.~K. }} \yr{2015}
  \at{Three-dimensional effects on flag flapping dynamics}.  \jt{Journal of
  Fluid Mechanics}  \bvol{783},  \pg{103--136}.

\bibitem[Beaumont {\em et~al.\/}(2024)Beaumont, Murer, Bogard \&
  Polidori]{beaumont2024aerodynamic}
{\sc \au{Beaumont, F. }, \au{Murer, S. }, \au{Bogard, F. } \& \au{Polidori, G.
  }} \yr{2024}  \at{The aerodynamic mechanisms of the formation flight of
  migratory birds: A narrative review}.  \jt{Applied Sciences}  \bvol{14}~(13),
   \pg{5402}.

\bibitem[Chen {\em et~al.\/}(2016)Chen, Fei, Chen, Chi \&
  Yang]{chen2016swimming}
{\sc \au{Chen, S. }, \au{Fei, Y.~J. }, \au{Chen, Y. }, \au{Chi, K. } \&
  \au{Yang, J. }} \yr{2016}  \at{The swimming patterns and energy-saving
  mechanism revealed from three fish in a school}.  \jt{Ocean Engineering}
  \bvol{122},  \pg{22--31}.

\bibitem[Deng {\em et~al.\/}(2007)Deng, Shao \& Yu]{deng2007hydrodynamic}
{\sc \au{Deng, J. }, \au{Shao, X.-M. } \& \au{Yu, Z.-S. }} \yr{2007}
  \at{Hydrodynamic studies on two traveling wavy foils in tandem arrangement}.
  \jt{Physics of fluids}  \bvol{19}~(11).

\bibitem[Farooq {\em et~al.\/}(2022)Farooq, Ghommem, Khalid \&
  Akhtar]{farooq2022numerical}
{\sc \au{Farooq, H. }, \au{Ghommem, M. }, \au{Khalid, M. S.~U. } \& \au{Akhtar,
  I. }} \yr{2022}  \at{Numerical investigation of hydrodynamic performance of
  flapping foils for energy harvesting}.  \jt{Ocean Engineering}  \bvol{260},
  \pg{112005}.

\bibitem[Fish \& Lauder(2006)]{fish2006passive}
{\sc \au{Fish, F. } \& \au{Lauder, G.~V. }} \yr{2006}  \at{Passive and active
  flow control by swimming fishes and mammals}.  \jt{Annu. Rev. Fluid Mech.}
  \bvol{38}~(1),  \pg{193--224}.

\bibitem[Floryan \& Rowley(2018)]{floryan2018clarifying}
{\sc \au{Floryan, D. } \& \au{Rowley, C.~W. }} \yr{2018}  \at{Clarifying the
  relationship between efficiency and resonance for flexible inertial
  swimmers}.  \jt{Journal of Fluid Mechanics}  \bvol{853},  \pg{271--300}.

\bibitem[Floryan {\em et~al.\/}(2017)Floryan, Van~Buren, Rowley \&
  Smits]{floryan2017scaling}
{\sc \au{Floryan, D. }, \au{Van~Buren, T. }, \au{Rowley, C.~W. } \& \au{Smits,
  A.~J. }} \yr{2017}  \at{Scaling the propulsive performance of heaving and
  pitching foils}.  \jt{Journal of Fluid Mechanics}  \bvol{822},
  \pg{386--397}.

\bibitem[Floryan {\em et~al.\/}(2018)Floryan, Van~Buren \&
  Smits]{floryan2018efficient}
{\sc \au{Floryan, D. }, \au{Van~Buren, T. } \& \au{Smits, A.~J. }} \yr{2018}
  \at{Efficient cruising for swimming and flying animals is dictated by fluid
  drag}.  \jt{Proceedings of the National Academy of Sciences}
  \bvol{115}~(32),  \pg{8116--8118}.

\bibitem[Furquan \& Mittal(2021)]{furquan2021multiple}
{\sc \au{Furquan, M. } \& \au{Mittal, S. }} \yr{2021}  \at{Multiple lock-ins in
  vortex-induced vibration of a filament}.  \jt{Journal of Fluid Mechanics}
  \bvol{916},  \pg{R1}.

\bibitem[Gazzola {\em et~al.\/}(2014)Gazzola, Argentina \&
  Mahadevan]{gazzola2014scaling}
{\sc \au{Gazzola, M. }, \au{Argentina, M. } \& \au{Mahadevan, L. }} \yr{2014}
  \at{Scaling macroscopic aquatic locomotion}.  \jt{Nature Physics}
  \bvol{10}~(10),  \pg{758--761}.

\bibitem[Gehrke \& Mulleners(2025)]{gehrke2025highly}
{\sc \au{Gehrke, A. } \& \au{Mulleners, K. }} \yr{2025}  \at{Highly deformable
  flapping membrane wings suppress the leading edge vortex in hover to perform
  better}.  \jt{Proceedings of the National Academy of Sciences}
  \bvol{122}~(6),  \pg{e2410833121}.

\bibitem[Han {\em et~al.\/}(2025)Han, Zhang, Zhang \& Huang]{han2025self}
{\sc \au{Han, P. }, \au{Zhang, J.-D. }, \au{Zhang, D. } \& \au{Huang, W.-X. }}
  \yr{2025}  \at{How a self-propelled fin gains hydrodynamic advantages behind
  a circular cylinder with vortex-induced vibrations}.  \jt{Journal of Fluid
  Mechanics}  \bvol{1011},  \pg{A26}.

\bibitem[Hedrick(2024)]{hedrick2024swimming}
{\sc \au{Hedrick, T.~L. }} \yr{2024}  \at{Swimming in a school shelters fish
  from turbulence}.  \jt{PLoS Biology}  \bvol{22}~(6),  \pg{e3002677}.

\bibitem[Huang {\em et~al.\/}(2021)Huang, Xia, Dai, Yang \& Wu]{huang2021fluid}
{\sc \au{Huang, G. }, \au{Xia, Y. }, \au{Dai, Y. }, \au{Yang, C. } \& \au{Wu,
  Y. }} \yr{2021}  \at{Fluid--structure interaction in piezoelectric energy
  harvesting of a membrane wing}.  \jt{Physics of Fluids}  \bvol{33}~(6).

\bibitem[Kumar {\em et~al.\/}(2025)Kumar, Seo \&
  Mittal]{kumar2025computational}
{\sc \au{Kumar, S. }, \au{Seo, J.-H. } \& \au{Mittal, R. }} \yr{2025}
  \at{Computational modelling and analysis of the coupled aero-structural
  dynamics in bat-inspired wings}.  \jt{Journal of Fluid Mechanics}
  \bvol{1010},  \pg{A53}.

\bibitem[Lauber {\em et~al.\/}(2023)Lauber, Weymouth \&
  Limbert]{lauber2023rapid}
{\sc \au{Lauber, M. }, \au{Weymouth, G.~D. } \& \au{Limbert, G. }} \yr{2023}
  \at{Rapid flapping and fibre-reinforced membrane wings are key to
  high-performance bat flight}.  \jt{Journal of the Royal Society Interface}
  \bvol{20}~(208),  \pg{20230466}.

\bibitem[Li {\em et~al.\/}(2021)Li, Jaiman \& Khoo]{li2021flow}
{\sc \au{Li, G. }, \au{Jaiman, R.~K. } \& \au{Khoo, B.~C. }} \yr{2021}
  \at{Flow-excited membrane instability at moderate reynolds numbers}.
  \jt{Journal of Fluid Mechanics}  \bvol{929},  \pg{A40}.

\bibitem[Li {\em et~al.\/}(2022)Li, Jaiman \& Khoo]{li2022aeroelastic}
{\sc \au{Li, G. }, \au{Jaiman, R.~K. } \& \au{Khoo, B.~C. }} \yr{2022}
  \at{Aeroelastic mode decomposition framework and mode selection mechanism in
  fluid--membrane interaction}.  \jt{Journal of Fluids and Structures}
  \bvol{108},  \pg{103428}.

\bibitem[Li \& Kumar~Jaiman(2023)]{li2023unsteady}
{\sc \au{Li, G. } \& \au{Kumar~Jaiman, R. }} \yr{2023}  \at{Unsteady
  aeroelastic characterization and scaling relations of flexible membrane
  wings}.  \jt{AIAA Journal}  \bvol{61}~(11),  \pg{5042--5060}.

\bibitem[Li {\em et~al.\/}(2019)Li, Law \& Jaiman]{li2019novel}
{\sc \au{Li, G. }, \au{Law, Y.~Z. } \& \au{Jaiman, R.~K. }} \yr{2019}  \at{A
  novel 3d variational aeroelastic framework for flexible multibody dynamics:
  Application to bat-like flapping dynamics}.  \jt{Computers \& Fluids}
  \bvol{180},  \pg{96--116}.

\bibitem[Li {\em et~al.\/}(2024)Li, Zhang, Lei, Wang, Jiang \&
  Liu]{li2024enhancing}
{\sc \au{Li, G. }, \au{Zhang, H. }, \au{Lei, B. }, \au{Wang, L. }, \au{Jiang,
  W. } \& \au{Liu, H. }} \yr{2024}  \at{Enhancing stability and performance of
  flexible membrane structures in wake flows through jet flow control}.
  \jt{Ocean Engineering}  \bvol{313},  \pg{119366}.

\bibitem[Ligman {\em et~al.\/}(2023)Ligman, Lund \&
  F{\"u}rth]{ligman2023comprehensive}
{\sc \au{Ligman, M. }, \au{Lund, J. } \& \au{F{\"u}rth, M. }} \yr{2023}  \at{A
  comprehensive review of hydrodynamic studies on fish schooling}.
  \jt{Bioinspiration \& Biomimetics}  \bvol{19}~(1),  \pg{011002}.

\bibitem[Liu {\em et~al.\/}(2024)Liu, Wang \& Liu]{liu2024vortices}
{\sc \au{Liu, H. }, \au{Wang, S. } \& \au{Liu, T. }} \yr{2024}  \at{Vortices
  and forces in biological flight: Insects, birds, and bats}.  \jt{Annual
  Review of Fluid Mechanics}  \bvol{56}~(1),  \pg{147--170}.

\bibitem[Maertens {\em et~al.\/}(2017)Maertens, Gao \&
  Triantafyllou]{maertens2017optimal}
{\sc \au{Maertens, A.~P. }, \au{Gao, A. } \& \au{Triantafyllou, M.~S. }}
  \yr{2017}  \at{Optimal undulatory swimming for a single fish-like body and
  for a pair of interacting swimmers}.  \jt{Journal of Fluid Mechanics}
  \bvol{813},  \pg{301--345}.

\bibitem[Mavroyiakoumou \& Alben(2020)]{mavroyiakoumou2020large}
{\sc \au{Mavroyiakoumou, C. } \& \au{Alben, S. }} \yr{2020}
  \at{Large-amplitude membrane flutter in inviscid flow}.  \jt{Journal of Fluid
  Mechanics}  \bvol{891},  \pg{A23}.

\bibitem[Mavroyiakoumou \& Alben(2022)]{mavroyiakoumou2022membrane}
{\sc \au{Mavroyiakoumou, C. } \& \au{Alben, S. }} \yr{2022}  \at{Membrane
  flutter in three-dimensional inviscid flow}.  \jt{Journal of Fluid Mechanics}
   \bvol{953},  \pg{A32}.

\bibitem[Mavroyiakoumou \& Alben(2025)]{mavroyiakoumou2025sail}
{\sc \au{Mavroyiakoumou, C. } \& \au{Alben, S. }} \yr{2025}  \at{Sail dynamics
  during tacking maneuvers}.  \jt{Physical Review Fluids}  \bvol{10}~(7),
  \pg{073901}.

\bibitem[Menon \& Mittal(2020)]{menon2020dynamic}
{\sc \au{Menon, K. } \& \au{Mittal, R. }} \yr{2020}  \at{Dynamic mode
  decomposition based analysis of flow over a sinusoidally pitching airfoil}.
  \jt{Journal of Fluids and Structures}  \bvol{94},  \pg{102886}.

\bibitem[Michelin {\em et~al.\/}(2008)Michelin, Smith \&
  Glover]{michelin2008vortex}
{\sc \au{Michelin, S. }, \au{Smith, S. G.~L. } \& \au{Glover, B.~J. }}
  \yr{2008}  \at{Vortex shedding model of a flapping flag}.  \jt{Journal of
  Fluid Mechanics}  \bvol{617},  \pg{1--10}.

\bibitem[Moore(2014)]{moore2014analytical}
{\sc \au{Moore, M. N.~J. }} \yr{2014}  \at{Analytical results on the role of
  flexibility in flapping propulsion}.  \jt{Journal of fluid mechanics}
  \bvol{757},  \pg{599--612}.

\bibitem[Newbolt {\em et~al.\/}(2024)Newbolt, Lewis, Bleu, Wu, Mavroyiakoumou,
  Ramananarivo \& Ristroph]{newbolt2024flow}
{\sc \au{Newbolt, J.~W. }, \au{Lewis, N. }, \au{Bleu, M. }, \au{Wu, J. },
  \au{Mavroyiakoumou, C. }, \au{Ramananarivo, S. } \& \au{Ristroph, L. }}
  \yr{2024}  \at{Flow interactions lead to self-organized flight formations
  disrupted by self-amplifying waves}.  \jt{Nature communications}
  \bvol{15}~(1),  \pg{3462}.

\bibitem[Parekh \& Jaiman(2025)]{parekh2025wake}
{\sc \au{Parekh, A.~R. } \& \au{Jaiman, R.~K. }} \yr{2025}  \at{Wake
  interference effects on flapping dynamics of elastic inverted foil}.
  \jt{Physical Review Fluids}  \bvol{10}~(1),  \pg{014702}.

\bibitem[Quinn {\em et~al.\/}(2014)Quinn, Moored, Dewey \&
  Smits]{quinn2014unsteady}
{\sc \au{Quinn, D.~B. }, \au{Moored, K.~W. }, \au{Dewey, P.~A. } \& \au{Smits,
  A.~J. }} \yr{2014}  \at{Unsteady propulsion near a solid boundary}.
  \jt{Journal of Fluid Mechanics}  \bvol{742},  \pg{152--170}.

\bibitem[Ramananarivo {\em et~al.\/}(2016)Ramananarivo, Fang, Oza, Zhang \&
  Ristroph]{ramananarivo2016flow}
{\sc \au{Ramananarivo, S. }, \au{Fang, F. }, \au{Oza, A. }, \au{Zhang, J. } \&
  \au{Ristroph, L. }} \yr{2016}  \at{Flow interactions lead to orderly
  formations of flapping wings in forward flight}.  \jt{Physical Review Fluids}
   \bvol{1}~(7),  \pg{071201}.

\bibitem[Ristroph \& Zhang(2008)]{ristroph2008anomalous}
{\sc \au{Ristroph, L. } \& \au{Zhang, J. }} \yr{2008}  \at{Anomalous
  hydrodynamic drafting of interacting flapping flags}.  \jt{Physical review
  letters}  \bvol{101}~(19),  \pg{194502}.

\bibitem[Sarpkaya(2004)]{sarpkaya2004critical}
{\sc \au{Sarpkaya, T. }} \yr{2004}  \at{A critical review of the intrinsic
  nature of vortex-induced vibrations}.  \jt{Journal of fluids and structures}
  \bvol{19}~(4),  \pg{389--447}.

\bibitem[Shao \& Pan(2011)]{shao2011hydrodynamics}
{\sc \au{Shao, X. } \& \au{Pan, D. }} \yr{2011}  \at{Hydrodynamics of a
  flapping foil in the wake of a d-section cylinder}.  \jt{Journal of
  Hydrodynamics, Ser. B}  \bvol{23}~(4),  \pg{422--430}.

\bibitem[Shinde \& Gaitonde(2021)]{shinde2021lagrangian}
{\sc \au{Shinde, V.~J. } \& \au{Gaitonde, D.~V. }} \yr{2021}  \at{Lagrangian
  approach for modal analysis of fluid flows}.  \jt{Journal of Fluid Mechanics}
   \bvol{928},  \pg{A35}.

\bibitem[Song {\em et~al.\/}(2016)Song, Fu, Dai, Zhang \&
  Chen]{song2016distribution}
{\sc \au{Song, L. }, \au{Fu, S. }, \au{Dai, S. }, \au{Zhang, M. } \& \au{Chen,
  Y. }} \yr{2016}  \at{Distribution of drag force coefficient along a flexible
  riser undergoing viv in sheared flow}.  \jt{Ocean Engineering}  \bvol{126},
  \pg{1--11}.

\bibitem[Theodorsen(1935)]{1935General}
{\sc \au{Theodorsen, T. }} \yr{1935}  \at{General theory of aerodynamic
  instability and the mecism of flutter}.  \jt{Naca} .

\bibitem[Tian~Dong {\em et~al.\/}(2021)Tian~Dong, Rui, Long, Yu \&
  Shuo]{tian2021research}
{\sc \au{Tian~Dong, Z. }, \au{Rui, W. }, \au{Long, C. }, \au{Yu, W. } \&
  \au{Shuo, W. }} \yr{2021}  \at{Research on energy-saving mechanism of fish
  schooling: A review}.  \jt{Acta Automatica Sinica}  \bvol{47}~(3),
  \pg{475--488}.

\bibitem[Tiomkin \& Jaworski(2022)]{tiomkin2022unsteady}
{\sc \au{Tiomkin, S. } \& \au{Jaworski, J.~W. }} \yr{2022}  \at{Unsteady
  aerodynamic theory for membrane wings}.  \jt{Journal of Fluid Mechanics}
  \bvol{948},  \pg{A33}.

\bibitem[Tiomkin \& Raveh(2021)]{tiomkin2021review}
{\sc \au{Tiomkin, S. } \& \au{Raveh, D.~E. }} \yr{2021}  \at{A review of
  membrane-wing aeroelasticity}.  \jt{Progress in Aerospace Sciences}
  \bvol{126},  \pg{100738}.

\bibitem[Triantafyllou \& Triantafyllou(1995)]{triantafyllou1995efficient}
{\sc \au{Triantafyllou, M.~S. } \& \au{Triantafyllou, G.~S. }} \yr{1995}
  \at{An efficient swimming machine}.  \jt{Scientific american}
  \bvol{272}~(3),  \pg{64--70}.

\bibitem[Turek \& Hron(2006)]{turek2006proposal}
{\sc \au{Turek, S. } \& \au{Hron, J. }} \yr{2006} {\em Proposal for numerical
  benchmarking of fluid-structure interaction between an elastic object and
  laminar incompressible flow\/}.  \publ{Springer}.

\bibitem[Tzezana \& Breuer(2019)]{tzezana2019thrust}
{\sc \au{Tzezana, G.~A. } \& \au{Breuer, K.~S. }} \yr{2019}  \at{Thrust, drag
  and wake structure in flapping compliant membrane wings}.  \jt{Journal of
  Fluid Mechanics}  \bvol{862},  \pg{871--888}.

\bibitem[Upfal {\em et~al.\/}(2025)Upfal, Zhu, Handy-Cardenas \&
  Breuer]{upfal2025shape}
{\sc \au{Upfal, I.~M. }, \au{Zhu, Y. }, \au{Handy-Cardenas, E. } \& \au{Breuer,
  K. }} \yr{2025}  \at{Shape-morphing membranes augment the performance of
  oscillating foil energy harvesting turbines}.  \jt{Physical Review Fluids}
  \bvol{10}~(3),  \pg{034702}.

\bibitem[Van~Buren {\em et~al.\/}(2017)Van~Buren, Floryan, Brunner, Senturk \&
  Smits]{van2017impact}
{\sc \au{Van~Buren, T. }, \au{Floryan, D. }, \au{Brunner, D. }, \au{Senturk, U.
  } \& \au{Smits, A. }} \yr{2017}  \at{Impact of trailing edge shape on the
  wake and propulsive performance of pitching panels}.  \jt{Physical Review
  Fluids}  \bvol{2}~(1),  \pg{014702}.

\bibitem[Waldman \& Breuer(2017)]{waldman2017camber}
{\sc \au{Waldman, R.~M. } \& \au{Breuer, K.~S. }} \yr{2017}  \at{Camber and
  aerodynamic performance of compliant membrane wings}.  \jt{Journal of Fluids
  and Structures}  \bvol{68},  \pg{390--402}.

\bibitem[Wang {\em et~al.\/}(2024)Wang, Du \& Sun]{wang2024thrust}
{\sc \au{Wang, Z. }, \au{Du, L. } \& \au{Sun, X. }} \yr{2024}  \at{Thrust
  enhancement of a flapping foil through interaction with a k{\'a}rm{\'a}n
  vortex street}.  \jt{Physics of Fluids}  \bvol{36}~(7).

\bibitem[Zhang {\em et~al.\/}(2025{\natexlab{{\em a\/}}})Zhang, Gao \&
  Zhu]{zhang2025flapping}
{\sc \au{Zhang, C. }, \au{Gao, A. } \& \au{Zhu, X. }} \yr{2025{\natexlab{{\em
  a\/}}}}  \at{Flapping dynamics of a compliant membrane in a uniform incoming
  flow}.  \jt{arXiv preprint arXiv:2508.00112} .

\bibitem[Zhang {\em et~al.\/}(2025{\natexlab{{\em b\/}}})Zhang, Zhang, Yin, Xia
  \& Tan]{zhang2025performance}
{\sc \au{Zhang, D. }, \au{Zhang, X. }, \au{Yin, P. }, \au{Xia, C. } \& \au{Tan,
  P. }} \yr{2025{\natexlab{{\em b\/}}}}  \at{Performance enhancement of
  piezoelectric energy harvesters based on flow-induced vibration via
  bio-inspired design}.  \jt{Physics of Fluids}  \bvol{37}~(8).

\bibitem[Zhang \& Lauder(2024)]{zhang2024energy}
{\sc \au{Zhang, Y. } \& \au{Lauder, G.~V. }} \yr{2024}  \at{Energy conservation
  by collective movement in schooling fish}.  \jt{Elife}  \bvol{12},
  \pg{RP90352}.

\end{thebibliography}

\section*{Appendix A. Variational formulation of governing equations}
\setcounter{equation}{0}
\setcounter{figure}{0}
\setcounter{table}{0}
\renewcommand{\theequation}{A.\arabic{equation}}
\renewcommand{\thefigure}{A.\arabic{figure}}
\renewcommand{\thetable}{A.\arabic{table}}
The fluid loads are calculated by the variational formulation of the Navier-Stokes equations in the fluid domain $\Omega^f$ at the time instant $t^{n+1}$, which is given as
\begin{eqnarray}
	\int_{\Omega^f(t^{n+1})} \rho^f \left( \partial_t {\boldsymbol{u}}^{f,n+\alpha^f_m}_h \bigg|_{\boldsymbol{\chi}} + ({\boldsymbol{u}}^{f,n+\alpha^f}_h - \boldsymbol{u}^{m,n+\alpha^f}_h) \cdot \nabla {\boldsymbol{u}}^{f,n+\alpha^f}_h \right) \cdot \boldsymbol{\psi}^f_h {\rm{d}\Omega} \nonumber\\
	+\int_{\Omega^f(t^{n+1})} {\boldsymbol{\sigma}}^{f,n+\alpha^f}_h:\nabla \boldsymbol{\psi}^f_h {\rm{d}\Omega}  \nonumber\\
	+ \sum_{e=1}^{n^f_{el}} \int_{\Omega^e} \tau_m (\rho^f ({\boldsymbol{u}}^{f,n+\alpha^f}_h-\boldsymbol{u}^{m,n+\alpha^f}_h) \cdot \nabla \boldsymbol{\psi}^f_h+\nabla q_h) \cdot \boldsymbol{\mathcal{R}}_m {\rm{d}\Omega^e} \nonumber\\
	-\int_{\Omega^f(t^{n+1})} q_h (\nabla \cdot {\boldsymbol{u}}^{f,n+\alpha^f}_h) {\rm{d}\Omega}  +\sum_{e=1}^{n^f_{el}} \int_{\Omega^e} \nabla \cdot \boldsymbol{\psi}^f_h \tau_c \boldsymbol{\mathcal{R}}_c {\rm{d}\Omega^e} \nonumber\\
	-\sum_{e=1}^{n^f_{el}} \int_{\Omega^e} \tau_m \boldsymbol{\psi}^f_h \cdot (\boldsymbol{\mathcal{R}}_m \cdot \nabla {\boldsymbol{u}}^{f,n+\alpha^f}_h) {\rm{d}\Omega^e} - \sum_{e=1}^{n^f_{el}} \int_{\Omega^e} \nabla \boldsymbol{\psi}^f_h : (\tau_m \boldsymbol{\mathcal{R}}_m \otimes \tau_m \boldsymbol{\mathcal{R}}_m) {\rm{d}\Omega^e} \nonumber\\
	= \int_{\Omega^f(t^{n+1})} \boldsymbol{b}^f(t^{n+\alpha^f}) \cdot \boldsymbol{\psi}^f_h {\rm{d}\Omega} + \int_{\Gamma^f_N} \boldsymbol{h}^f \cdot \boldsymbol{\psi}^f_h {\rm{d} \Gamma} ,
	\label{eq:eqGA6} 
\end{eqnarray}
where $\boldsymbol{u}^{f}_h$ and $\boldsymbol{u}^{m}_h$ denote fluid and mesh velocities. The integration coefficients of the generalized-$\alpha$ algorithm are defined as $\alpha^f$ and $\alpha^f_m$. Unsteady fluid flows with a density of $\rho^f$ produce body forces $\boldsymbol{b}^f$ on immersed structures. $\boldsymbol{\psi}^f_h$ and $q_h$ are weighting functions of the fluid velocity $\boldsymbol{u}^{f}_h$ and pressure $p_h$ in the Petrov-Galerkin finite element method. The second line represents the Galerkin term for the viscous stress term ${\boldsymbol{\sigma}}^{f}_h$. The Neumann condition $\boldsymbol{h}^f$ is applied at the boundary of the computational fluid domain. $\boldsymbol{\mathcal{R}}_c$ and $\boldsymbol{\mathcal{R}}_m$ denote the element-wise residuals of the continuity and momentum equations. The stabilization parameters for these two equations are represented by $\tau_c$ and $\tau_m$. 

The kinematic flapping motion of flexible foils is described by the flexible multibody structure equations. The variational form of the structure equation for the $i$-th multibody component $\Omega^s_i$ can be given as
\begin{eqnarray}
	\int_{t_n}^{t^{n+1}} \left( \int_{\Omega^s_i} \rho^s \frac{\partial^2 \boldsymbol{d}^s_h}{\partial t^2} \cdot \boldsymbol{\psi}^s_h  \text{d}\Omega +   \int_{\Omega^s_i} \boldsymbol{P}^s : \boldsymbol{\nabla} \boldsymbol{\psi}^s_h \text{d}\Omega \right) \text{d} t	\nonumber\\
	=  \int_{t_n}^{t^{n+1}} \left( \int_{\Omega^s_i} \boldsymbol{b}^s \cdot \boldsymbol{\psi}^s_h \text{d}\Omega + \int_{\Gamma^s_i} \boldsymbol{h}^s \cdot \boldsymbol{\psi}^s_h \text{d}\Gamma  \right) \text{d} t ,
\end{eqnarray}
where the material density and the displacement of the flexible foil are defined as $\rho^s$ and $\boldsymbol{d}^s_h$. The first Piola-Kirchhoff stress tensor and the structural body force are denoted by $\boldsymbol{P}^s$ and $\boldsymbol{b}^s$. Herein, $\boldsymbol{\psi}^s_h$ and $\boldsymbol{h}^s$ represent weighting functions of structure displacement and the Neumann condition of the $i$-th multibody structure. 

A partitioned iterative scheme based on the proposed NIFC approach is employed to couple the fluid and structure equations. In the predictor-corrector process, the predicted structural displacements and the corrected fluid forces are solved in an iterative manner. In the first step of the coupling procedure, the structural displacements at each structural node are calculated from the flexible multibody structural equations. Then, structural displacements are transferred to the surface nodes in the fluid domain along the non-matching fluid-structure interface by using the RBF method. The continuity condition of the displacement and velocity between two domains should be satisfied to avoid gap or overlap of the two types of meshes, which are given as
\begin{equation}
	\boldsymbol{d}^{s}=\boldsymbol{d}^{m} = \boldsymbol{d}^f \quad \forall \boldsymbol{x}^s \in \Gamma^{fs},
\end{equation}
\begin{equation}
	\boldsymbol{u}^{f}= \boldsymbol{u}^{m} = \boldsymbol{u}^s \quad \forall \boldsymbol{x}^f \in \Gamma^{fs},
\end{equation}

In the third step, the meshes in the fluid domain are updated following the continuity condition of the body-fitted meshes on the surface through the RBF remeshing approach. Then, the fluid forces acting on the flexible foils are computed from the Navier-Stokes equations for the remeshed configuration. In the last step, the fluid forces are corrected by the NIFC scheme to achieve numerical stability under large added-mass effects. The corrected fluid forces are transferred to the structural surfaces by using the RBF method. The continuity of the shear stress should be satisfied along the interface $\Gamma^{fs}_i$, which is given as
\begin{equation}
	\int_{\boldsymbol{\varphi}^s(\gamma,t)}{\boldsymbol{\sigma}}^f(\boldsymbol{x}^f,t) \cdot \boldsymbol{n}^f \text{d}\Gamma+\int_{\gamma} \boldsymbol{P}^s(\boldsymbol{X}^s,t) \cdot \boldsymbol{n}^s \text{d}\Gamma=0  \quad \quad \forall \gamma \in \Gamma^{fs}_i ,
\end{equation}
where $\boldsymbol{\varphi}^s(\gamma,t)$ denotes the associated fluid domain corresponding to the $i$-th structure at time instant $t$. The unit normal vectors of the interface in the fluid and structural domains are defined as $\boldsymbol{n}^f$ and $\boldsymbol{n}^s$. 

\section*{Appendix B. Experimental validation}
\setcounter{equation}{0}
\setcounter{figure}{0}
\setcounter{table}{0}
\renewcommand{\theequation}{B.\arabic{equation}}
\renewcommand{\thefigure}{B.\arabic{figure}}
\renewcommand{\thetable}{B.\arabic{table}}

\subsection*{B1: Flapping dynamics of pitching flexible foil}
\begin{figure}
	\centering
	\subfloat[]{\includegraphics[width=0.4 \textwidth]{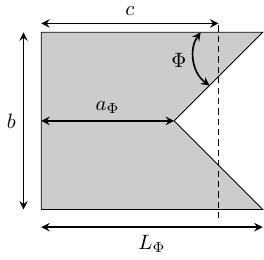}\label{plate_geoa}}
	\quad
	\quad
	\quad
	\subfloat[]{\includegraphics[width=0.5 \textwidth]{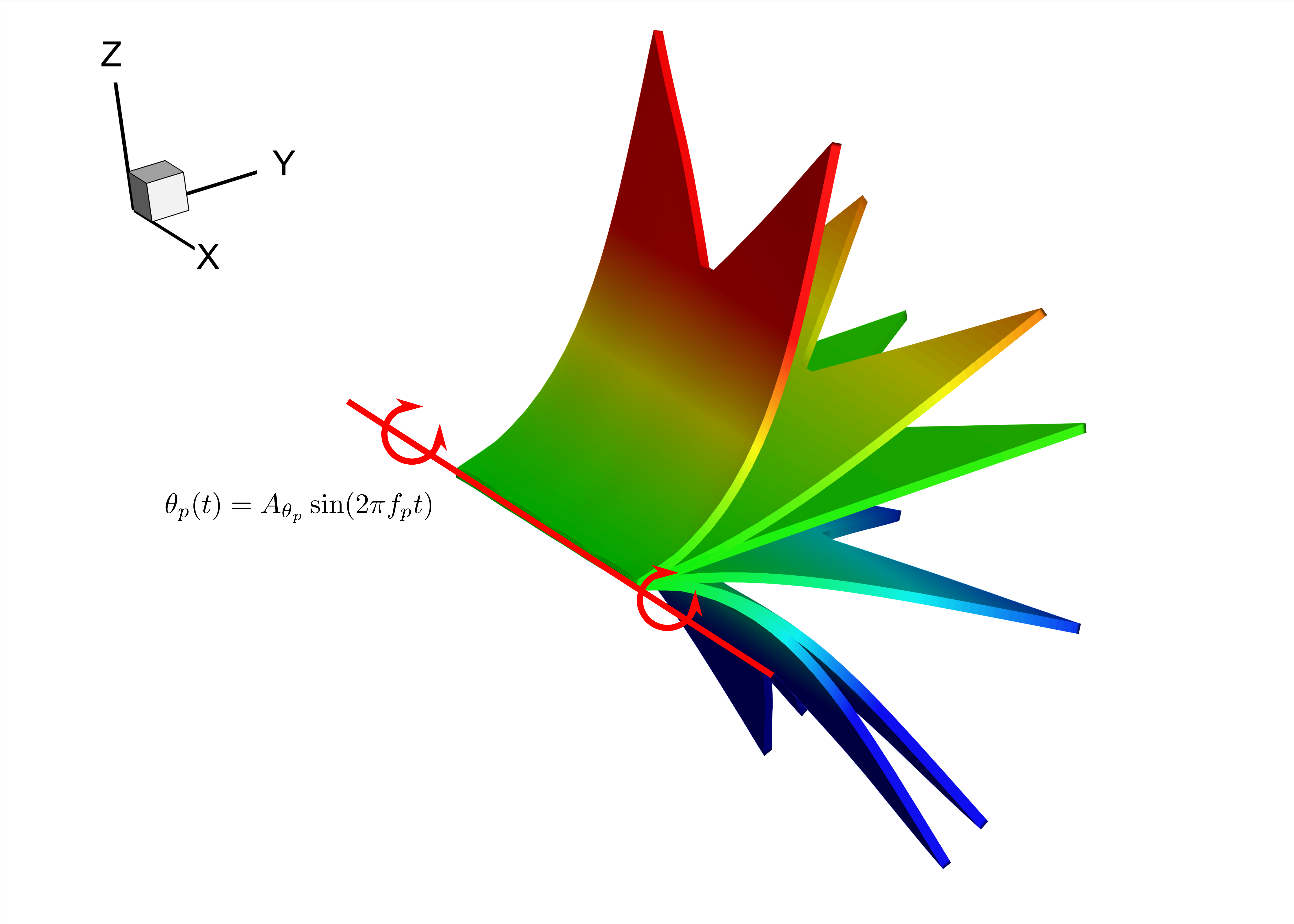}\label{plate_geob}}
	\caption{\label{plate_geo} Illustration of (a) flexible foil geometry and (b) pitching motion applied at leading edge.}
\end{figure}

\begin{figure}
	\centering
	\subfloat[]{\includegraphics[width=0.5 \textwidth]{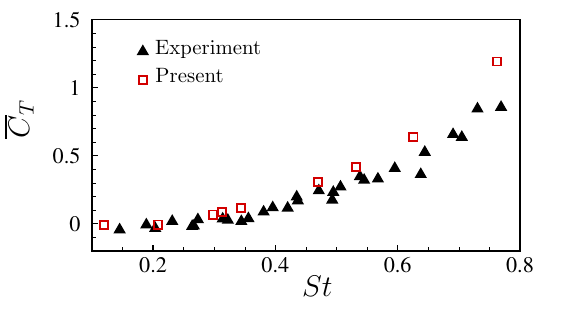}\label{validationa}}
	\subfloat[]{\includegraphics[width=0.5 \textwidth]{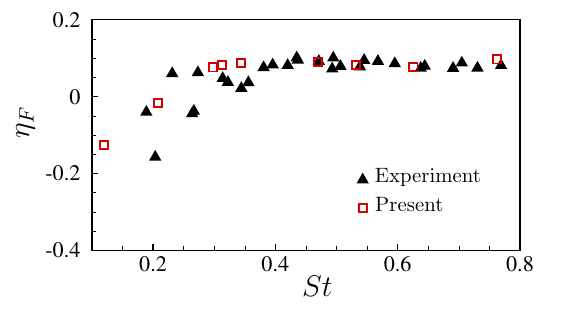}\label{validationb}}
	\caption{\label{validation} Comparison of (a) time-averaged trust coefficient and (b) propulsive efficiency between experimental results and present simulation data.}
\end{figure}

A thin flexible foil immersed in uniform flows is considered for validation purposes. A pitching motion is applied along the leading edge of the foil. The propulsive performance of the flapping foil is simulated by our high-fidelity FSI solver and compared to the experimental measurements. The geometry and sizes of the thin foil are built from the water tunnel experiment performed by \cite{van2017impact}, which is shown in \reffig{plate_geo}. The width of the foil is $b$=0.1 m and the mean chord length is set to $c$=0.1 m. The concave angle at the trailing edge is $\Phi$=$45^\circ
$. The thickness is $h$=2.54 $\times$ $10^{-3}$ m and the Young's modulus is $E^s$=3.1 $\times$ $10^9$ N/m$^2$. The dimensionless parameters that govern the flapping dynamics are set to $Re$=10000 and $m^*$=0.03. The amplitude of the pitching angle is given as $A_{\theta_p}$=$9^\circ$. The flapping dynamics of the flexible foil are simulated within the range of $St \in [0.12, 0.76]$. A comparison of the time-averaged thrust coefficient $\overline{C}_T$ and the propulsive efficiency $\eta_F$ between the experimental data and our simulation results is presented in \reffig{validation}. The flapping dynamics of the flexible foil are well predicted by our FSI solver.

\subsection*{B2: Flapping dynamics of thin flexible foil behind a circular cylinder}
\begin{figure}
	\centering
	\subfloat[]{\includegraphics[width=0.5 \textwidth]{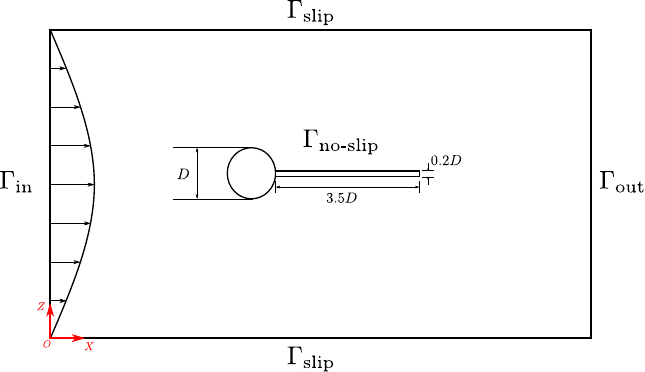}\label{cylinder_validationa}}
	\subfloat[]{\includegraphics[width=0.5 \textwidth]{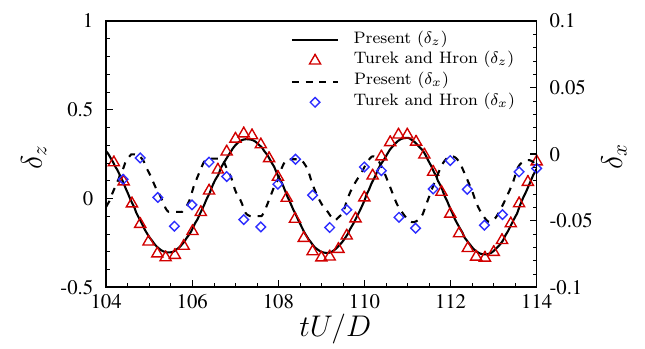}\label{cylinder_validationb}}
	\caption{\label{cylinder_validation} (a) Schematic of computational fluid domain and (b) time-varying foil tip displacements along the vertical and horizontal directions.}
\end{figure}

The flapping dynamics of a thin foil attached to a circular cylinder are simulated by our FSI solver. A comparison with the benchmark data of FSI-III setup is performed for validation purposes. The ratios of the thickness and length of the flexible foil relative to the cylinder diameter $D$ are 0.2 and 3.5 respectively. In the FSI-III setup, the Reynolds number and mass ratio are given as $Re$=200 and $m^*$=1.0. A schematic of the computational fluid domain is plotted in \reffig{cylinder_validation} \subref{cylinder_validationa}. We apply a parabolic velocity profile along the inlet boundary. Traction-free boundary condition is applied along the outlet boundary. The slip wall condition is employed on the side walls and the no-slip wall condition is applied on the foil surface. A comparison of foil tip displacements along the vertical and horizontal directions between the reference data \citep{turek2006proposal} and our simulation results is presented in \reffig{cylinder_validation} \subref{cylinder_validationb}. It can be noticed that the flapping dynamics of the foil is well captured by our FSI solver.

\section*{Appendix C. Convergence study}
\setcounter{equation}{0}
\setcounter{figure}{0}
\setcounter{table}{0}
\renewcommand{\theequation}{C.\arabic{equation}}
\renewcommand{\thefigure}{C.\arabic{figure}}
\renewcommand{\thetable}{C.\arabic{table}}

\begin{figure}
	\centering
	\subfloat[]{\includegraphics[width=0.5 \textwidth]{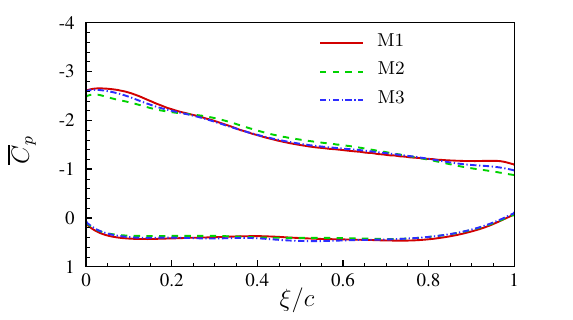}\label{convergencea}}
	\subfloat[]{\includegraphics[width=0.5 \textwidth]{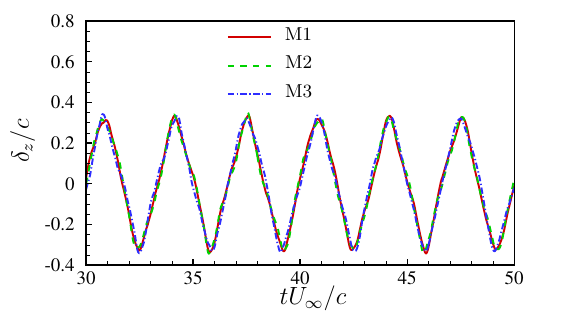}\label{convergenceb}}
	\caption{\label{convergence} Comparison of (a) time-averaged pressure coefficient distribution along the foil surface and (b) time-varying structure displacement at the middle position of the foil for three mesh sets.}
\end{figure}

Before proceeding to investigating the flapping dynamics of flexible foils in disturbed flows, we perform a mesh convergence study to select appropriate mesh resolutions for numerical simulations. Three different mesh sets, namely M1, M2 and M3, are built for the fluid and structure domains, respectively. In the mesh convergence study, moderate flapping amplitude $A^*$=0.25 and frequency $f^*$=0.3 are selected at the SWF mode. The computational fluid domain is discretized by 84995, 153641, 211679 unstructured triangular elements for three mesh sets, respectively. The element numbers of the flexible foil are set to 30, 50, 80, respectively. A comparison of the aerodynamic performance of the heaving flexible foil is summarized in \reftab{tab:mesh convergence}. The aerodynamic forces of M3 are selected as a reference to calculate the relative percentage differences. The maximum difference between M1 and M3 is observed for the r.m.s value of the drag fluctuation more than $5\%$. The absolute differences of force characteristics between M2 and M3 are all less than $1\%$. We further compare the time-averaged pressure coefficient distribution along the foil and time-varying structure displacement at the middle position for three mesh sets in \reffig{convergence}. The coupled dynamics of M2 show a reasonable agreement with those of M3. Therefore, M2 is selected in further numerical simulations with adequate mesh resolution.

\begin{table}
	\begin{center}
		\begin{tabular}{ccccccc}
			Mesh & M1    &    M2    &   M3 \\[3pt]
			Mean lift $\overline{C}_L$ & 1.9904 ($1.2154\%$)  & 1.9680 ($0.0763\%$) & 1.9665 \\
			Mean drag $\overline{C}_D$  & 0.3281 ($0.6442\%$)  & 0.3279 ($0.5828\%$) & 0.3260 \\
			Mean lift-to-drag ratio $\overline{C_L/C_D}$  & 6.0664 ($0.5675\%$) & 6.0018 ($-0.5036\%$) & 6.0322 \\
			r.m.s. lift fluctuation ${C_L^{\prime}}^{rms}$  & 2.0333  ($3.2814\%$)  &  1.9662 ($-0.1270\%$) & 1.9687 \\
			r.m.s. drag fluctuation ${C_D^{\prime}}^{rms}$  & 0.2950 ($5.8106\%$)  & 0.2764 ($-0.8608\%$) & 0.2788 \\
		\end{tabular}
		\caption{Mesh convergence of a heaving flexible foils immersed in the wake flows of a stationary circular cylinder at $A^*$=0.25 and $f^*$=0.3. The percentage discrepancies in the brackets are calculated with respect to the finest mesh M3.}{\label{tab:mesh convergence}}
	\end{center}
\end{table}

\end{document}